\documentclass[12pt]{article}
\usepackage{xcolor}
\usepackage{amsfonts}
\usepackage{amssymb}
\usepackage{amsmath}
\usepackage{amsthm}
\usepackage{graphicx}
\usepackage{natbib}
\usepackage{rotating}
\usepackage{setspace}
\usepackage{arydshln}
\usepackage{stackengine}
\usepackage{epstopdf}
\usepackage{verbatim}
\usepackage{float}
\usepackage[utf8]{inputenc}
\usepackage{booktabs}
\usepackage[font={small}]{caption}
\usepackage{pdflscape}
\usepackage{cmap}
\usepackage{varwidth}
\usepackage[flushleft]{threeparttable}
\usepackage{multirow}
\usepackage{multicol}
\usepackage{float}
\usepackage[flushleft]{threeparttable}
\usepackage{tikz}
\usetikzlibrary{arrows,positioning}
\usepackage{lipsum}

\usepackage{caption}
\usepackage{subcaption}
\usepackage{float}

\usepackage{xcolor,colortbl}
\definecolor{Cinza}{gray}{0.85}
\definecolor{Azul}{RGB}{102,175,255}
\definecolor{section.color}{HTML}{292B2E}
\definecolor{subsection.color}{HTML}{31363A}
\definecolor{subsubsection.color}{HTML}{AF5F30}
\definecolor{cor1}{HTML}{F37878}
\definecolor{cor2}{HTML}{91BDFC}

\usepackage{hyperref}
\usepackage{cleveref}
\usepackage{geometry}
\begin{document}

\begin{titlepage}
\title{Lockdown effects in US states: an artificial counterfactual approach}

\author{
    Carlos B. Carneiro\\
	\hspace{7mm}University of Brasilia \hspace{8mm}
	\and
	Iúri H. Ferreira\\
	Pontificial Catholic University of Rio de Janeiro
	\and
    Marcelo C. Medeiros\footnote{Updated versions of this paper will be available on the corresponding author's webpage.}\\
	Pontificial Catholic University of Rio de Janeiro
	\and
	Henrique F. Pires\\
	Pontificial Catholic University of Rio de Janeiro
	\and
	Eduardo Zilberman\\
	Pontificial Catholic University of Rio de Janeiro
	}

\date{\small{First version: June 2020 \\ This version: January 2021}}

\maketitle
\thispagestyle{empty}

\onehalfspacing

\begin{abstract}

We adopt an artificial counterfactual approach to assess the impact of lockdowns on the short-run evolution of the number of cases and deaths in some US states. To do so, we explore the different timing in which US states adopted lockdown policies, and divide them among treated and control groups. For each treated state, we construct an artificial counterfactual. On average, and in the very short-run,  the counterfactual accumulated number of cases would be two times larger if lockdown policies were not implemented.

\bigskip

\noindent
\textbf{Keywords:} Covid-19 cases, lockdown effects, mobility, ArCo, synthetic control.
\vspace{5mm}
\end{abstract}
\end{titlepage}
\clearpage

\section{Introduction}
\label{sec:introduction}
The evolution of the Covid-19 has been posing several challenges to policymakers. Decisions have to be made in a timely fashion, without much undisputed evidence to support them. Being a new disease, and despite the enormous research effort to understand it, estimates of the transmission, recovery and death rates remain uncertain. Nevertheless, these are key pieces of information to assess potential pressures on the health system capacity, as well as the need of a lockdown policy and its intensity if implemented.

Not surprisingly, similar regions have implemented different strategies regarding lockdowns. The leading example in the media is the looser social distancing policy in Sweden versus strict policies in its Scandinavian peers. By informally comparing the evolution of the pandemics in Sweden and Denmark (or Norway), many commentators argue that several Covid-19 cases and deaths in Sweden would be avoided in the short-run were a strict lockdown in place.\footnote{\cite{juranek2020} explore this case study to construct proper counterfactuals. Hospitalizations and ICU patients would be much higher in Denmark and Norway were Sweden's more lenient measures adopted. \cite{andersen2020} argue that despite the divergence in deaths in Sweden relative to Denmark, at least in the very short-run, there was not a large difference in the aggregate spending drop due to the stricter lockdown strategy in Denmark.}

Aiming to provide a quantitative assessment on the short-run effects of lockdowns, this paper takes this exercise seriously in the context of US states. Given that the timing US states adopted lockdown policies differs among them, we adopt techniques based on synthetic control (SC) approach of \citet{aAjG2003} and \citet{aAaDjH2010} to assess the impact of lockdowns on the short-run evolution of the number of cases (and deaths) in the treated US states.\footnote{Throughout the main text in this paper, we focus on the number of cases. Results concerning the number of deaths are relegated to the Appendix. The timing of most lockdowns was soon enough such that there is not enough in-sample observations of deaths to apply the synthetic control method, so we use an alternative methodology we explain below.} More specifically, we consider an extension of the original SC method called Artificial Counterfactual (ArCo) which was put forward by \citet{CarvalhoMasiniMedeiros2018ArCo}. Due to the nonstationary nature of the data, the correction of \citet{rMmcM2019} is necessary.

It is hard to downplay the importance of finding out the effects of lockdown policies, especially now that several countries are experiencing even harsher second waves of Covid-19. Our results point to a substantial short-run taming of the cumulative number cases due to the adoption of lockdown policies. On average, for treated states, the counterfactual accumulated number of cases, according to the method adopted here, would be two times larger were lockdown policies not implemented.

The decision to implement a lockdown policy is not taken out of the blue. It might complement or substitute other types of containment policies implemented in control or treated states, such as, for example, mask mandates. Hence, in principle, our estimates are better interpreted as capturing the effect of a ``combo" of policies that include lockdowns, relative to another ``combo" of policies that do not include them. Nonetheless, to address some confounding effects, we use a simple causal model similar to \cite{vChKpS2021} to claim that a sizable part of the estimated effects is arguably attributable to lockdown policies.

A key feature of our approach is that it is purely data-driven. In the beginning of the crisis, the majority of papers written by economists to evaluate the effectiveness of lockdowns relied on epidemiological models for analysis, including the most recent ones that incorporate behavioral responses.\footnote{Descriptions of epidemiological models and simulations concerning the evolution of the Covid-19 pandemic can be found in \cite{atkeson2020will} and \cite{berger2020seir}. \cite{alvarez2020simple}, \cite{bethune2020covid}, \cite{callum2020coronavirus}, \cite{eichenbaum2020macroeconomics}, among many others, incorporate behavioral responses and evaluate several containment policies.}  These models are hard to discipline quantitatively. Many calibrated parameters remain uncertain,\footnote{See, for example, \cite{atkeson2020death} on the uncertainty regarding estimates of the fatality rate.} and models that incorporate behavioral responses need time to mature and agree on a reliable set of ingredients and moments to be matched.

Model-free approaches like ours or \cite{medeiros2020forecast} should complement policy discussions or forecasting exercises based on those models, especially from a quantitative point of view. There are related papers using state or county level US data.\footnote{There are also related papers for other countries. For example, \cite{fang2020china} for China.} At least one of them, \cite{california}, uses a synthetic control approach but it is restricted solely to California. Other papers, such as \cite{brzezinski2020}, \cite{dave} and \cite{sears2020stayinghome}, use variations in the timing of statewide adoption of containment policies, and difference-in-differences models to document substantial reductions in mobility and improvements of health outcomes.  The key identification assumption in these papers is that variations in the timing are random after controlling for covariates. \cite{brzezinski2020} also consider an instrumental-variable approach. \cite{fowler2020stayinghome} and \cite{barrot2020} follow similar empirical strategies but at county level, and also find  substantial reductions in cases and fatalities in counties that adopted stay-at-home orders and state-mandated
business closures, respectively. Our analysis, that rests on alternative identification assumption and method, should be seen as complementary. 

The paper is organized as follows. Section \ref{sec:data} describes the data, while Section \ref{sec:empstrategy} presents the empirical strategy. The results are discussed in Section \ref{sec:results}. In Section \ref{sec:confouding}, we discuss our estimates and address some confounding effects. Finally, Section \ref{sec:conclusion} concludes the paper. Additional results are included in the Appendix.

\section{Data}
\label{sec:data}

Data on Covid-19 (confirmed) cases are obtained from the repository at the Johns Hopkins University Center for Systems Science and Engineering (JHU CSSE). We consider the cumulative cases for a subset of the 50 US states and the District of Columbia. Instead of using the chronological time across the states, we consider the epidemiological time, which means that the day one in a given state is the day that the first Covid-19 case was confirmed there.

The econometric approach adopted here relies on the fact that some states adopted a lockdown strategy (the treatment), whereas others did not adopt social distancing measures (control group) and are used to construct the counterfactual.\footnote{The timing of those policies at each state were obtained, and double checked, in several press articles, e.g., https://www.businessinsider.com/us-map-stay-at-home-orders-lockdowns-2020-3 and https://www.nytimes.com/interactive/2020/us/coronavirus-stay-at-home-order.html.} Lockdown strategies include a mix of state-wide non-pharmaceutical measures aiming to limit social interactions, such as restrictions on non-essential activities and requirements that residents stay at home.

\section{Empirical strategy}
\label{sec:empstrategy}

In this section, we describe how we assign states to control and treatment groups, and then, describe the method used to construct the counterfactuals.

\subsection{Treated and non-treated states}
Aiming to balance control and treatment states, and at the same time obtain enough observations to estimate properly the model before the lockdown policy was implemented, we divide US states into three groups.

For a state to be included in the analysis, a state-wide lockdown policy must be established at least twenty days after the first case. We assume that whenever an individual becomes infected, it takes an average of ten days to show up as a confirmed case in the statistics.\footnote{This assumption is motivated by the incubation period of the virus. According to the World Health Organization, the ``[...] the incubation period for COVID-19, which is the time between exposure to the virus (becoming infected) and symptom onset, is on average 5-6 days, however can be up to 14 days." See https://www.who.int/docs/default-source/coronaviruse/situation-reports/20200402-sitrep-73-covid-19.pdf.} Hence, the in-sample period used to estimate the synthetic control (``before" the lockdown policy) for each treated state (to be defined below) is the number of days between the tenth day after the first confirmed case and the tenth day after the lockdown strategy was implemented. We choose to start the in-sample from the tenth day as a way to smooth the initial volatility of the data. 

We adopt a criteria that a state must have at least twenty observations in the in-sample period to be included in the analysis. This criteria excludes states that adopted a state-wide lockdown strategy too early, such as Connecticut, New Jersey, Ohio, among others. These are the unmarked states in Table \ref{tab:groups}, which reports the dates of the first case and lockdown policy, as well the difference in days between them, and also helps visualize the three groups of states.

The remaining states must be divided into treated and control groups. The idea is to find a synthetic control for each of the treated states. The group of potential controls should consist of states that adopted a lockdown policy too late (or never adopted), such that counterfactuals are not contaminated by lockdown policies implemented in those states. At the same time, and for a similar reasoning, the lockdown strategies adopted in treated states must be in place during the period of analysis.\footnote{In the Appendix \ref{app:reopen}, Table \ref{tab:table_reopen} shows the reopen dates for the treated states.}

\begin{table}[H]
\centering
\caption{Number of Days from First Case until Lockdown for each State}
\label{tab:groups}
\resizebox{\columnwidth}{!}{
\begin{tabular}{|l|ccc|l|ccc|}
  \hline
State & First Case & Lockdown ($T_0+10$) & Days Diff. & State & First Case & Lockdown ($T_0+10$) & Days Diff. \\
  \hline
  \hline
\cellcolor{cor2}{Alabama} & \cellcolor{cor2}{03/13/2020} & \cellcolor{cor2}{04/14/2020}  & \cellcolor{cor2}{32}  & \cellcolor{cor2}{Mississippi}  & \cellcolor{cor2}{03/12/2020}  & \cellcolor{cor2}{04/13/2020}  & \cellcolor{cor2}{32}  \\ 
  Alaska & 03/13/2020 & 04/07/2020 & 25 & \cellcolor{cor2}{Missouri}  & \cellcolor{cor2}{03/08/2020}  & \cellcolor{cor2}{04/16/2020}  & \cellcolor{cor2}{39}  \\ 
  \cellcolor{cor1}{Arizona}  & \cellcolor{cor1}{01/26/2020}  & \cellcolor{cor1}{04/10/2020}  & \cellcolor{cor1}{75}  & Montana & 03/13/2020 & 04/07/2020 & 25 \\ 
  \cellcolor{cor1}{Arkansas}  & \cellcolor{cor1}{03/13/2020}  & \cellcolor{cor1}{-}  & \cellcolor{cor1}{-}  & \cellcolor{cor1}{Nebraska}  & \cellcolor{cor1}{03/06/2020}  & \cellcolor{cor1}{-}  & \cellcolor{cor1}{-}  \\ 
  \cellcolor{cor1}{California}  & \cellcolor{cor1}{01/26/2020}  & \cellcolor{cor1}{03/29/2020}  & \cellcolor{cor1}{63}  & \cellcolor{cor2}{Nevada}  & \cellcolor{cor2}{03/05/2020}  & \cellcolor{cor2}{04/11/2020}  & \cellcolor{cor2}{37}  \\ 
  \cellcolor{cor2}{Colorado}  & \cellcolor{cor2}{03/06/2020}  & \cellcolor{cor2}{04/05/2020}  & \cellcolor{cor2}{30}  & \cellcolor{cor2}{New Hampshire}  & \cellcolor{cor2}{03/02/2020}  & \cellcolor{cor2}{04/06/2020}  & \cellcolor{cor2}{35}  \\ 
  Connecticut & 03/10/2020 & 04/02/2020 & 23 & New Jersey & 03/05/2020 & 03/31/2020 & 26 \\ 
  Delaware & 03/11/2020 & 04/03/2020 & 23 & New Mexico & 03/11/2020 & 04/03/2020 & 23 \\ 
  DC & 03/16/2020 & 04/03/2020 & 18 & \cellcolor{cor2}{New York}  & \cellcolor{cor2}{03/02/2020}  & \cellcolor{cor2}{04/01/2020}  & \cellcolor{cor2}{30}  \\ 
  \cellcolor{cor2}{Florida}  & \cellcolor{cor2}{03/02/2020}  & \cellcolor{cor2}{04/11/2020}  & \cellcolor{cor2}{40}  & \cellcolor{cor2}{North Carolina}  & \cellcolor{cor2}{03/03/2020}  & \cellcolor{cor2}{04/09/2020}  & \cellcolor{cor2}{37}  \\ 
  \cellcolor{cor2}{Georgia}  & \cellcolor{cor2}{03/03/2020}  & \cellcolor{cor2}{04/13/2020}  & \cellcolor{cor2}{41}  & \cellcolor{cor1}{North Dakota}  & \cellcolor{cor1}{03/12/2020}  & \cellcolor{cor1}{-}  & \cellcolor{cor1}{-}  \\ 
  Hawaii & 03/07/2020 & 04/01/2020 & 25 & Ohio & 03/10/2020 & 04/02/2020 & 23 \\ 
  Idaho & 03/13/2020 & 04/04/2020 & 22 & \cellcolor{cor2}{Oregon}  & \cellcolor{cor2}{02/29/2020}  & \cellcolor{cor2}{04/02/2020}  & \cellcolor{cor2}{33}  \\ 
  \cellcolor{cor1}{Illinois}  & \cellcolor{cor1}{01/24/2020}  & \cellcolor{cor1}{03/31/2020}  & \cellcolor{cor1}{67}  & \cellcolor{cor2}{Pennsylvania}  & \cellcolor{cor2}{03/06/2020}  & \cellcolor{cor2}{04/11/2020}  & \cellcolor{cor2}{36}  \\ 
  Indiana & 03/06/2020 & 04/02/2020 & 27 & \cellcolor{cor2}{Rhode Island}  & \cellcolor{cor2}{03/01/2020}  & \cellcolor{cor2}{04/07/2020}  & \cellcolor{cor2}{37}  \\ 
  \cellcolor{cor1}{Iowa}  & \cellcolor{cor1}{03/09/2020}  & \cellcolor{cor1}{-}  & \cellcolor{cor1}{-}  & \cellcolor{cor2}{South Carolina}  & \cellcolor{cor2}{03/07/2020}  & \cellcolor{cor2}{04/17/2020}  & \cellcolor{cor2}{41}  \\ 
  \cellcolor{cor2}{Kansas}  & \cellcolor{cor2}{03/08/2020}  & \cellcolor{cor2}{04/09/2020}  & \cellcolor{cor2}{32}  & \cellcolor{cor1}{South Dakota}  & \cellcolor{cor1}{03/11/2020}  & \cellcolor{cor1}{-}  & \cellcolor{cor1}{-}  \\ 
  \cellcolor{cor2}{Kentucky}  & \cellcolor{cor2}{03/06/2020}  & \cellcolor{cor2}{04/05/2020}  & \cellcolor{cor2}{30}  & \cellcolor{cor2}{Tennessee}  & \cellcolor{cor2}{03/05/2020}  & \cellcolor{cor2}{04/10/2020}  & \cellcolor{cor2}{36}  \\ 
  Louisiana & 03/11/2020 & 04/02/2020 & 22 & \cellcolor{cor2}{Texas}  & \cellcolor{cor2}{03/05/2020}  & \cellcolor{cor2}{04/12/2020}  & \cellcolor{cor2}{38}  \\ 
  \cellcolor{cor2}{Maine}  & \cellcolor{cor2}{03/12/2020}  & \cellcolor{cor2}{04/12/2020}  & \cellcolor{cor2}{31}  & Vermont & 03/08/2020 & 04/04/2020 & 27 \\ 
  \cellcolor{cor2}{Maryland}  & \cellcolor{cor2}{03/06/2020}  & \cellcolor{cor2}{04/09/2020}  & \cellcolor{cor2}{34}  & Virginia & 03/08/2020 & 04/04/2020 & 27 \\ 
  \cellcolor{cor1}{Massachusetts}  & \cellcolor{cor1}{02/01/2020}  & \cellcolor{cor1}{04/02/2020}  & \cellcolor{cor1}{61}  & \cellcolor{cor1}{Washington}  & \cellcolor{cor1}{01/22/2020}  & \cellcolor{cor1}{04/02/2020}  & \cellcolor{cor1}{71}  \\ 
  Michigan & 03/11/2020 & 04/03/2020 & 23 & West Virginia & 03/18/2020 & 04/03/2020 & 16 \\ 
  Minnesota & 03/06/2020 & 04/04/2020 & 29 & Wisconsin & 03/10/2020 & 04/04/2020 & 25 \\ 
   \hline
\end{tabular}}
\label{tab:table_states_color}
\end{table}
\vspace{-8mm}
\begin{figure}[H]
    \centering
    \includegraphics[scale = .55]{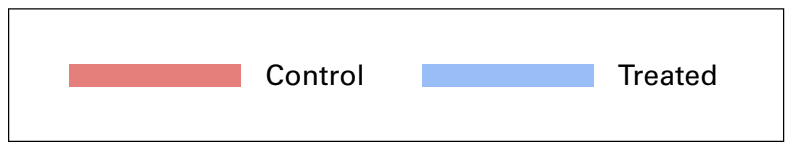}
\end{figure}

Fortunately, there are horizons that can balance both goals: enough states to build the synthetic controls and a relative extensive period to construct the counterfactuals. In particular, we restrict the analysis up to the 58th epidemiological day. This figure accommodates at least ten control states to build the synthetic controls,\footnote{That is, to be in the control group, whenever a lockdown policy was implemented in a given control state, it was implemented at least 48th days after the first epidemiological day. Given the aforementioned assumption, its effects on Covid-19 confirmed cases only show up in the statistics ten days later, on average.} at the same time it maximizes the out-of-sample days to run the counterfactuals. In this sense, our analysis concerns the very short-run impact of lockdowns, up to nearly three weeks.

The treated states are marked in blue in Table \ref{tab:groups}, and include twenty states: Alabama, Colorado, Florida, Georgia, Kansas, Kentucky, Maine, Maryland, Mississippi, Missouri, Nevada, New Hampshire, New York, North Carolina, Oregon, Pennsylvania, Rhode Island, South Carolina, Tennessee, and Texas. The potential control states are marked in red, and include ten states: Arizona, Arkansas, California, Illinois, Iowa, Massachusetts, Nebraska, North Dakota, South Dakota, and Washington. Nonetheless, due to the lack of variation within the in-sample period, we exclude four states from this control pool as we explain below.

Importantly, Oklahoma, Utah and Wyoming only implemented partial lockdowns (not reported in the table). Therefore, they are hard to classify as either treated or control states. We opt to exclude them from the analysis.

\begin{figure}[H]
    \centering
    \caption{(Log) Cumulative cases for each State in treated and control groups.}
    \includegraphics[scale=0.32]{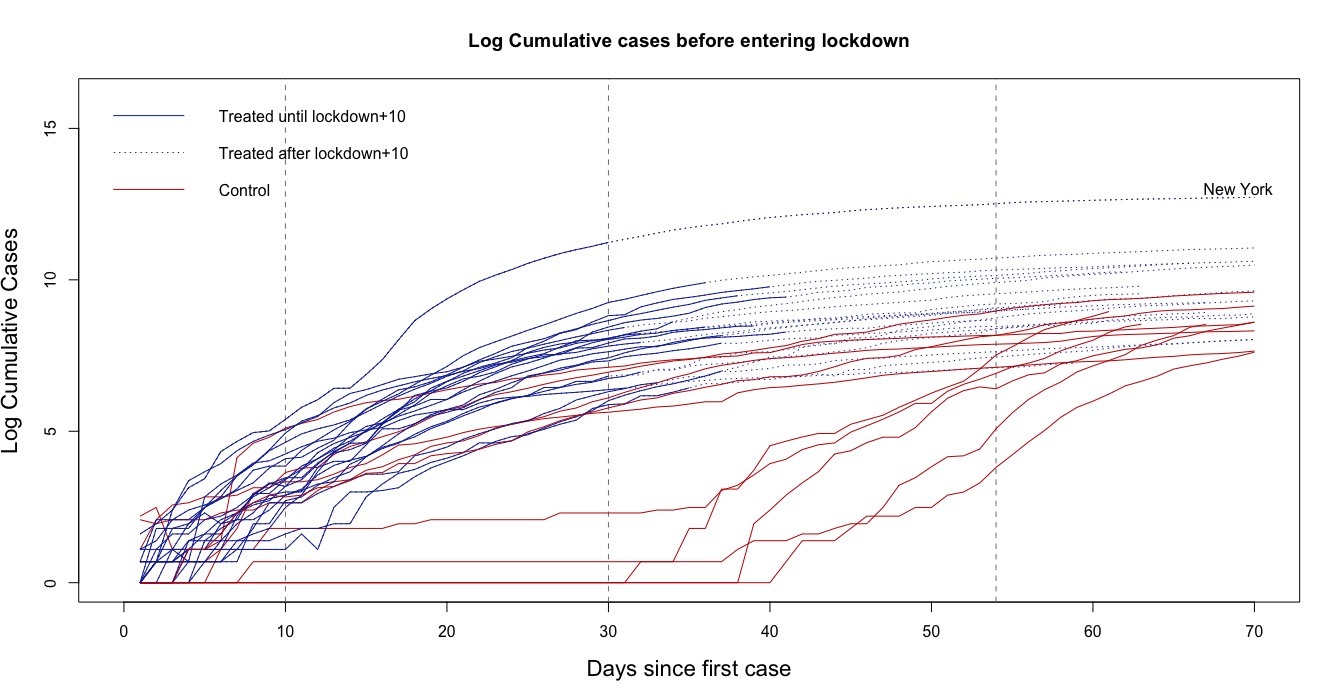}
    \label{fig:states_cum_cases}
\end{figure}

Figure \ref{fig:states_cum_cases} illustrates the empirical strategy, which is formalized in the next subsection. It plots the evolution of (log) cumulative cases along the epidemiological time. The first vertical dashed line represents the tenth day after the first confirmed case. The in-sample period is represented in between the first and second vertical dashed lines, which mark the tenth day and the following twenty days, respectively. Similarly, the out-of-sample period is in between the second and third vertical dashed lines, which mark the 31th and 58th epidemiological day, respectively.

Blue lines represent the treated states, whereas the red ones the potential control states. The turning points from blue full- to dashed-lines represent the days lockdowns were implemented (plus ten days) in treated states. Note that New York is clearly an outlier among the treated states, exhibiting a huge amount of cases (more on that below). We use the red lines to build synthetic controls for each full blue-line up to the turning point, and then construct counterfactuals by simulating the synthetic controls forward up to the 58th day. The idea is to compare counterfactuals with the blue dashed-lines that capture actual cases, and obtain the effect of lockdowns.\footnote{A lockdown policy might complement or substitute other types of containment policies implemented in control or treated states. Hence, our empirical strategy is arguably  capturing the effect of a ``combo" of policies that include lockdowns, relative to another ``combo" of policies that do not include them. We further discuss below how to disentangle the role of lockdown policies from alternative policies.}

As Figure \ref{fig:states_cum_cases} highlights, some states display lack of variation within the in-sample period. Just to give an example, Washington had had only one confirmed case for the first 36 days since its first confirmed Covid-19 infection. Therefore, we exclude it from the control group. For similar reasons, we also exclude  Arizona, Illinois, and Massachusetts from the control pool. The analysis ended up relying on six control states.

\subsection{Estimation}

We propose a two-step approach using the artificial counterfactual (ArCo) method introduced by \citet{CarvalhoMasiniMedeiros2018ArCo} with the correction of \citet{rMmcM2019} to estimate the number of cases for each US state.

Let $t=10,11,\ldots,58$ represents the number of days after the first confirmed case of Covid-19 in a given state. Define $y_{t}$ as the natural logarithm of the number of confirmed cases $t$ days after the outbreak of Covid-19 in this specific treated state, and $\boldsymbol{x}_t$ contains the logarithm of the number of cases for $p$ control states $t$ days after the first case, as well as a logarithmic trend, $\log(t)$. The inclusion of the trend is important to capture the shape of the curve.

The model is estimated as follows. We use the weighted least absolute and shrinkage operator (WLASSO) as described in \citet{rMmcM2019} to select the control states that will be used to estimate counterfactuals. The goal of the WLASSO is to balance the trade-off between bias and variance and is an useful tool to select the relevant peers in an environment with very few data points. The estimator is given as:
\begin{equation}\label{eq:lasso1}
    \widehat{\boldsymbol{\omega}} = \arg\underset{\boldsymbol{\omega}}{\min}
\left[\frac{1}{L-10} \sum_{t = 10}^{L}\left(y_t - \boldsymbol{x'}_t \boldsymbol{\omega}\right)^2 + \lambda \sum_{j=1}^p \kappa_j|\beta_j| \right],
\end{equation}
where $\kappa_j=|x_{j,L}|$, $j=1,\ldots,p-1$, and $\kappa_p=1$. $L$ is, for each state, the number of days from the first reported case until the lockdown plus ten extra days, and $\lambda>0$ is the penalty parameter which is selected by the Bayesian Information Criterion (BIC), in accordance with \citet{MedeirosMendes2016L1}. The weight correction in the WLASSO is necessary in order to control for the nonstationarity of the data; see \citet{rMmcM2019} for a detailed discussion.

The counterfactual for $t=L+1,\ldots,$ is computed as $\widehat{y}_t=\boldsymbol{x}_t'\widehat{\boldsymbol{\omega}}$. We also report 95\% confidence intervals based on the resampling procedure proposed in \citet{rMmcM2019}.

\subsection{Constructing a counterfactual to deaths} \label{sec:counter_deaths}

We are interested in examining the effects of lockdown policies not only on the number of cases, but also on the number of deaths. However, we cannot implement the strategy described above because there is not enough variation in deaths for the in-sample period. Some states, for instance, implemented a state-wide lockdown policies before the first confirmed death.

Thus, we propose an alternative method. We consider a counterfactual state for the number of deaths based on the counterfactual estimated for the number of cases. This is not straightforward as in the traditional synthetic control method because the ArCo methodology described above includes an intercept in the estimation, which is measured in the log of the number of cases, and the counterfactual is not only a convex combination of other states. Intuitively, the methodology described above chooses a combination of states that is at a fixed distance from the treated unit at the in-sample period and not a convex combination of states that matches exactly the actual number of cases. The intercept controls for all time-invariant characteristics that define the counterfactual.

Then, we proceed as follows. Let $y_{st}$ be the number of accumulated deaths in state $s$ at the day $t$. Also, let $\boldsymbol{\beta}^s$ be the vector of estimated coefficients for the state $s$ as in expression \eqref{eq:lasso1} above and used to construct the counterfactual for cases. In addition, let $\boldsymbol{y}_t$ be a vector of the number of deaths for all states in the control pool at time $t$. We define the counterfactual number of deaths in that state as
\begin{equation*}
    y_{st}^C=\boldsymbol{y}_t \boldsymbol{\beta}^s-\boldsymbol{y}_{\bar{t}} \boldsymbol{\beta}^s+y_{s\bar{t}},
\end{equation*}
 where $\bar{t}$ is the day that state $s$ implemented the lockdown policies. That is, we maintain the weights estimated above and adjust the intercept so that the counterfactual series for deaths matches the number of actual observed deaths in the beginning of the quarantine. For the sake of exposition, we relegate the results on cumulative deaths to Appendix \ref{app:deaths}.

\section{Results}\label{sec:results}

\begin{figure}[H]
\centering
\caption{ArCo estimates counterfactual without state lockdown ($T_0+10$)}
\begin{subfigure}[b]{0.87\textwidth}
    \centering
    \includegraphics[trim={0 1.3cm 0 1.6cm}, clip, scale = .15, width=1\linewidth]{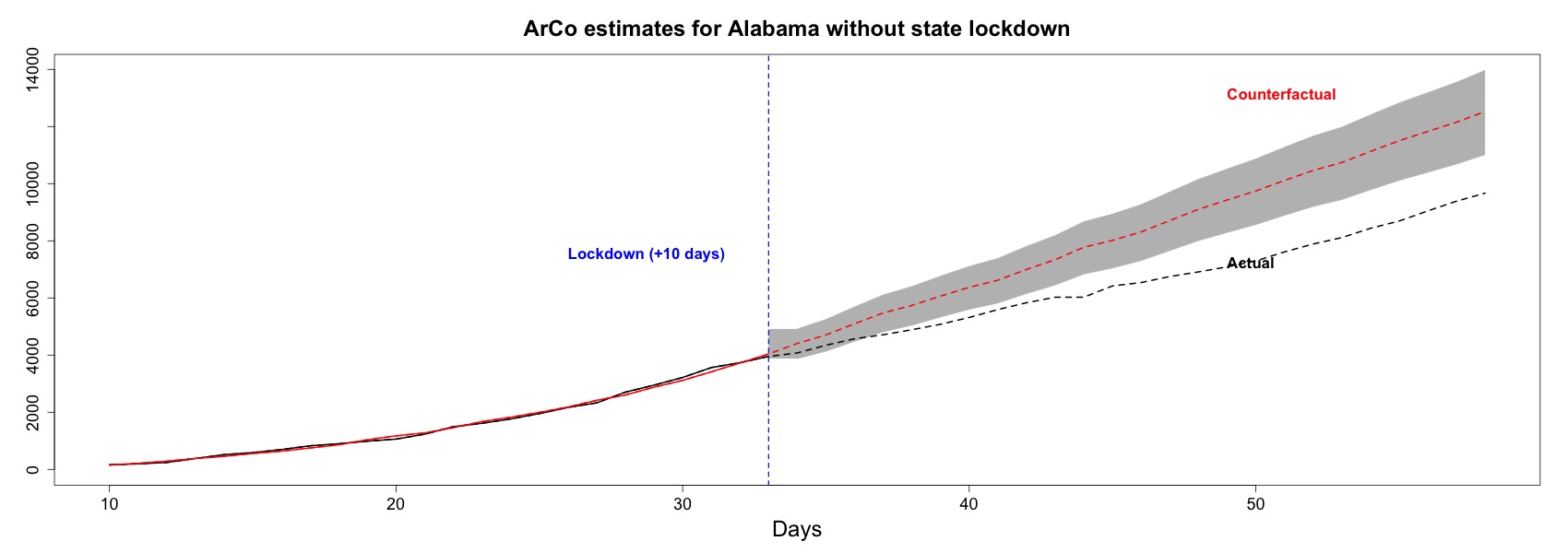}
    \caption{Alabama}
    \label{fig:al_results}
\end{subfigure}%

\begin{subfigure}[b]{0.87\textwidth}
    \centering
    \includegraphics[trim={0 1.3cm 0 1.6cm}, clip, scale = .15, width=1\linewidth]{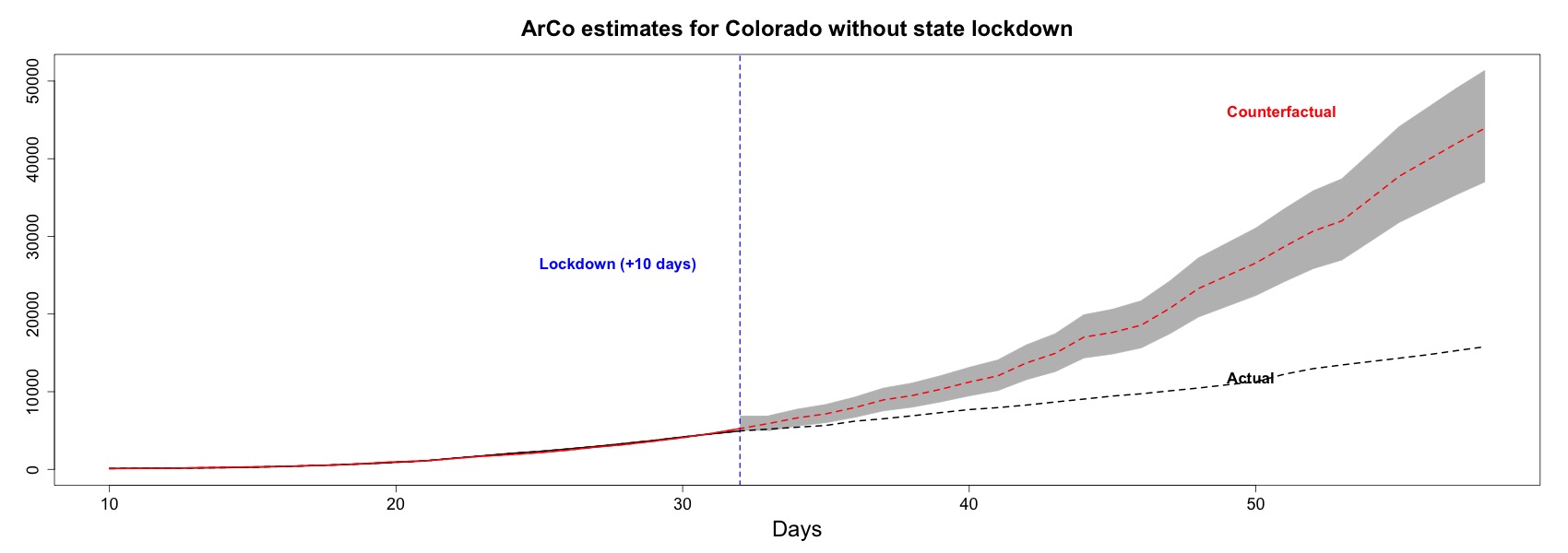}
    \caption{Colorado}
    \label{fig:co_results}
\end{subfigure}%

\begin{subfigure}[b]{0.87\textwidth}
    \centering
    \includegraphics[trim={0 1.3cm 0 1.6cm}, clip, scale = .15, width=1\linewidth]{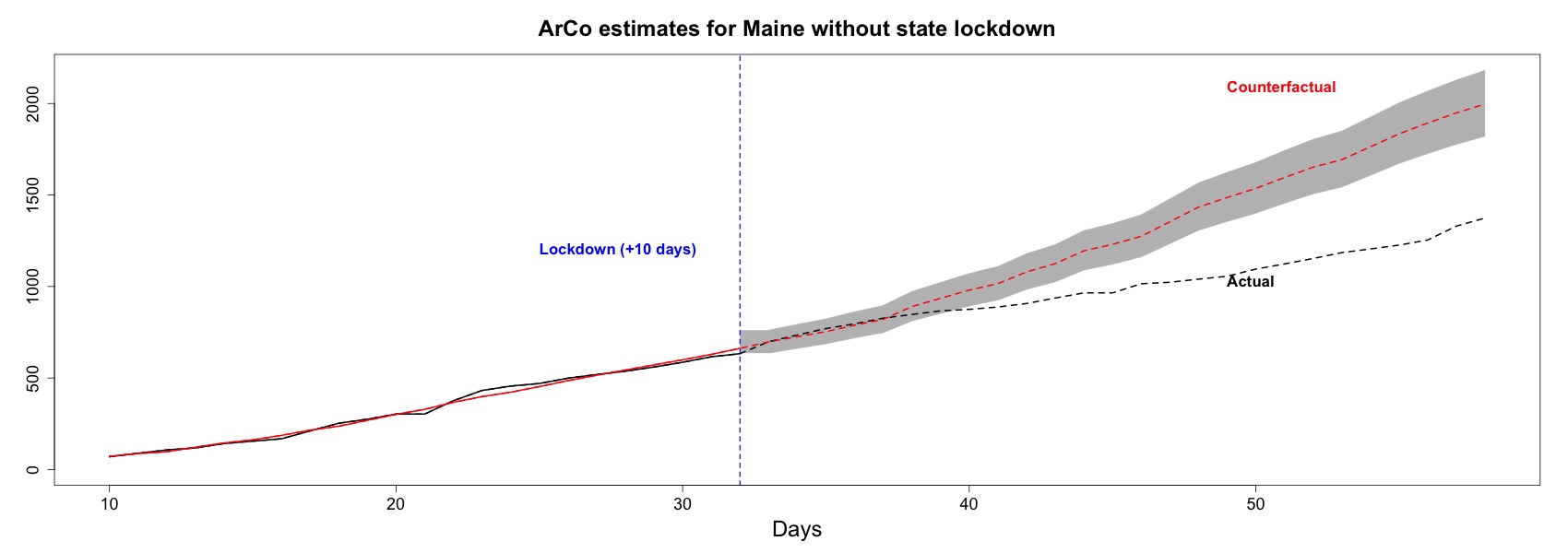}
    \caption{Maine}
    \label{fig:me_results}
\end{subfigure}%

\label{fig:arco_results}
\end{figure}

To illustrate how the method works, Figure \ref{fig:arco_results} presents the ArCo counterfactuals for the states of Alabama, Colorado, and Maine. The timing of the policy intervention ($T_0 + 10$) corresponds to the lockdown date plus ten days. The gray area represents 95\% confidence intervals.

The counterfactual analysis makes it clear the importance of lockdown policies in mitigating the acceleration of the number of Covid-19 confirmed cases in the treated states. As shown in Figure \ref{fig:al_results}, for example, our results point to a substantial increase in the number of cases in Alabama if it had not adopted an early lockdown. Similarly, Figures \ref{fig:co_results} and \ref{fig:me_results} reveal the same behavior for the cumulative curves in the other selected states.  Counterfactuals are constructed with the estimated weights and cumulative cases of the six states that compose the control group. These weights are reported in Table \ref{tab:table_coef} in Appendix \ref{app:statistics}. In  Appendix \ref{app:cases}, we present similar counterfactual plots for the remaining treated states. Similar results apply for most of the treated states.

\begin{figure}[H]
    \centering
    \caption{Ratio between estimated and actual cumulative cases ($T = \{10,...,58\}$)}
    \includegraphics[trim={0 0 0 2cm}, clip, scale = 0.6]{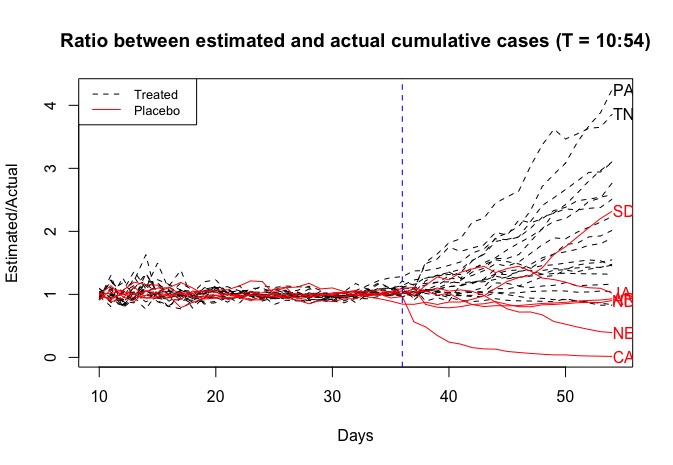}
    \label{fig:ratio_cases}
\end{figure}

In order to assure that the proposed methodology is producing proper counterfactual analysis, we generate placebo results by producing a ``synthetic control" for each control state using the remaining control states as donor pool. Results are displayed in Figure \ref{fig:ratio_cases}, which shows the ratio of the estimated counterfactual cumulative cases to the actual ones for treated states except New York (black lines), and non-treated states (red lines). We assume that the epidemiological day of the placebo intervention is $T_0 = 36$, marked by the vertical dashed line, which is the median (and the mean) timing of the policy interventions in the treated states.

It is reassuring that for half of the placebo counterfactuals, these ratios fluctuate around one, whereas for the majority of treated states ratios grew above one at some point (likely around the actual timing of policy intervention). The latter result means that lockdown policies were effective to tame the spread of the virus, whereas the former suggests that results are not driven by chance.

Regarding South Dakota, the only placebo counterfactual that reached a ratio well above one, by using Google Mobility Data (described in Appendix \ref{ssec:google_mobility}), we show that mobility in residential areas increased whereas mobility in outdoor areas decreased substantially once compared to the period before the pandemic (see Figures \ref{fig:median_residential} and \ref{fig:median_outdoors} in Appendix \ref{ssec:google_mobility}). This is suggestive that South Dakota's population endogenously decided to stay more at home, and avoided environments prone to the risk of contamination. At the time, a proper lockdown policy was not necessary, and South Dakota's non-conformity to the placebo test does not seem to invalidate our approach.

In contrast, for Nebraska and California, the counterfactuals are pointing to a smaller number of cases than the actual ones, which goes against finding that lockdowns were effective to reduce cases of Covid-19. The case of California is quite emblematic, as the number of cases during the estimation window remained very small and with very low variation. However, the number of cases started to grow at a fast rate much after the cut-off date. The state of Nebraska displays a similar pattern. 

To gauge the quantitative impact of lockdown policies, for each state, whether treated or control used as placebo, we compute the ratio of the counterfactual estimated cumulative cases (``without" a lockdown strategy in place) to actual ones on the 58th epidemiological day, which is the last day used to compute the counterfactual. Table \ref{tab:arco_statistics} reports the mean and median of the ratios across states, whereas Table \ref{tab:arco_forecast} in Appendix \ref{app:statistics} reports these ratios for each state. The first row corresponds the case in which controls are used as placebos, whereas the second considers the treated states only. As we discuss below, New York is clearly an outlier, whose ratio reached an implausible value of 16.5 as reported in Table \ref{tab:arco_forecast}. Hence, our preferred specification is displayed in the third row which excludes New York from the pool of treated states. We also compute other two versions of these ratios using the lower bound (lb) and upper bound (up) of the 95\% confidence interval in the numerator.

\begin{table}[H]
\centering
\caption{ArCo estimates (58th day)}
\begin{tabular}{|l|cccccc|}
  \hline
 & Mean ArCo & Med ArCo & Mean lb & Med lb & Mean ub & Med ub \\
  \hline
  \hline
Control & 1.04 & 0.92 & 0.91 & 0.80 & 1.18 & 1.03 \\
Treated & 3.08 & 2.28 & 2.46 & 1.91 & 3.67 & 2.63 \\
Treated (-NY) & 2.37 & 2.08 & 1.99 & 1.72 & 2.75 & 2.32 \\
   \hline
\end{tabular}
\label{tab:arco_statistics}
\end{table}

The ratios are clearly above one for the treated units, whether New York is excluded or not. According to our preferred specification, counterfactual estimates suggest that the number of cases would be nearly two times larger were lockdown policies absent. Again, it is reassuring that among the controls used as placebo, these average ratios remain around one.

Of course, a lockdown policy might complement or substitute other types of containment policies implemented in control or treated states (e.g., mask mandates). Hence, our estimates are arguably better interpreted as capturing the effect of a ``combo" of policies that include lockdowns, relative to another ``combo" of policies that do not include them. Nonetheless, in the next section, we use a simple casual model to argue that a sizable part of the counterfactual is attributable to lockdown policies.

Regarding the effects of lockdowns on cumulative deaths, we present the results for all treated states in Appendix \ref{app:deaths}. For some states, the counterfactual cumulative deaths exhibit similar patterns to those regarding cumulative cases. But, for many other states, they are not statistically significant at least for the first days after the policy implementation. One possible explanation is that there is a delay between cases and deaths, as the latter is a consequence of the former. Hence, deaths only show up in the official statistics days after cases. Perhaps, if we could estimate counterfactuals for longer periods, the synthetic accumulated deaths would further decouple from the actual ones. In addition, since weights on the controls are estimated considering the (log) cumulative number of cases, the counterfactuals for cumulative deaths are arguably noisier.

\subsection{Outlier: New York}\label{sec:ny}

\begin{figure}[H]
\centering
\caption{New York: ArCo Estimates Counterfactual Without State Lockdown ($T_0+10$)}
    \label{fig:ny_grate}

    \includegraphics[width=0.85\linewidth]{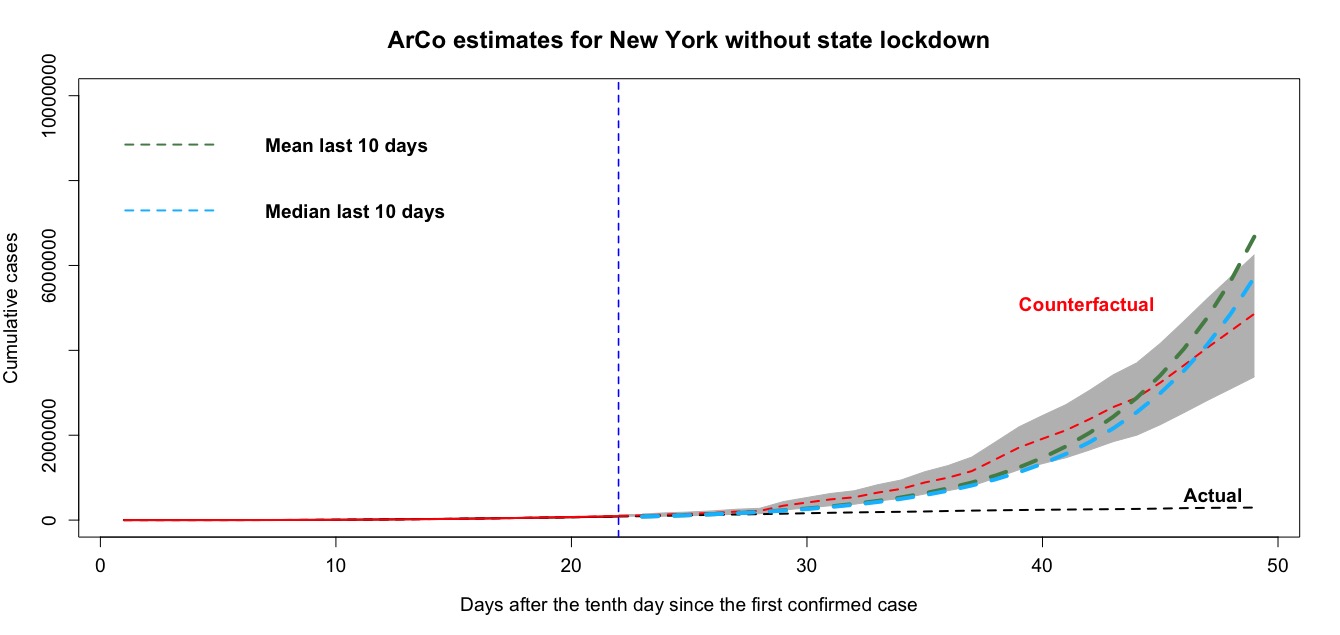}

\end{figure}

As discussed above and presented in Table \ref{tab:arco_forecast} in Appendix \ref{app:statistics}, we obtain an implausible ratio (of counterfactuals to actual cumulative cases) of 16.5 to New York. This section zooms on this state. In particular, Figure \ref{fig:ny_grate} displays the estimated cumulative number of cases for New York ``without" lockdown, as well as extrapolations of the cumulative number of cases based on the mean and median growth rate of the last ten days of the in-sample period.

As reported in Table \ref{tab:table_states_color}, among the treated states, New York was the fastest one to react to the pandemic, and  established a state-wide lockdown policy only 20 days after the first case. Figure \ref{fig:ny_grate} extrapolates the last in-sample observations by using both the observed mean and median growth rates for the last ten days, which yields a similar pattern to the result obtained by applying the ArCo approach. Due to the progression of the virus, particularly in New York City, the in-sample observed rates are quite high once compared to other states as illustrated in Figure \ref{fig:states_cum_cases}, which can be explained not only by the dynamics of the city but also by its high population density.  Hence, New York is clearly an outlier and might not be amenable to our synthetic control approach, which justifies reporting results excluding New York.

\section{Confounding effects}\label{sec:confouding}

Several factors may act as possible confounders to the estimaded effects of lockdowns.

First, individuals may react to the pandemic and change their behavior endogenously independent of the adoption of stricter lockdown rules imposed by the authorities. To the extent that this endogenous change of behavior would be similar across control and treated states, this is less of a concern. After all, we  would like to report the impact of lockdown policies above and beyond individual responses to the pandemic that would occur in the absence of lockdowns.

Second, lockdown is not the only policy in the menu. Control states may not implement a lockdown, but may enact other alternative policies, such as, for example, mandatory mask-wearing or massive campaigning for people to stay at home, that may contain the pandemic evolution. This would introduce a negative bias in our estimates, suggesting even more sizable effects of lockdowns. Alternatively, lockdown policies in treated states may be designed altogether with other containment measures. In this case, our results should be better interpreted as the average effect of a ``combo" of policies that include a lockdown strategy.

In order to guide the interpretation of the empirical estimates, and try to isolate the role of lockdowns, we describe a simple causal model similar to \citet{vChKpS2021}. The model helps to organize ideas on how lockdown policies affect the variables of interest and how they interact with confounders.  We argue below that, through the lens of this simple model and some auxiliary evidence, lockdowns (rather than other confounders or alternative policies) explain  a sizable part of our estimated effects.

The model incorporates the interactions between the following variables: (i) $Y_{s,t+l}$, which are the cases of or deaths by Covid-19 in state $s$ and period $t+l$; (ii) $L_{st}$, an indicator variable that a state adopted a lockdown policy in state $s$ and period $t$; (iii) $P_{st}$, another indicator variable of alternative policies implemented; (iv) $I_{st}$, which is available information to individuals that maybe be useful to affect behavior and/or contain the pandemic;\footnote{In the \citet{vChKpS2021} model, this variable generates inter-temporal dependence between periods. In our model, this variable is modeled differently as an alternative mechanism through which policies can contain the pandemic.} (v) $B_{st}$  summarizes the relevant behavior of individuals  such as adherence to social distancing, use of masks, etc; and (vi) $U_{st}$ is the set of confounders that might affect the determination of policies, individuals' behavior, and the pandemic evolution. We represent these interactions through the Direct Acyclic Graph (DAG) below.\footnote{See \citet{jP1995} and \citet{jP2009} for a discussion of DAGs and causal models.}

\begin{figure}[H]
\centering
\caption{Simple Causal Direct Acyclic Graph (DAG)}
\begin{tikzpicture}
\node (U) {$U_{st}$};
\node [below of=U, xshift=2cm] (P) {$P_{st}$};
\node[right of=U, yshift=2cm] (L) {$L_{st}$};
\node [right of=L, xshift=1cm] (I) {$I_{st}$};
\node[below of=I] (B) {$B_{st}$};
\node[right of=B, xshift=3cm] (Y) {$Y_{s,t+l}$};
\draw[->] (U) --  (P);
\draw[->] (U) --  (L);
\draw[->] (P) --  (L);
\draw[->] (L) --  (I);
\draw[->] (L) --  (B);
\draw[->] (B) --  (Y);
\draw [->] (I) -- (B);
\draw[->] (I) --  (Y);
\draw[->] (P) --  (Y);
\draw[->] (U.south) .. controls +(down:4cm) and +(right:0cm) .. (Y.south);
\end{tikzpicture}
\vspace{-2cm}
\end{figure}
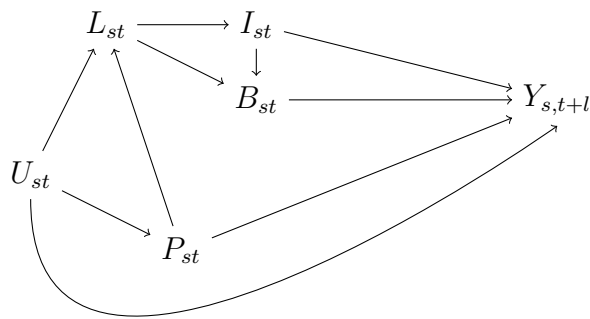

The sequence of events is the following. First, every period, potential confounders are determined. Second, public policies ($L_{st}$ and $P_{st}$) are set, and note that $P_{st}$ already encodes the transmission mechanisms (e.g., behavioral responses) through which policies other than lockdowns affect cases or deaths. Third, conditional on confounders and policies, individuals' information is updated. Fourth, individuals behavior are determined by confounders, policies and information. Finally, the variable of interest ($Y_{s,t+l}$) is determined.

Importantly, we assume that a lockdown policy, $L_{st}$, does not directly affect the number cases or deaths, $Y_{s,t+l}$. In fact, a lockdown policy only has effects on $Y_{s,t+l}$ to the extent that it affects some mediating variables, such as individuals' behavior (above and beyond endogenous responses to the pandemic in the absence of lockdowns) or available information.

Ideally, we would like to identify the causal effect of a lockdown policy  ($L_{st}$) on the pandemic evolution ($Y_{st+l}$). By using the back-door criteria suggested by \citet{jP1993}, we can envision two threats to the identification of the causal effect of interest.

First, non-observed variables ($U_{st}$) affect the probability of lockdown adoption and the evolution of the pandemic simultaneously. This is the traditional omitted variable bias in public policy evaluation. We discuss the problem of omitted variable bias below in the end of this section.

Second, as mentioned above, the potential additional problem related to the simultaneous implementation of alternative policies. Local authorities could adopt other containment policies $P_{st}$ as a substitute to the lockdown policy $L_{st}$ in control states, or as a complement in the treated ones. Hence, the adoption of alternative policies may affect the probability of implementing a lockdown and, simultaneously, affect individuals' behavior and the number of Covid-19 cases and deaths. Thus, we need to control for this possibility to identify the causal effect of interest.


The model also incorporates the possibility of endogenous responses of individuals to the evolution of the pandemic (or the accumulation of information in terms of the model). However, this is a mechanism through which our treatment acts. Therefore, according to the front-door criteria in \cite{jP1993}, we should not control for these variables, $I_{st}$ and $B_{st}$ (more on that below).

In what follows we discuss how we can control for some of the threats to the identification scheme described above. We also discuss the nature of the treatment.

\subsection{Mandatory mask-wearing}

Consider the implementation of simultaneous policies. Beyond policies such as the closure of schools, firms, and services included in $L_{st}$, mandatory mask-wearing is arguably the most important alternative implemented policy. We focus, here, on this alternative policy.

If the state is in lockdown, the mandatory mask-wearing is less of a concern as social contact and mobility are substantially curtailed (as we document in the next subsection). Regarding control states that did not adopt a lockdown strategy, in order to deal with policy simultaneity, we leverage on the temporal mismatch between policies.

In the beginning of the pandemic, lockdown policies were widely suggested by international institutions. In contrast, mandatory mask-wearing was not encouraged by the World Health Organization (WHO), which only changed its recommendation in the beginning of July. Note, however, that we restrict the sample to the first 56 days after the first Covid-19 case in each state. Thus, the last calendar day  in the sample is May 11 (Arkansas).

In Table \ref{T:mask}, we report the dates a mandatory mask-wearing policy was adopted in control states.\footnote{We manually collected data for these dates from state-level executive orders. For an example of these executive orders, see: https://governor.arkansas.gov/images/uploads/executiveOrders/EO\_20-43.pdf.} Note that these dates are not within our sample. Indeed, the earliest adoption of such policy was in June 26 in both Illinois and Washington.

\begin{table}[H]
    \centering
    \caption{Implementation dates for mandatory mask wearing}
    \label{T:mask}
    \begin{tabular}{|c c c c|}
    \hline
    State & Date & State & Date \\
    \hline
        Arizona & No & Massachusetts & 11/06/2020 \\
        Arkansas & 07/20/2020 & Nebraska & No\\
        California & 06/29/2020 & North Dakota & 11/14/2020 \\
        Illinois & 06/26/2020 & South Dakota & No\\
        Iowa & 11/17/2020 & Washington & 06/26/2020\\
        \hline
    \end{tabular}
\end{table}

We conclude that the most important alternative policy was not implemented until the end of the period considered in this paper, and that lockdowns implemented during this period were exogenous to this policy. Hence, mandatory mask wearing among control states does not seem to attenuate the effect of interest. Below, we change the casual DAG above to accommodate this insight.

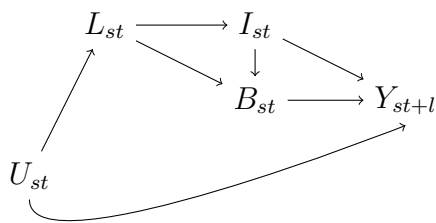
\begin{figure}[H]
\caption{Causal Direct Acyclic Graph (DAG) with timing of restrictions}
\label{F:DAG}
\centering
\begin{tikzpicture}
\node (U) {$U_{st}$};
\node[right of=U, yshift=2cm] (L) {$L_{st}$};
\node [right of=L, xshift=1cm] (I) {$I_{st}$};
\node[below of=I] (B) {$B_{st}$};
\node[right of=B, xshift=1cm] (Y) {$Y_{st+l}$};

\draw[->] (U) --  (L);
\draw[->] (L) --  (I);
\draw[->] (L) --  (B);
\draw[->] (B) --  (Y);
\draw [->] (I) -- (B);
\draw[->] (I) --  (Y);
\draw[->] (U.south) .. controls +(down:1cm) and +(right:0cm) .. (Y.south);
\end{tikzpicture}
\vspace{-1cm}
\end{figure}

\subsection{The nature of the treatment}

Before discussing the other threat to identification of causal effects, i.e. the omitted variable bias, we explore the nature of the treatment we are considering. What a lockdown policy does? Why would it affect the variables of interest?

The causal diagram in Figure \ref{F:DAG} suggests two mechanisms. First, a lockdown affects individuals' behavior by reducing mobility. Second, the policy might affect the available information. For instance, its implementation can increase the awareness of individuals about the pandemic and provide incentives to further changes in behavior.

\begin{figure}[H]
    \centering
    \caption{Changes in Google searches about the pandemic relative to lockdown announcement}
    \label{F:google0}
    \includegraphics[scale=0.90]{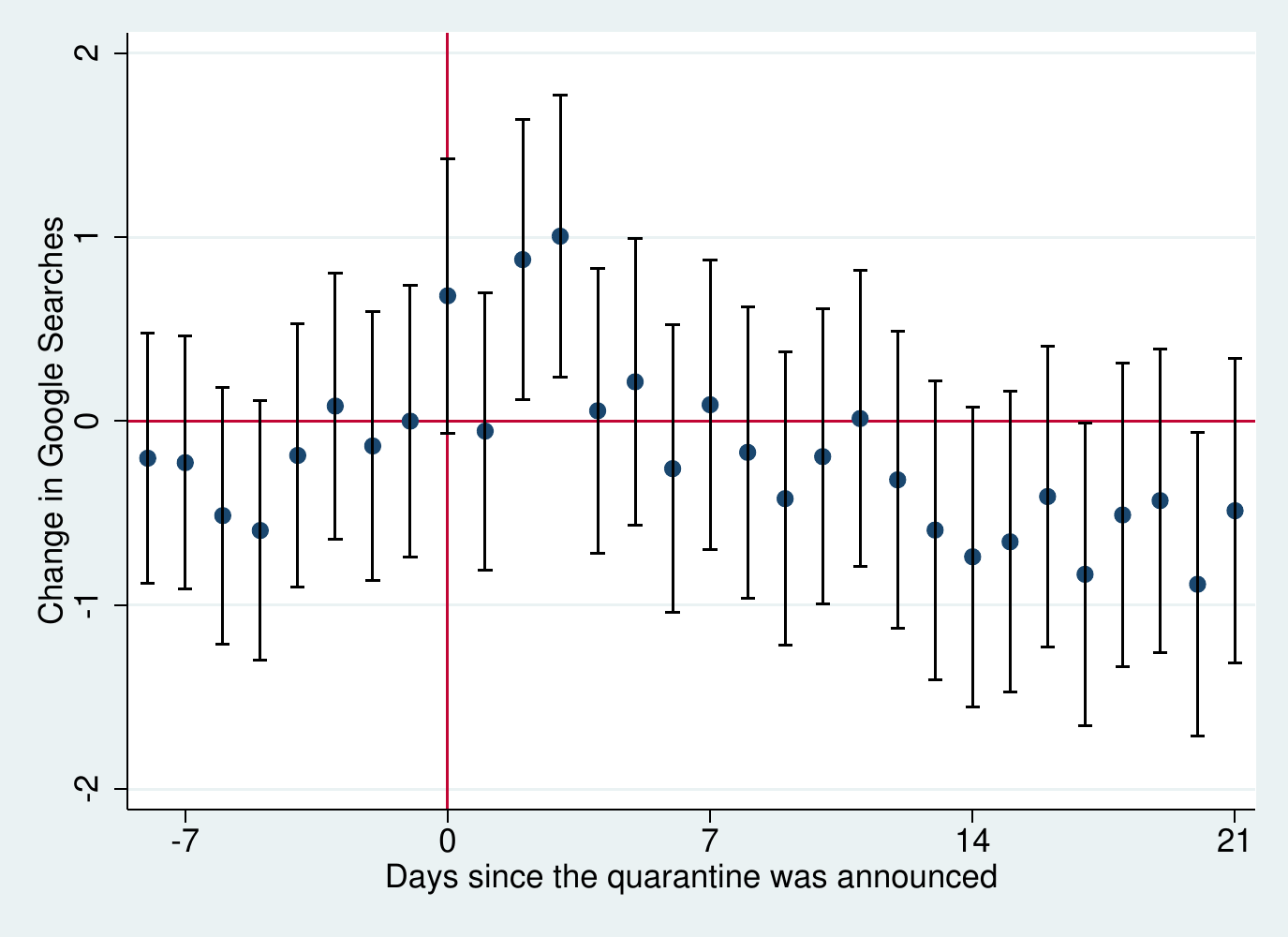}
\end{figure}

Despite considering the theoretical possibility of an informational transmission channel, some auxiliary empirical evidence suggests that its relevance is quite limited. Figure \ref{F:google0} plots the number of (changes in) pandemic-related Google searches in the days immediately before and after the policy is implemented, obtained through an event-study design for the treated states.\footnote{We collected data on total Google searches for terms related to the pandemic for all states in the sample. The search terms include ``pandemic" and ``Covid-19". Google analyses a sample of total searches and makes available the relative amount of searches at each point in time. The results are normalized to a fraction of the highest number of searches in time for each state. In order to make the data comparable across states, we use the \cite{SD2014} methodology.} We also plot the 95\% confidence interval for each estimate. Note there is a small increase in searches a few days after the lockdown announcement, but its magnitude is very limited and it vanishes almost immediately.

\begin{figure}[H]
    \centering
        \caption{Lockdown treatment effects on searches for traditional (top panel) and alternative (bottom panel) methods of fighting the pandemic}
    \label{F:google1}\includegraphics[scale=1]{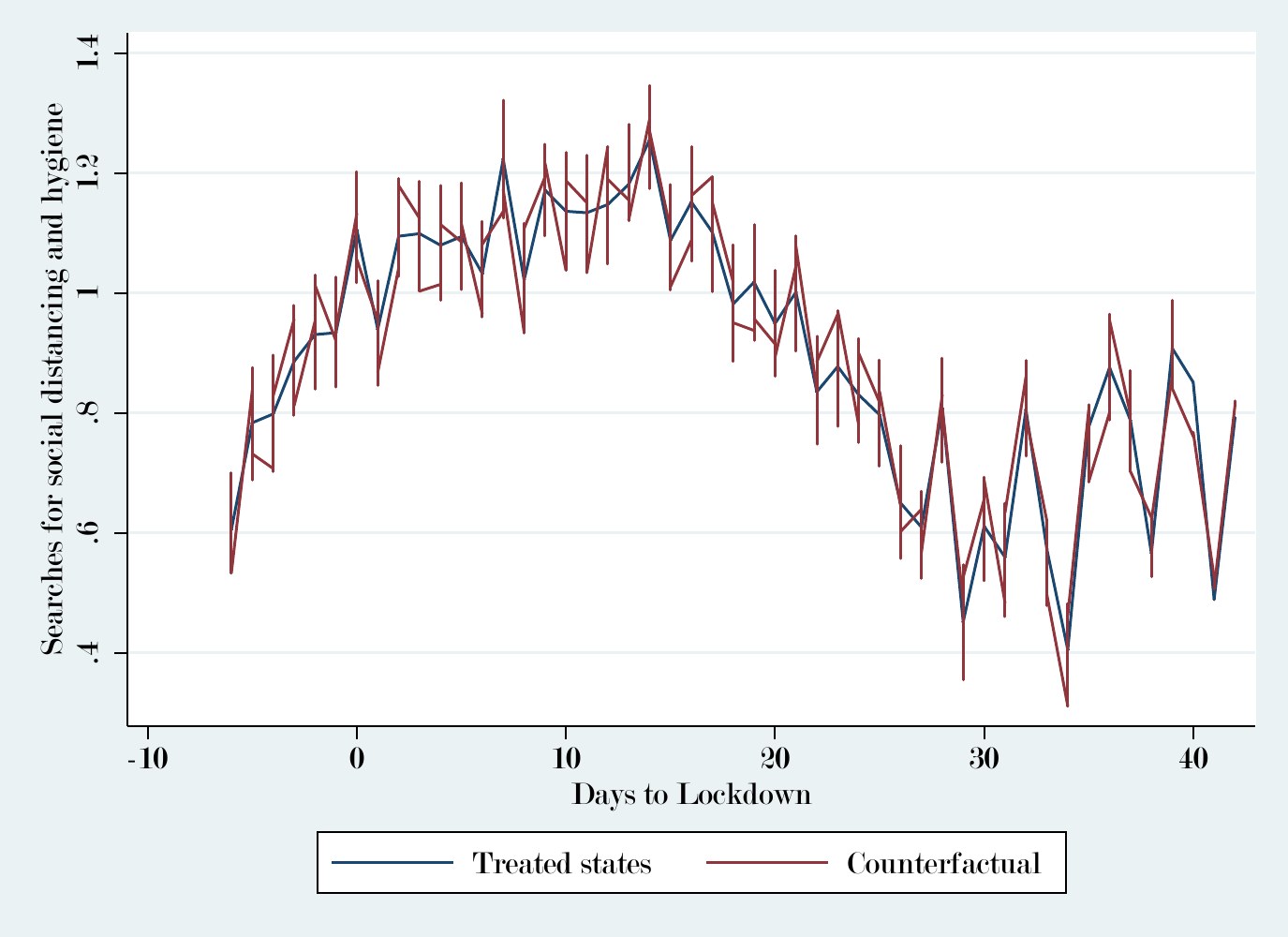}\\
        \includegraphics[scale=1]{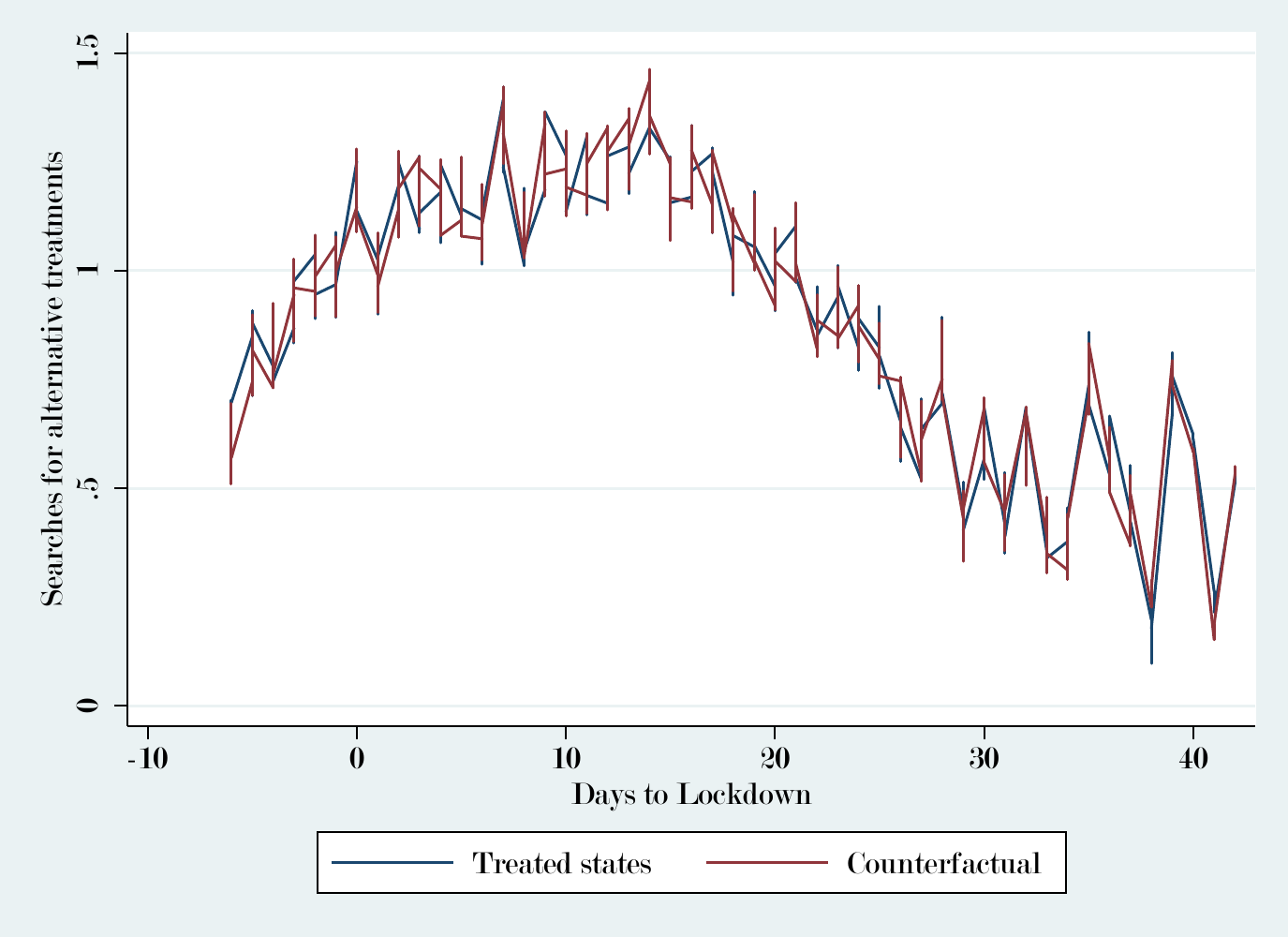}

\end{figure}

 We also examine if lockdown measures impact Google searchers related to traditional and alternative methods to fight the pandemic. The search terms for traditional methods include ``social distancing", ``mask use" and ``washing hands". The search terms for alternative treatment include ``zinc", ``hydroxychloroquine", and ``Covid alternative treatments".\footnote{Again, we use \cite{SD2014} methodology to standardize the data.}  Results are reported in Figure \ref{F:google1}. The top (bottom) panel plots searchers for traditional (alternative) methods in both treated and control sates.

We find little evidence that the search patterns are systematically different between treatment and control groups, or that the lockdown implementation affected these search patterns. Hence, information seem to evolve in an aggregate way and not to be affected by policies. We  further rewrite the causal diagram to incorporate this insight.
\begin{figure}[H]
\caption{Causal Direct Acyclic Graph (DAG) with timing and mechanisms restrictions}
\label{F:DAG_final}
\centering
\begin{tikzpicture}
\node (U) {$U_{st}$};
\node[right of=U, yshift=2cm] (L) {$L_{st}$};
\node [right of=L, xshift=1cm] (I) {$I_{t}$};
\node[below of=I] (B) {$B_{st}$};
\node[right of=B, xshift=1cm] (Y) {$Y_{st+l}$};

\draw[->] (U) --  (L);
\draw[->] (U) --  (B);
\draw[->] (L) --  (B);
\draw[->] (B) --  (Y);
\draw [->] (I) -- (B);
\draw[->] (I) --  (Y);
\draw[->] (U.south) .. controls +(down:1cm) and +(right:0cm) .. (Y.south);
\end{tikzpicture}
\vspace{-0.5cm}
\end{figure}
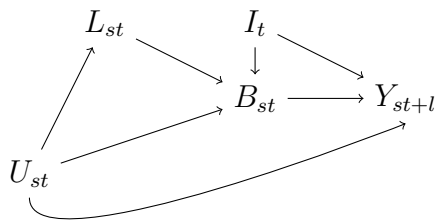

By eliminating the casual link between $L_{st}$ and $I_t$, this simplification allows a  straightforward interpretation of the treatment. The lockdown policy affects pandemic evolution mainly through its effects on behavior. To confirm this, we evaluate the impact of lockdown policies in mobility, captured by Google Mobility Data (described in Appendix \ref{ssec:google_mobility}), above and beyond the effects that would have happened endogenously.

\begin{figure}[H]
\centering
\caption{Lockdown effects on outside (top panel) and residential (bottom panel) mobility}
\label{F:mob}
\includegraphics[scale=1]{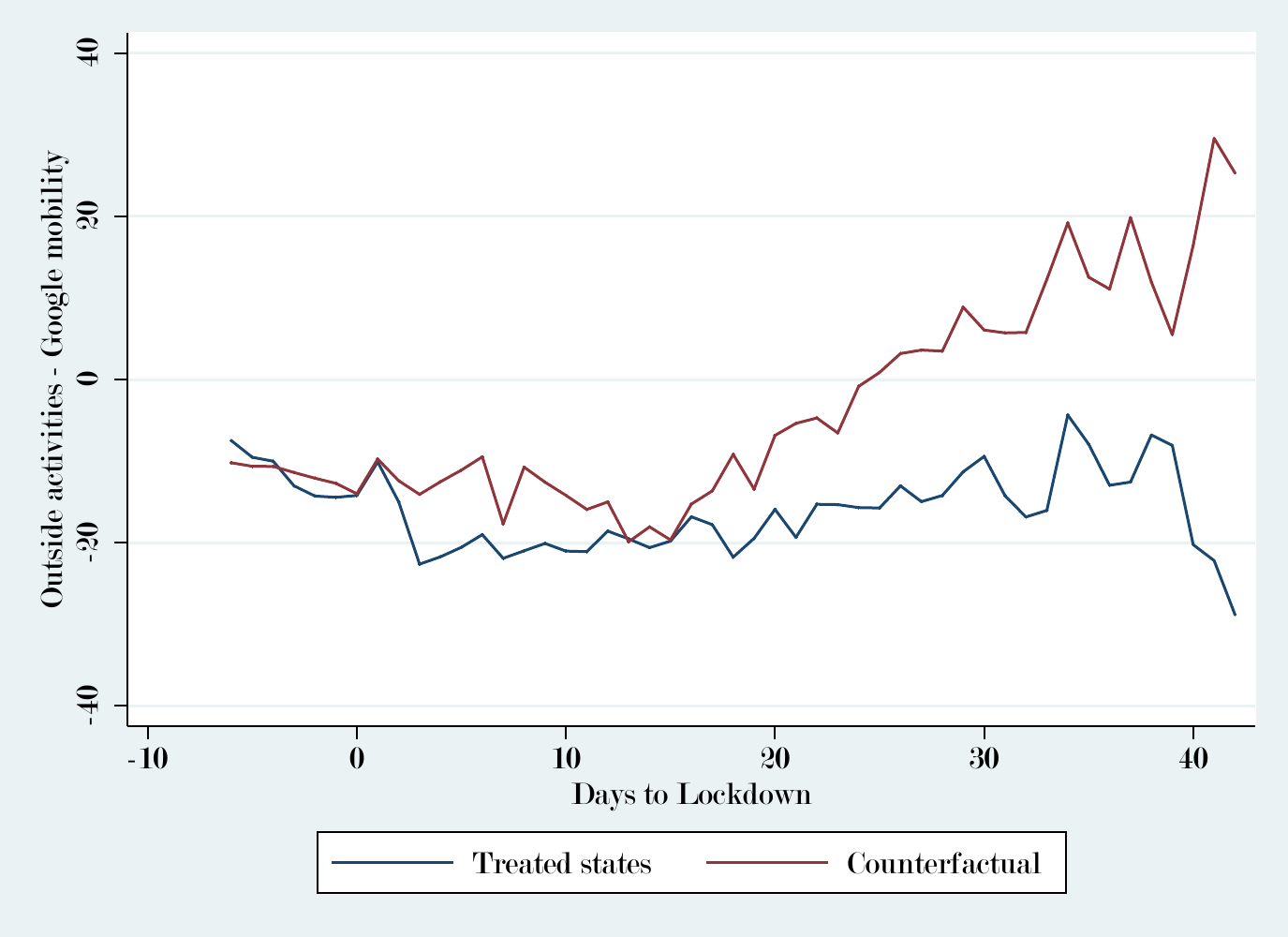}
\includegraphics[scale=1]{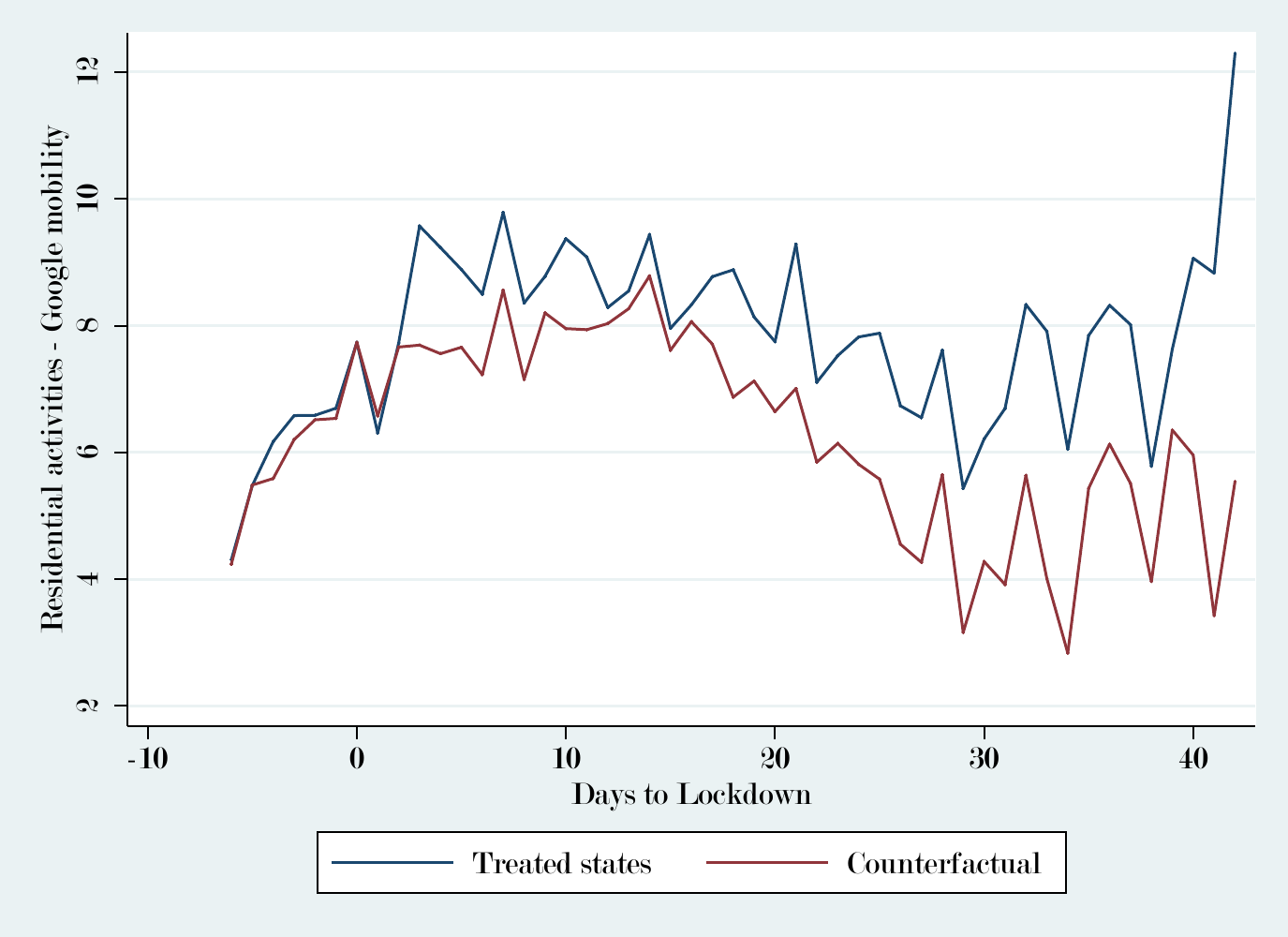}
\end{figure}

Similar to the way we compute the counterfactual to cumulative deaths, we compute the counterfactual  to mobility in residential and outdoor areas. Figure \ref{F:mob} plots the average  measures of mobility that in fact realized in treated states (blue lines), and the average counterfactual measures were lockdowns not implemented there (red lines). We consider the same control states above and we use the same weights as in the main estimates. The top panel considers outside activities, whereas the bottom panel considers residential activities.

As the figure makes it clear, lockdown policies affected the pandemic evolution mainly through its effects on behavior, captured by this substantial decrease (increase) in outside (residential) mobility in treated states relative to the counterfactuals (that is, above and beyond endogenous behavioral responses in the absence of lockdowns).

\subsection{Omitted variable bias and estimation}

In this subsection we discuss the problem of omitted variable bias. To do so, we write a linear model based on the DAG in Figure \ref{F:DAG_final}. The variable of interest depends on available information ($I_{st}$), individuals' behavior ($B_{st}$) and confounders ($U_{st}$). That is,
\[
Y_{s,t+l}=\pi B_{st}+\mu I_{st}+\delta U_{st}+\epsilon^Y_{st},
\]
where $\pi$, $\mu$, and $\delta$ are unknown parameters and $\epsilon^Y_{st}$ is a zero-mean random term.

The behavior of individuals also depends on the lockdown policies implemented ($L_{st}$) and can be expressed as
\[
B_{st} = \alpha L_{st}+\eta I_{st}+\epsilon^B_{st},
\]
where $\alpha$ and $\eta$ are unknown parameters and $\epsilon^B_{st}$ is a zero-mean random term.

As a consequence, we can write the reduced-form of the model above as
\[
Y_{s,t+l}=\beta L_{st}+\gamma I_{st}+\delta U_{st}+\epsilon_{st},
\]
where $\beta=\alpha\pi$, $\gamma=\pi\eta+\mu$, and $\epsilon_{st}=\pi\epsilon^B_{st} + \epsilon^Y_{st}$.

Since we do not observe $U_{st}$ and these confounders are correlated with $L_{st}$, a simple ordinary least-squares (OLS) estimation of the reduced-form equation  does not identify $\beta$.  The ArCo methodology used in this paper helps to mitigate the effects of potential confounders.

Indeed, for each state, we find a (non-necessarily convex) combination  $\boldsymbol{w}$ of states in the control pool that most closely matches the number of cases in the treated state. Thus, we propose the following estimator,
\[
\widehat{\boldsymbol{\beta}}_{s,t+l}=Y_{s,{t+l}}-\omega_0-\boldsymbol{\omega}'\boldsymbol{Y}^C_{t+l},
\]
where $\boldsymbol{Y}^C_{t}$  is the vector of cases for the control states. Also, note that
\[
\widehat{\beta}_{s,t+l}=\beta+\delta(U_{st}-\omega_0-\boldsymbol{\omega}'\boldsymbol{U}^C_{t}),
\]
where $\boldsymbol{U}^C_t$ is the vector of non-observed confounders for the control group. It is clear that if
\[
U_{st}=\omega_0+\boldsymbol{\omega}'\boldsymbol{U}^C_{t},
\]
then the proposed estimator recovers the effect of the lockdown policy on the number of registered cases. That is, the additional identification hypothesis is that the combination of estimated weights does not only reproduce the trends in the in-sample period but also the non-observed relevant variables. Finally, note that $\beta$ recovers precisely the effect of lockdowns through the behavioral channel.

\section{Conclusion}
\label{sec:conclusion}

In this paper, as opposed to most of the early and incipient literature on the lockdown effects during the Covid-19 crisis, we consider a purely data-driven approach to assess the impact of lockdowns on the short-run evolution of the number of cases and deaths in some US states. Also, as opposed to some recent papers that use a difference-in-difference approach, we adopt a variant of the synthetic control approach. On average, according to the synthetic controls, the counterfactual accumulated number of cases would be two times larger were lockdown policies not implemented in treated states.

\clearpage

\appendix
\setcounter{figure}{0} \renewcommand{\thefigure}{\thesection.\arabic{figure}}
\setcounter{table}{0} \renewcommand{\thetable}{\thesection.\arabic{table}}
\setcounter{equation}{0} \renewcommand{\theequation}{\thesection.\arabic{equation}}

\section{Appendix: Additional analyzes and results}
\label{sec:appendix}

\subsection{Reopen Dates}\label{app:reopen}

In the first two columns of Table \ref{tab:table_reopen} we show the date of the first confirmed case in every treated state we analyze and its reopen date (plus ten days), whenever available at the time we started to circulate this paper.\footnote{In other words, it represents the tenth day after the day the state-wide lockdown policy was supposed to end. It is possible that these dates changed.}  In the third column, we show the difference (in days) from the first confirmed case and the reopen date plus ten days.

\begin{table}[!h]
\centering
\caption{Number of Days from First Case until Reopen for each Treatment State}
\begin{tabular}{|l|ccc|}
  \hline
State & First Case & Reopen (+10) & Days Diff. \\
  \hline
  \hline
  Alabama & 03/13/2020 & 05/10/2020 & 58 \\
  Colorado & 03/06/2020 & 05/07/2020 & 62 \\
  Florida & 03/02/2020 & 05/14/2020 & 73 \\
  Georgia & 03/03/2020 & 05/10/2020 & 68 \\
  Kansas & 03/08/2020 & 05/14/2020 & 67 \\
  Kentucky & 03/06/2020 & - &  - \\
  Maine & 03/12/2020 & 05/10/2020 & 59 \\
  Maryland & 03/06/2020 & 06/11/2020 & 97 \\
  Mississippi & 03/12/2020 & 05/17/2020 & 66 \\
  Missouri & 03/08/2020 & 05/14/2020 & 67 \\
  Nevada & 03/05/2020 & 05/19/2020 & 75 \\
  New Hampshire & 03/02/2020 & 05/21/2020 & 80 \\
  New York & 03/02/2020 & 05/25/2020 & 84 \\
  North Carolina & 03/03/2020 & 05/18/2020  & 76  \\
  Oregon & 02/29/2020 & 05/25/2020 & 86  \\
  Pennsylvania & 03/06/2020 & 06/08/2020 & 94 \\
  Rhode Island & 03/01/2020 & 05/18/2020 & 78 \\
  South Carolina & 03/07/2020 & 05/14/2020 & 68  \\
  Tennessee & 03/05/2020 & 05/07/2020 & 63 \\
  Texas & 03/05/2020 & 05/10/2020 & 66 \\
   \hline
\end{tabular}
\label{tab:table_reopen}
\end{table}

These figures illustrate why we had to limit our sample size to only 58 epidemiological days. For example, if we had used 60 days in our analysis, we would have to exclude Alabama and Maine from our treated states, given that they would not be in a state-wide lockdown in the last days of the out-of-sample period.

\clearpage

\subsection{Cumulative Cases: ArCo Statistics}\label{app:statistics}

We report in the first seven rows of Table \ref{tab:table_coef} the coefficients estimated by the LASSO model for each treated state for the $T_0$+10 in-sample period, where $T_0$ is the epidemiological day the lockdown was implemented in a given treated state. The last two rows display the mean and the median (across the out-of-sample period) of the ratio between the actual cumulative cases and the counterfactual cases for every state.

\begin{table}[H]
\caption{LASSO Coefficients}
\resizebox{\columnwidth}{!}{\begin{tabular}{lcccccccccccccccccccc}
\multicolumn{21}{c}{LASSO Coefficients ($T_0$+10)}                                                                                                                                                  \\ \cline{2-21} 
                                        & AL   & CO   & FL    & GA    & KS    & KY    & ME   & MD    & MS   & MO    & NV    & NH    & NY    & NC    & OR    & PA    & RI    & SC    & TN    & TX    \\ \hline
\multicolumn{1}{l|}{(Intercept)}        & 3.49 & 0.95 & -1.93 & -0.47 & -0.31 & -1.45 & 2.49 & -0.30 & 3.38 & -0.62 & -0.31 & -3.36 & -2.00 & -2.56 & -1.10 & -0.75 & -7.31 & 0.66  & -1.78 & -0.32 \\
\multicolumn{1}{l|}{Arkansas}           & 0.00 & 0.00 & 0.00  & 0.00  & 0.00  & 0.00  & 0.00 & 0.00  & 0.00 & 0.00  & 0.00  & 0.00  & 0.00  & 0.00  & 0.00  & 0.00  & 1.48  & 0.00  & 0.00  & 0.00  \\
\multicolumn{1}{l|}{California}         & 0.00 & 0.00 & -0.25 & -0.14 & 0.00  & 0.00  & 0.00 & 0.00  & 0.00 & -0.36 & -0.36 & 0.00  & 0.00  & 0.00  & 0.00  & 0.00  & 0.21  & -0.20 & 0.00  & 0.00  \\
\multicolumn{1}{l|}{Iowa}               & 0.21 & 0.82 & 0.91  & 0.68  & 0.77  & 0.78  & 0.31 & 0.60  & 0.18 & 0.54  & 0.35  & 0.00  & 0.79  & 0.88  & 0.44  & 1.12  & 0.52  & 0.45  & 0.46  & 1.00  \\
\multicolumn{1}{l|}{Nebraska}           & 0.00 & 0.00 & 0.41  & 0.24  & 0.00  & 0.08  & 0.00 & 0.20  & 0.03 & -0.16 & 0.38  & 0.20  & 0.00  & 0.00  & 0.56  & 0.22  & 0.00  & 0.21  & -0.09 & 0.00  \\
\multicolumn{1}{l|}{North Dakota}       & 0.00 & 0.00 & 0.32  & 0.15  & 0.00  & 0.39  & 0.20 & 0.00  & 0.10 & 0.70  & 0.45  & 1.02  & 1.40  & 0.61  & 0.00  & 0.00  & -0.73 & 0.26  & 1.23  & 0.14  \\
\multicolumn{1}{l|}{South Dakota}       & 0.29 & 0.27 & 0.00  & 0.20  & 0.16  & 0.00  & 0.00 & 0.35  & 0.18 & 0.00  & 0.00  & 0.36  & 0.00  & 0.00  & 0.16  & 0.00  & 0.40  & 0.00  & -0.14 & 0.09  \\
\multicolumn{1}{l|}{Log($t$)}           & 0.43 & 0.00 & 0.24  & 0.32  & 0.23  & 0.07  & 0.21 & 0.28  & 0.41 & 0.88  & 0.42  & 0.00  & -     & 0.13  & 0.00  & 0.29  & 0.08  & 0.53  & 0.33  & 0.19  \\ \hline
\multicolumn{1}{l|}{Mean $y/\hat{y}$ (OSS)}   & 0.81 & 0.57 & 0.63  & 0.76  & 0.73  & 0.68  & 0.82 & 0.71  & 0.97 & 1.12  & 1.12  & 0.51  & 0.33  & 0.48  & 0.57  & 0.46  & 0.76  & 0.89  & 0.39  & 0.54  \\
\multicolumn{1}{l|}{Median $y/\hat{y}$ (OSS)} & 0.78 & 0.53 & 0.58  & 0.74  & 0.68  & 0.68  & 0.80 & 0.70  & 0.97 & 1.09  & 1.12  & 0.46  & 0.22  & 0.41  & 0.55  & 0.40  & 0.72  & 0.87  & 0.31  & 0.48  \\ \hline
\end{tabular}}
\label{tab:table_coef}
\end{table}

With only two exceptions (Missouri and Nevada), every state has an out-of-sample mean and median of the observed-to-predicted ratio below one. This means that, on average, the realized cumulative cases were smaller than the counterfactual, which highlight that lockdowns had a meaningful impact on slowing down the Covid-19 spread in these states.

\subsubsection{ArCo forecasts for every state}

Table \ref{tab:arco_forecast} reports the ratio of the counterfactual cumulative cases to the actual ones on the 58th day after the first confirmed case in each state. It also reports the lower and upper limits of the 95\% confidence interval. Among the 20 treated states, the ratio is larger than one in 18 of them. For Missouri and Nevada, there is no evidence on the effectiveness of lockdown policies. For Mississipi and South Carolina, the impacts of lockdowns are only modest. Note that New York is clearly an outlier, with such ratio around 16.5. We discuss this case in the main text.

In contrast, among the non-treated states, we obtain ratios close to one for three out of six cases. We assume that the cut-off of the placebo intervention is $T_0 = 36$,  which is the median (and the mean) timing of the policy interventions in the treated states. As discussed in the main text  and in Appendix \ref{ssec:google_mobility}, South Dakota, which displays a ratio well above one, experienced a large reduction in outside mobility even without official  lockdown measures. California and Nebraska, which display ratios below one, had very few Covid-19 confirmed cases during the period before the cut-off.   

\begin{table}[!h]
\centering
\caption{Counterfactual/actual cases ratio for every State (58th day)}
\begin{tabular}{|l|cccc|}
  \hline
State & ArCo forecast & ArCo lb & ArCo ub & Treated \\
  \hline
  \hline
  Alabama & 1.30 & 1.14 & 1.44 & Yes \\ 
  Colorado & 2.78 & 2.34 & 3.25 & Yes \\ 
  Florida & 2.48 & 2.11 & 2.94 & Yes \\ 
  Georgia & 1.68 & 1.43 & 2.10 & Yes \\ 
  Kansas & 1.49 & 1.34 & 1.69 & Yes \\ 
  Kentucky & 2.68 & 2.31 & 2.99 & Yes \\ 
  Maine & 1.45 & 1.33 & 1.59 & Yes \\ 
  Maryland & 2.08 & 1.72 & 2.32 & Yes \\ 
  Mississippi & 1.03 & 0.98 & 1.08 & Yes \\ 
  Missouri & 0.74 & 0.61 & 0.86 & Yes \\ 
  Nevada & 0.82 & 0.63 & 1.05 & Yes \\ 
  New Hampshire & 2.90 & 2.15 & 3.55 & Yes \\ 
  New York & 16.48 & 11.44 & 21.18 & Yes \\
  North Carolina & 3.69 & 2.95 & 4.28 & Yes \\ 
  Oregon & 3.96 & 3.21 & 4.33 & Yes \\ 
  Pennsylvania & 5.63 & 4.99 & 6.41 & Yes \\ 
  Rhode Island & 1.72 & 1.47 & 2.06 & Yes \\ 
  South Carolina & 1.18 & 1.03 & 1.31 & Yes \\ 
  Tennessee & 4.25 & 3.56 & 4.95 & Yes \\ 
  Texas & 3.19 & 2.56 & 4.00 & Yes \\ 
  \hline
  \hline
  Arkansas & 0.97 & 0.93 & 1.03 & No \\ 
  California & 0.01 & 0.01 & 0.01 & No \\ 
  Iowa & 0.93 & 0.81 & 1.02 & No \\ 
  Nebraska & 0.28 & 0.23 & 0.32 & No \\
  North Dakota & 0.92 & 0.78 & 1.09 & No \\ 
  South Dakota & 3.13 & 2.72 & 3.59 & No \\   
   \hline
\end{tabular}
\label{tab:arco_forecast}
\end{table}


\clearpage

\subsection{Cumulative cases: ArCo estimates for treated states} \label{app:cases}

\begin{figure}[!htbp]
    \caption{ArCo Estimates for Alabama  (Cumulative Cases)}
    \centering
    \includegraphics[width=\textwidth]{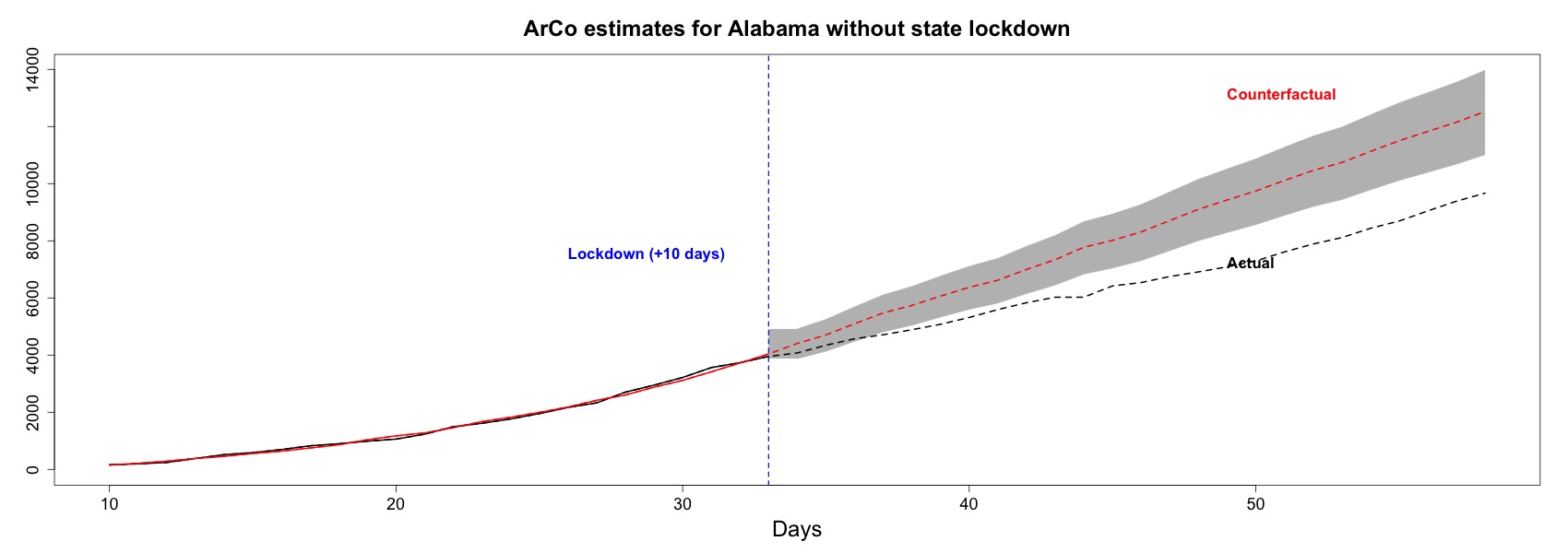}
    \label{fig:cs_all_AL}
\end{figure}

\begin{figure}[!htbp]
    \caption{ArCo Estimates for Colorado  (Cumulative Cases)}
    \centering
    \includegraphics[width=\textwidth]{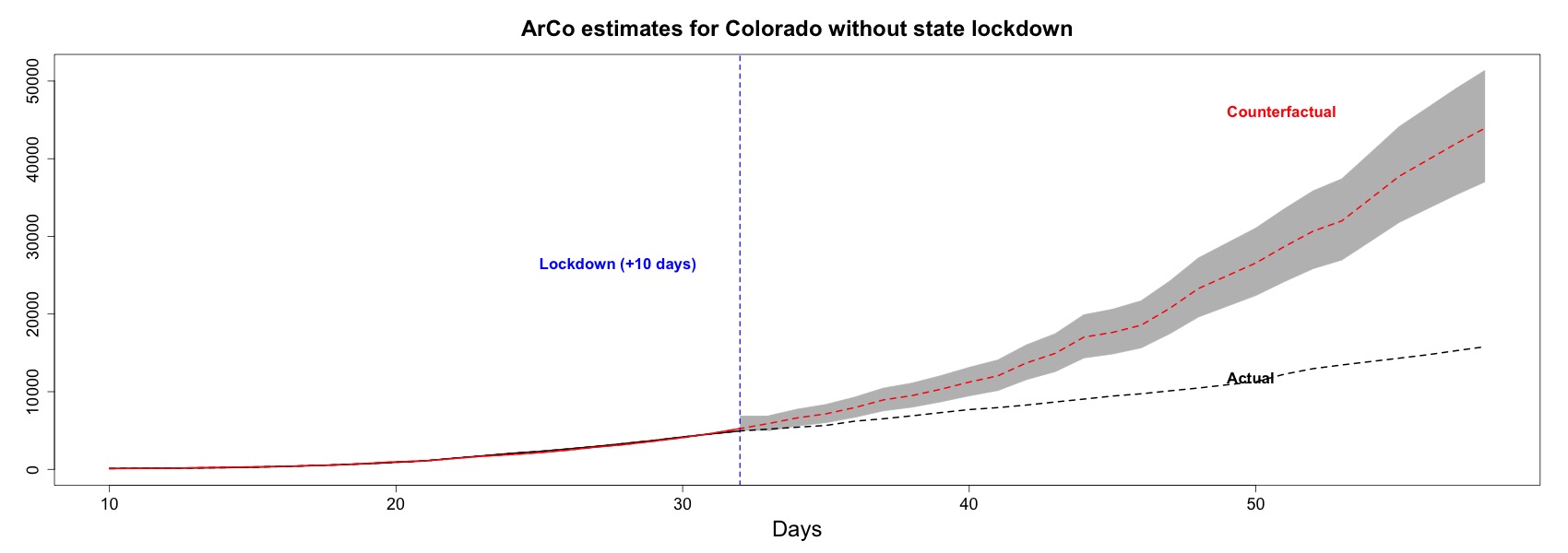}
    \label{fig:cs_all_CO}
\end{figure}

\begin{figure}[!htbp]
    \centering
    \caption{ArCo Estimates for Florida  (Cumulative Cases)}
    \includegraphics[width=\textwidth]{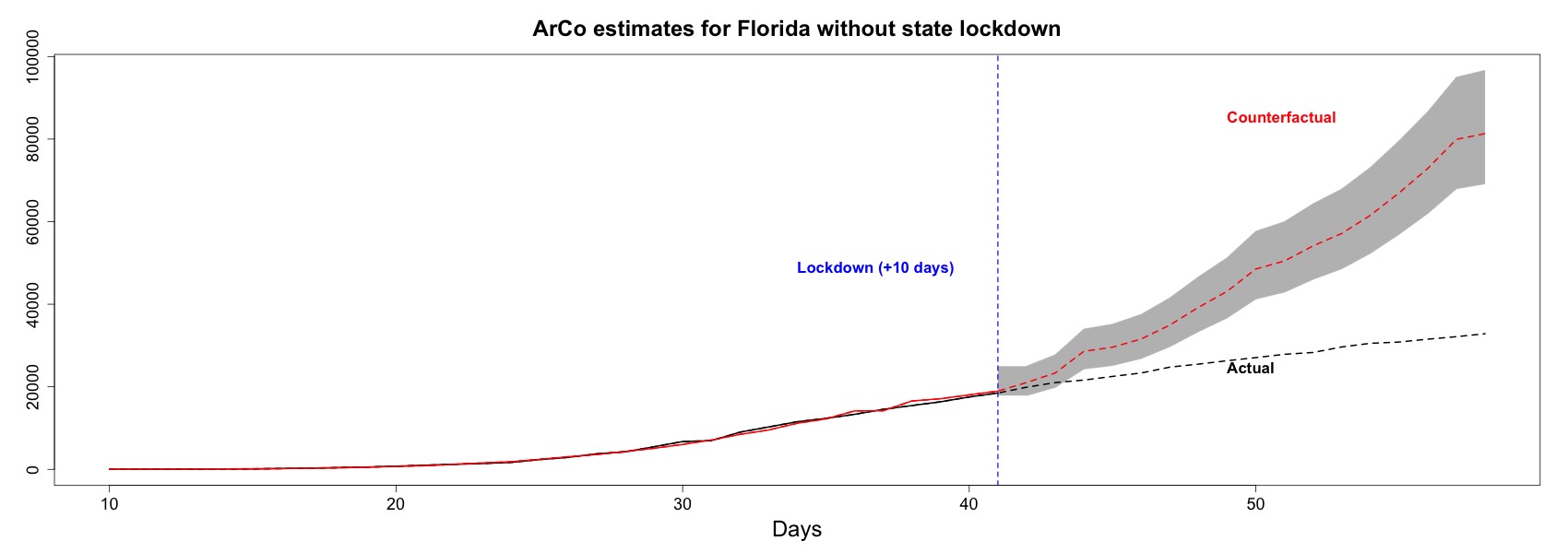}
    \label{fig:cs_all_FL}
\end{figure}

\begin{figure}[!htbp]
    \centering
    \caption{ArCo Estimates for Georgia  (Cumulative Cases)}
     \includegraphics[width=\textwidth]{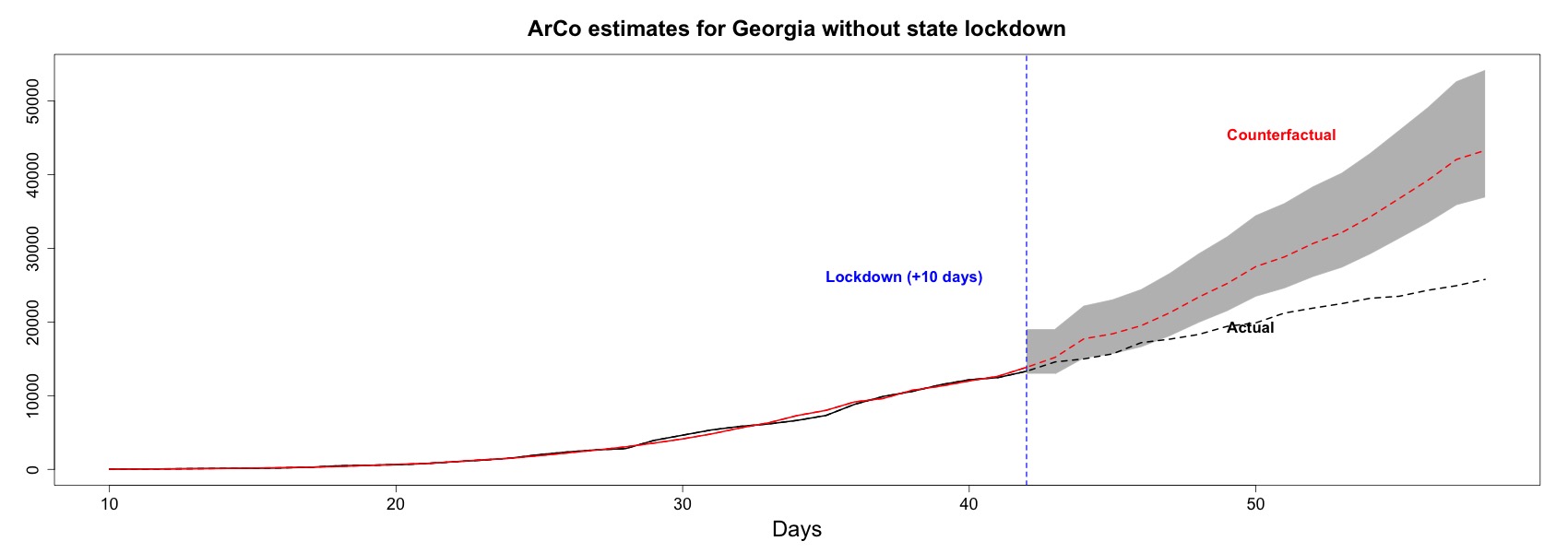}
    \label{fig:cs_all_GA}
\end{figure}

\begin{figure}[!htbp]
    \centering
    \caption{ArCo Estimates for Kansas  (Cumulative Cases)}
    \includegraphics[width=\textwidth]{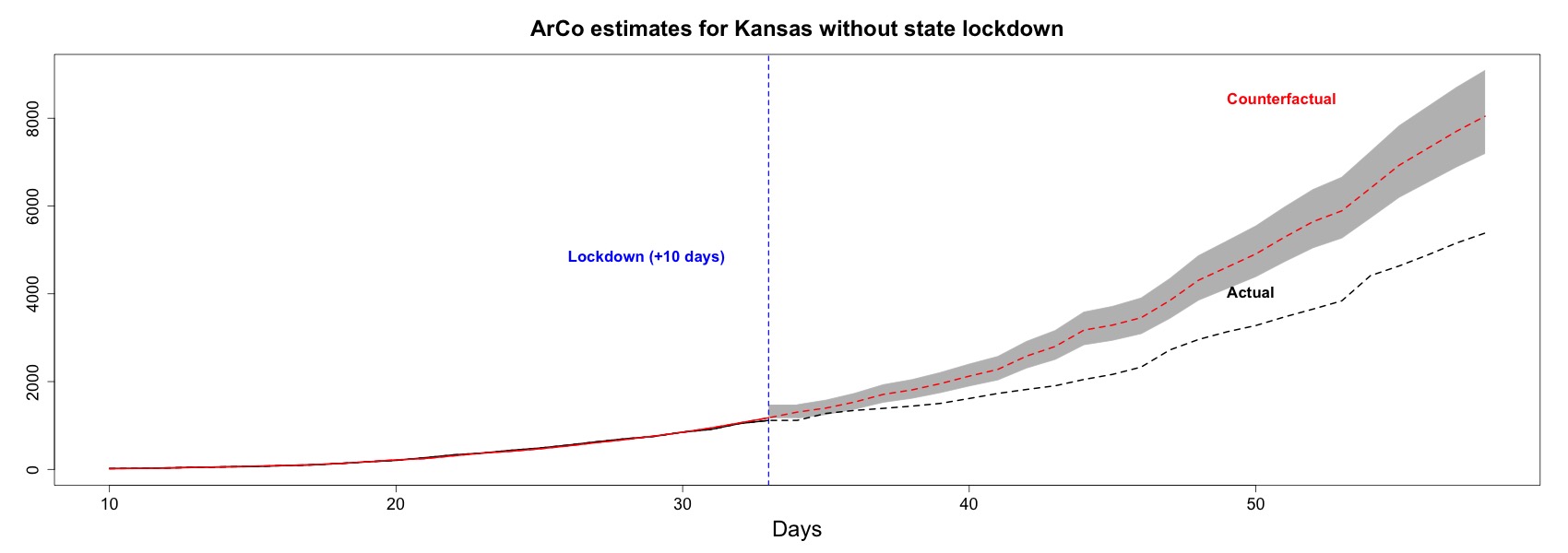}
    \label{fig:cs_all_KS}
\end{figure}

\begin{figure}[!htbp]
    \centering
    \caption{ArCo Estimates for Kentucky  (Cumulative Cases)}
    \includegraphics[width=\textwidth]{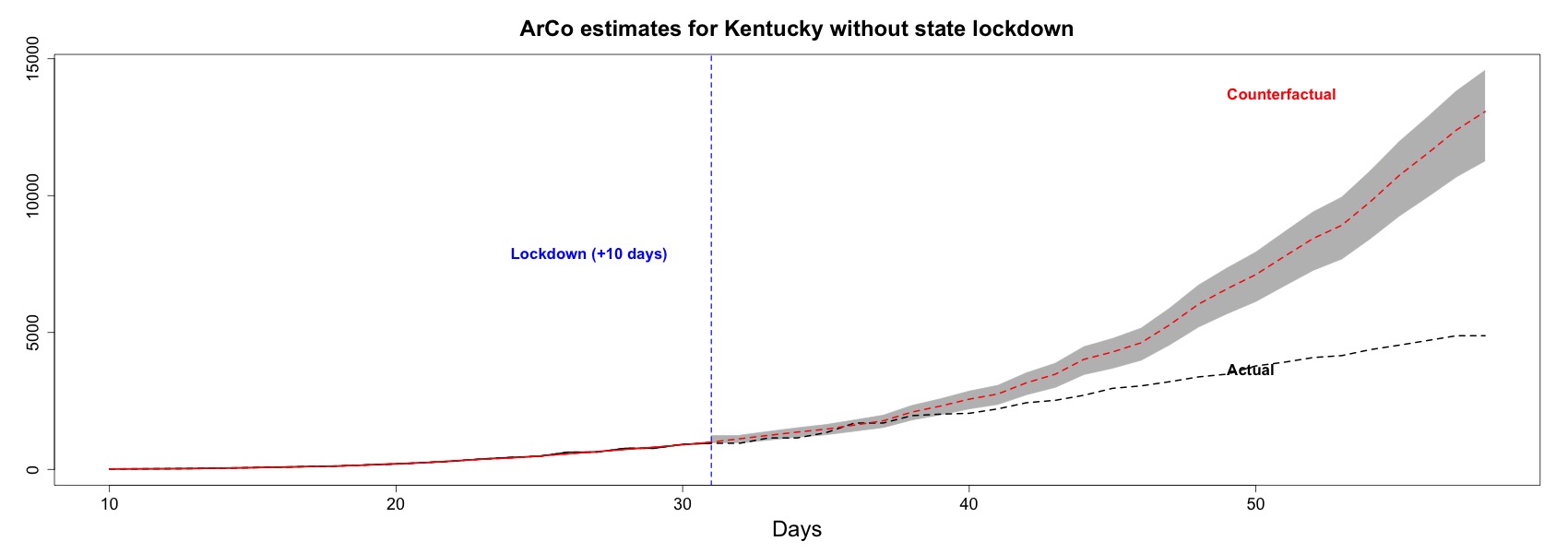}
    \label{fig:cs_all_KY}
\end{figure}

\begin{figure}[!htbp]
    \centering
    \caption{ArCo Estimates for Maryland  (Cumulative Cases)}
    \includegraphics[width=\textwidth]{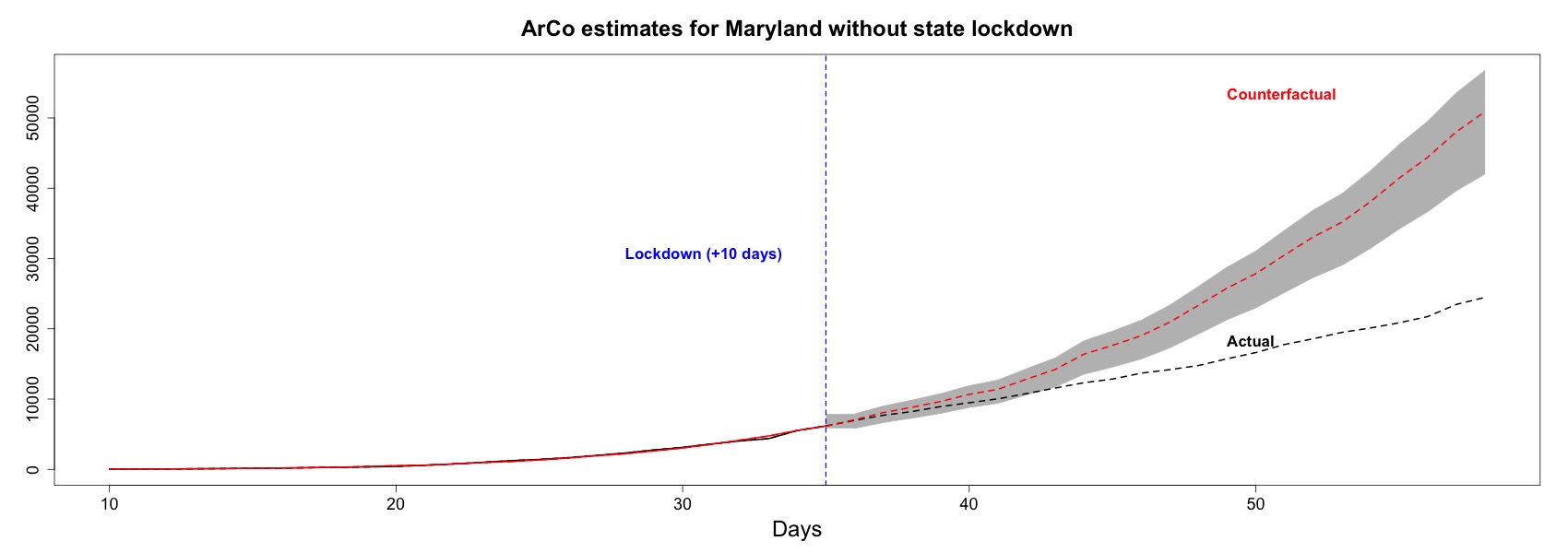}
    \label{fig:cs_all_MD}
\end{figure}

\begin{figure}[!htbp]
    \centering
    \caption{ArCo Estimates for Maine  (Cumulative Cases)}
    \includegraphics[width=\textwidth]{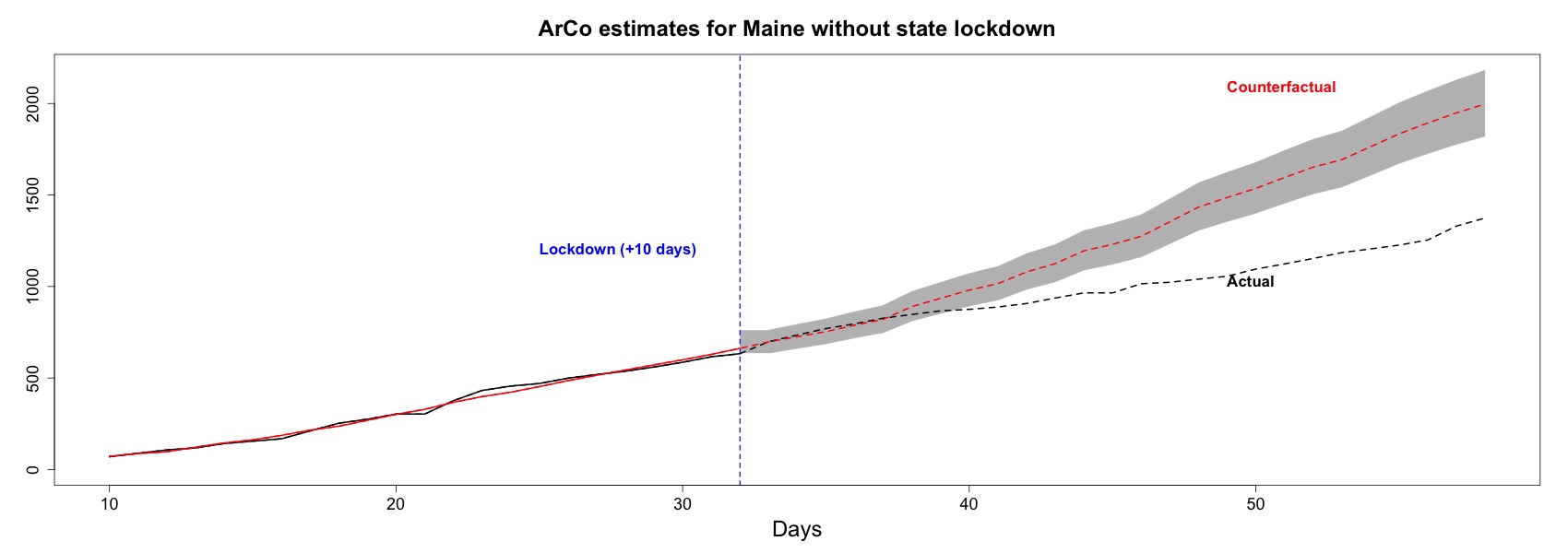}
    \label{fig:cs_all_ME}
\end{figure}

\begin{figure}[!htbp]
    \centering
    \caption{ArCo Estimates for Missouri  (Cumulative Cases)}
    \includegraphics[width=\textwidth]{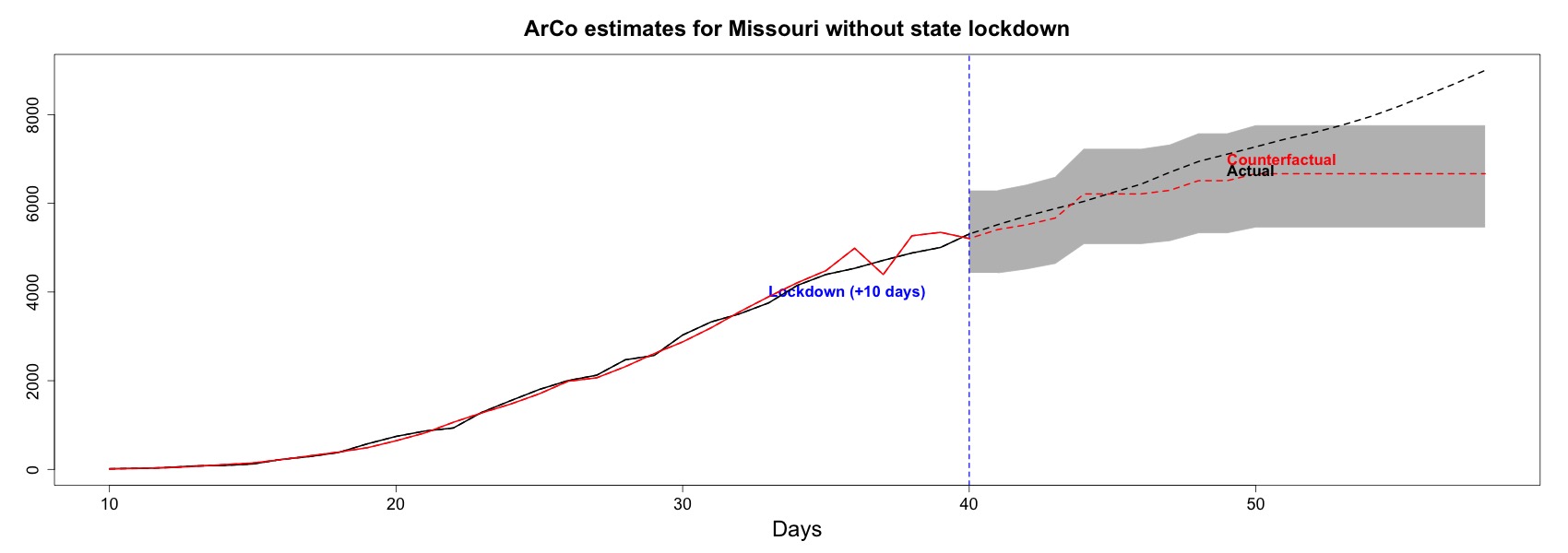}
    \label{fig:cs_all_MO}
\end{figure}

\begin{figure}[!htbp]
    \centering
    \caption{ArCo Estimates for Mississippi  (Cumulative Cases)}
    \includegraphics[width=\textwidth]{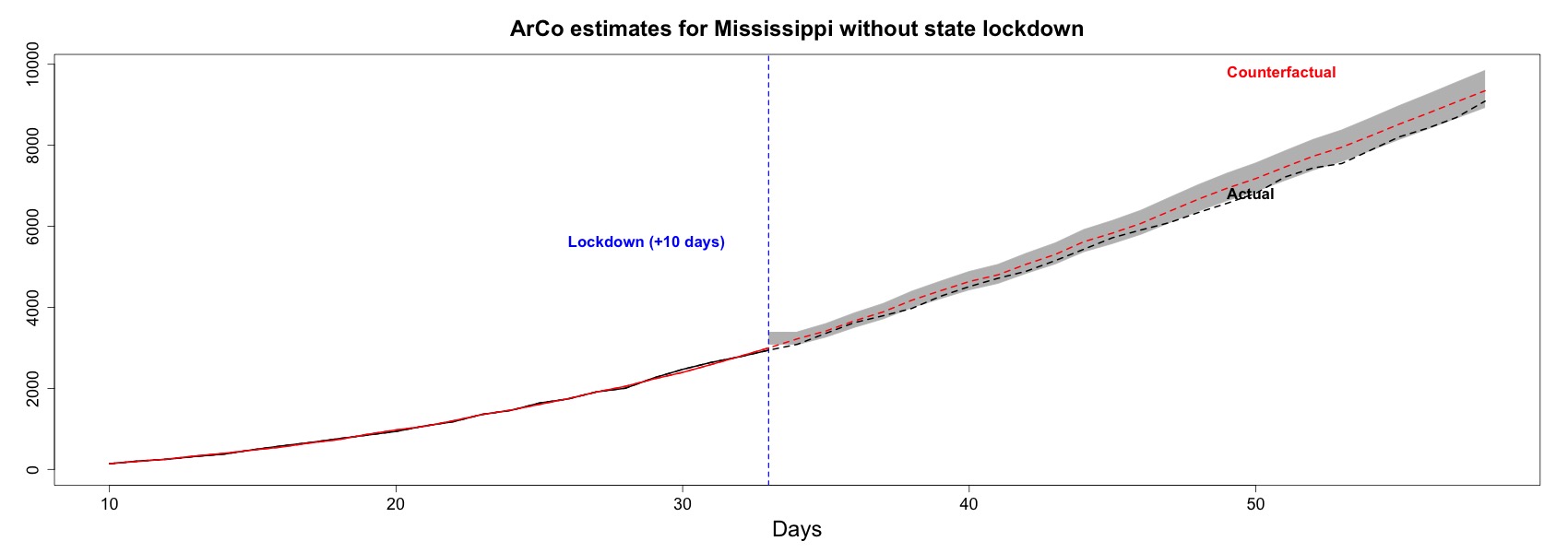}
    \label{fig:cs_all_MS}
\end{figure}

\begin{figure}[!htbp]
    \centering
    \caption{ArCo Estimates for North Carolina  (Cumulative Cases)}
    \includegraphics[width=\textwidth]{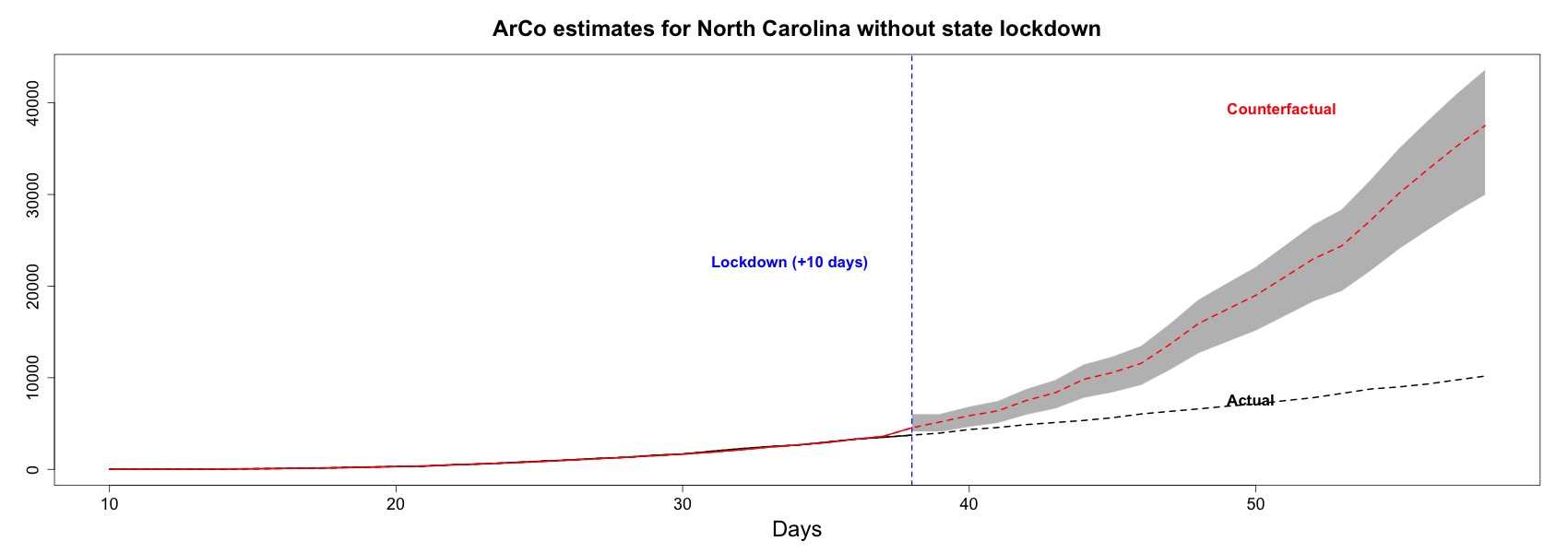}
    \label{fig:cs_all_NC}
\end{figure}

\begin{figure}[!htbp]
    \centering
    \caption{ArCo Estimates for Nevada  (Cumulative Cases)}
    \includegraphics[width=\textwidth]{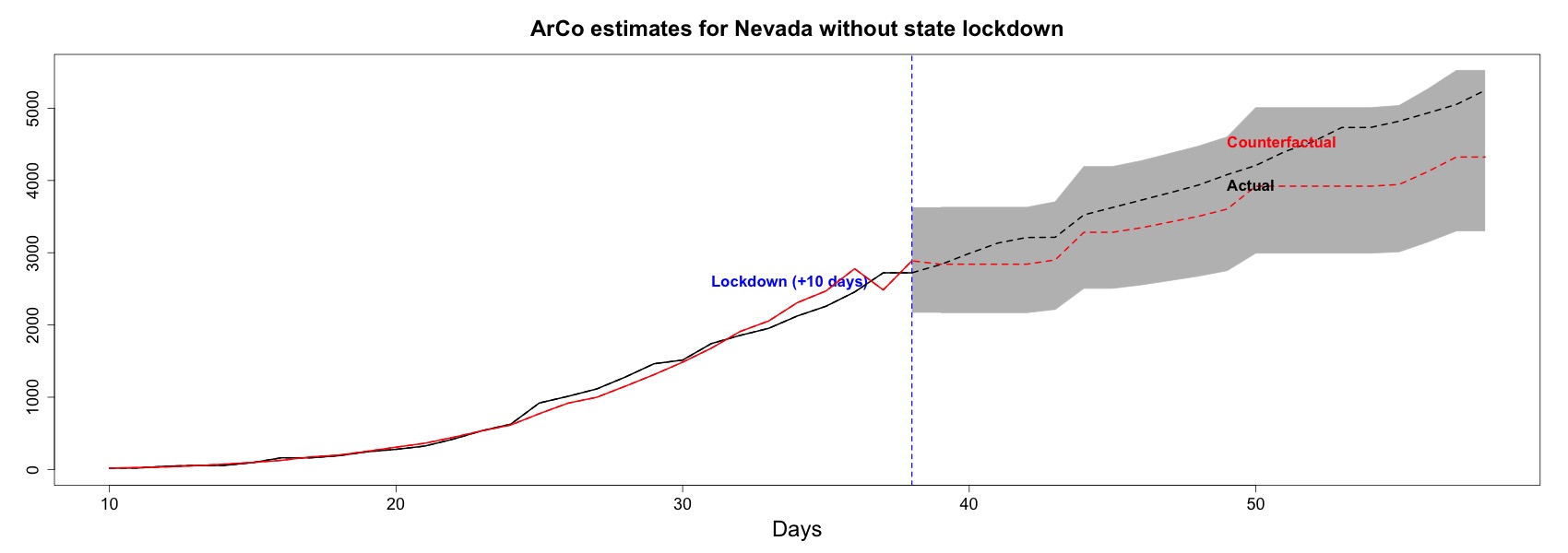}
    \label{fig:cs_all_NV}
\end{figure}

\begin{figure}[!htbp]
    \centering
    \caption{ArCo Estimates for Oregon  (Cumulative Cases)}
    \includegraphics[width=\textwidth]{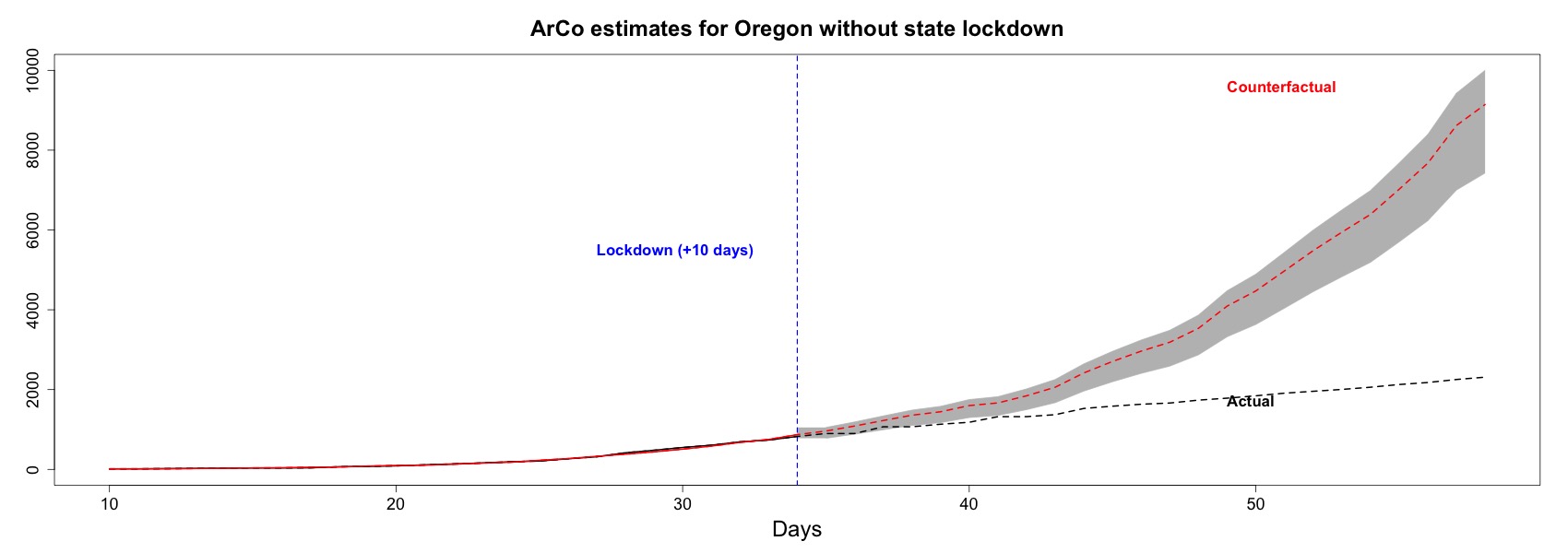}
    \label{fig:cs_all_OR}
\end{figure}

\begin{figure}[!htbp]
    \centering
    \caption{ArCo Estimates for Pennsylvania  (Cumulative Cases)}
    \includegraphics[width=\textwidth]{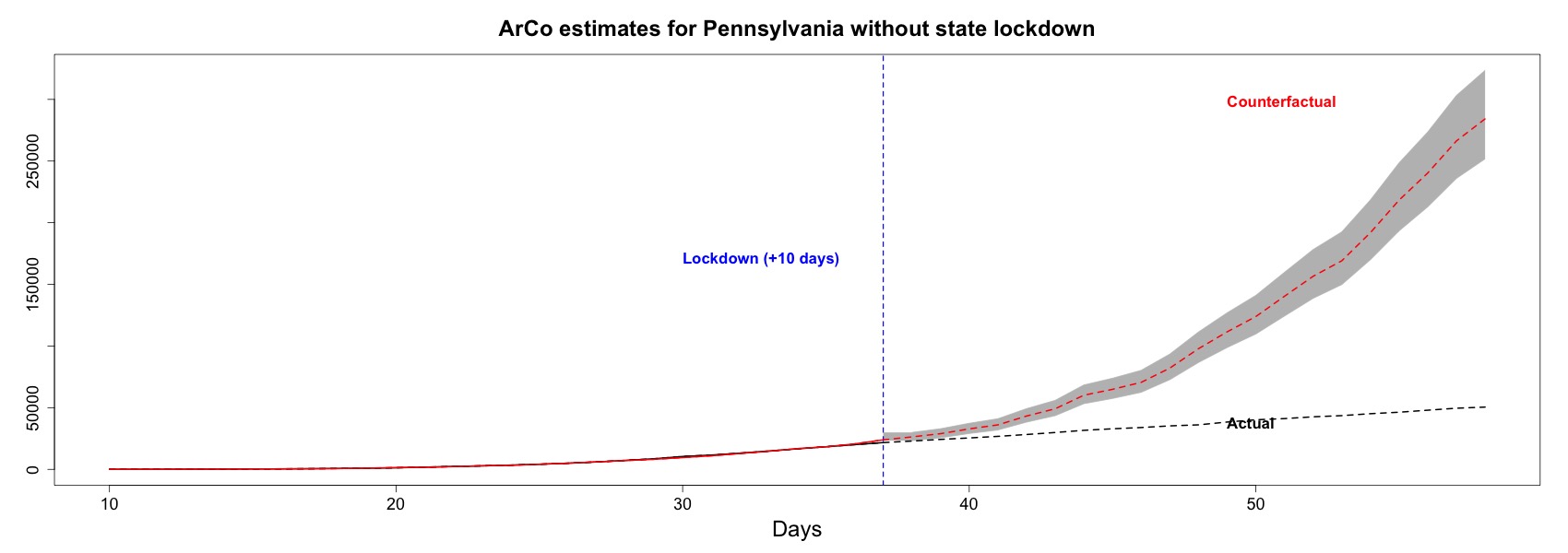}
    \label{fig:cs_all_PA}
\end{figure}

\begin{figure}[!htbp]
    \centering
    \caption{ArCo Estimates for Rhode Island  (Cumulative Cases)}
    \includegraphics[width=\textwidth]{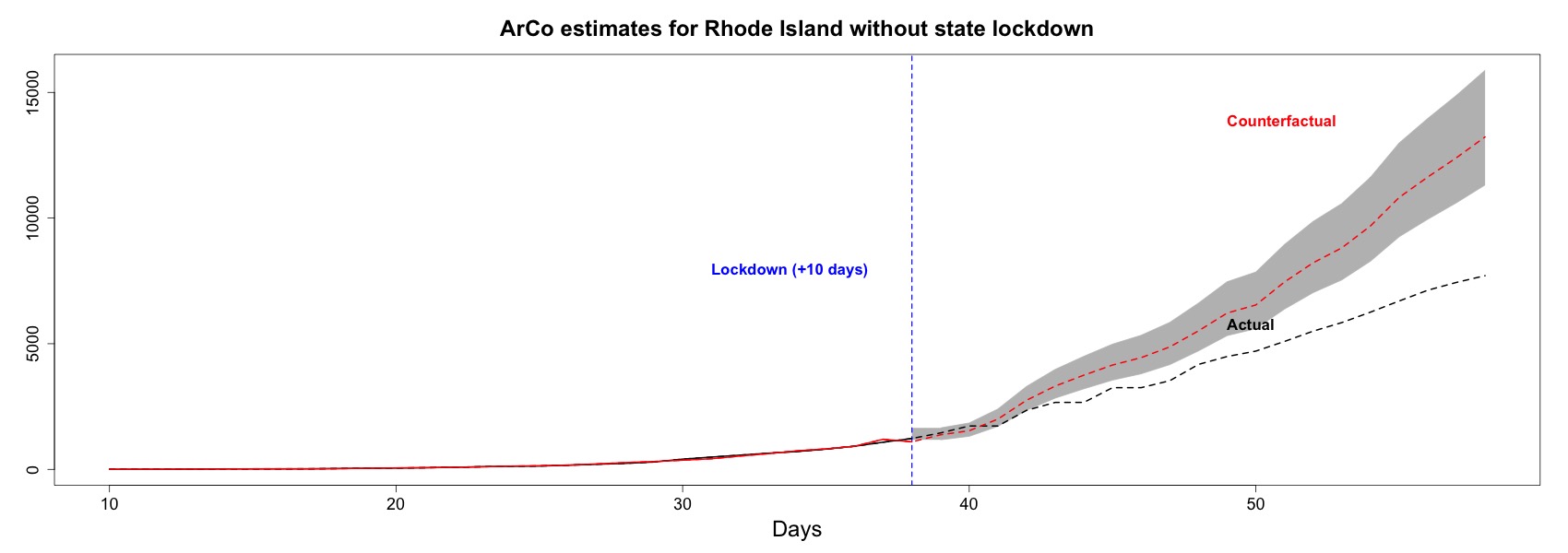}
    \label{fig:cs_all_RI}
\end{figure}

\begin{figure}[!htbp]
    \centering
    \caption{ArCo Estimates for South Carolina  (Cumulative Cases)}
    \includegraphics[width=\textwidth]{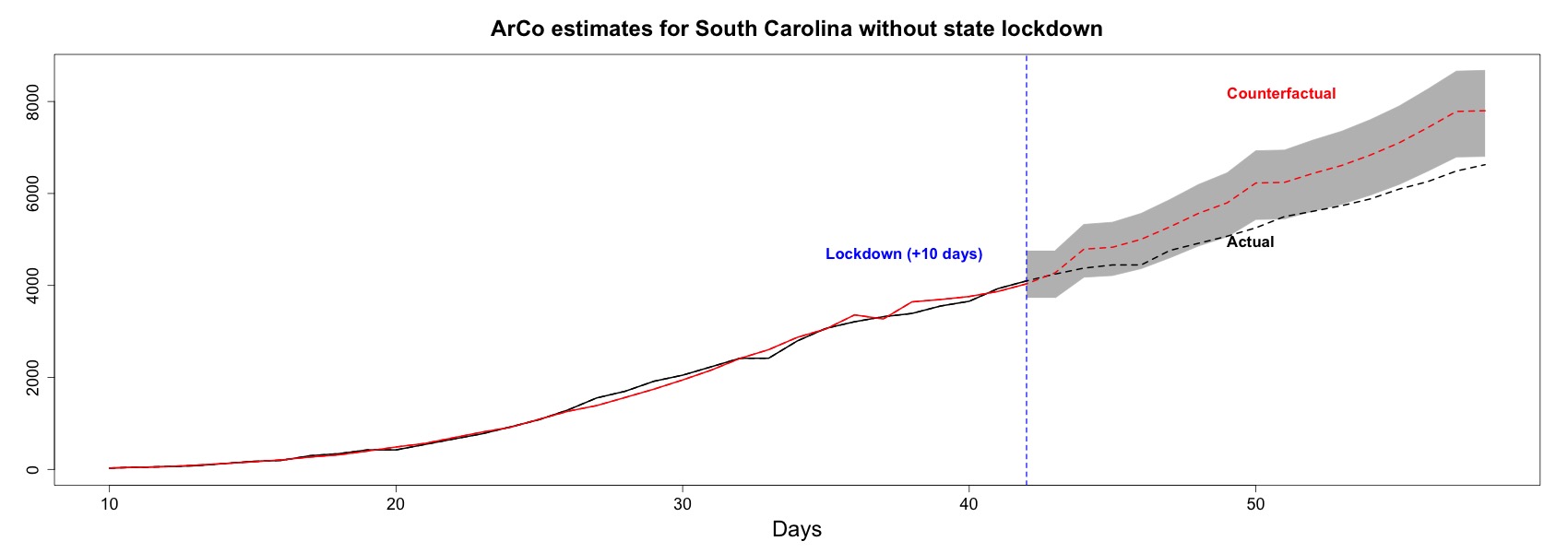}
    \label{fig:cs_all_SC}
\end{figure}

\begin{figure}[!htbp]
    \centering
    \caption{ArCo Estimates for Tennessee  (Cumulative Cases)}
    \includegraphics[width=\textwidth]{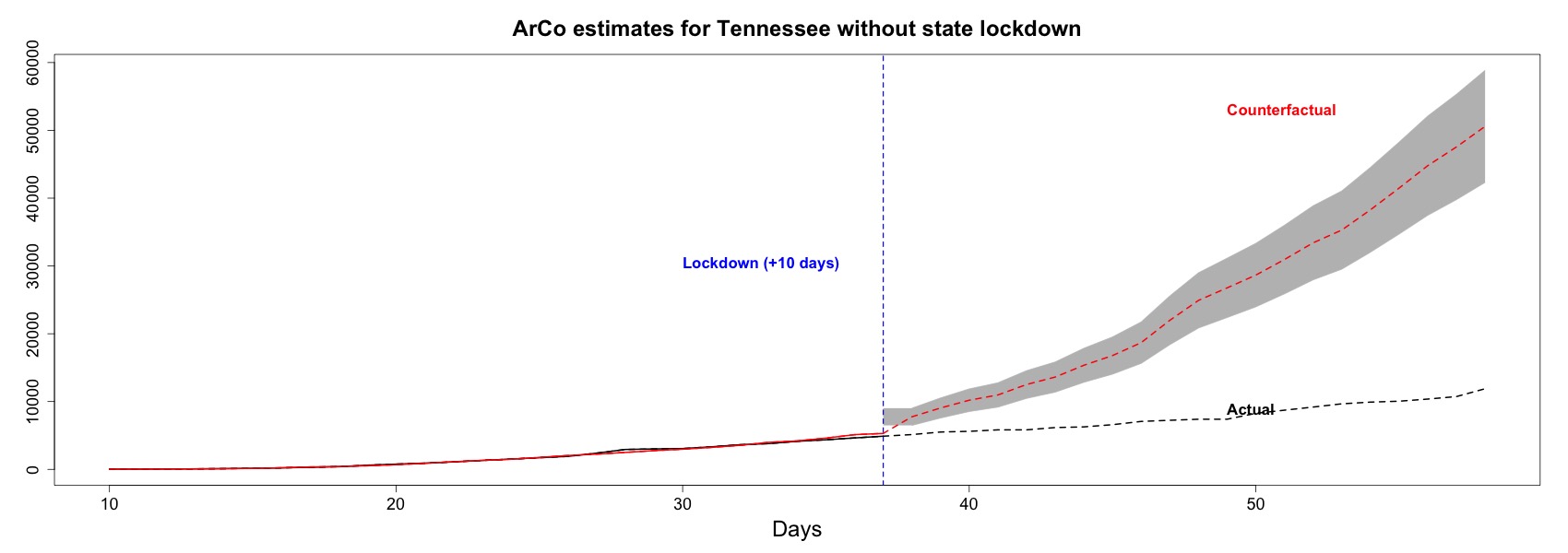}
    \label{fig:cs_all_TN}
\end{figure}

\begin{figure}[!htbp]
    \centering
    \caption{ArCo Estimates for Texas  (Cumulative Cases)}
    \includegraphics[width=\textwidth]{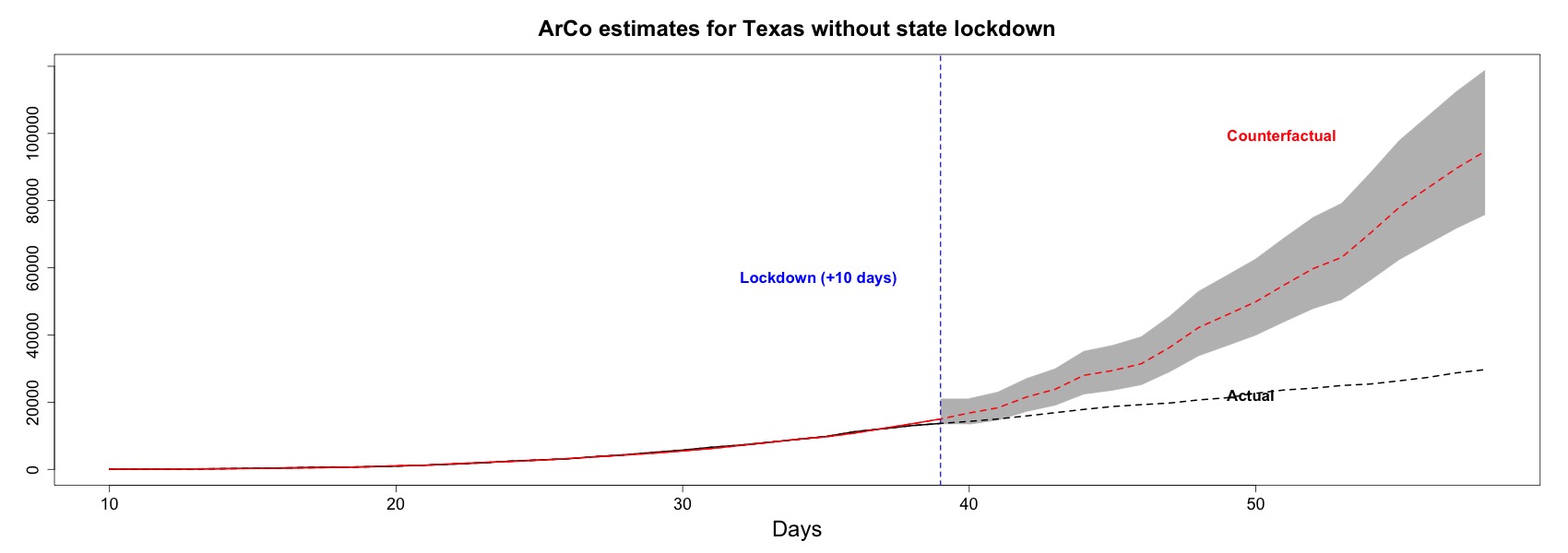}
    \label{fig:cs_all_TX}
\end{figure}

In this section, in Figures \ref{fig:cs_all_AL}--\ref{fig:cs_all_TX}, we report the counterfactual estimates for all states that adopted lockdown strategies. With the exception of Missouri, Mississippi, Nevada, and South Carolina, lockdown measures were effective in reducing the number of confirmed cases in the very short-run.

\begin{figure}[!htbp]
    \centering
    \caption{ArCo Estimates for New Hampshire  (Cumulative Cases)}
    \includegraphics[width=\textwidth]{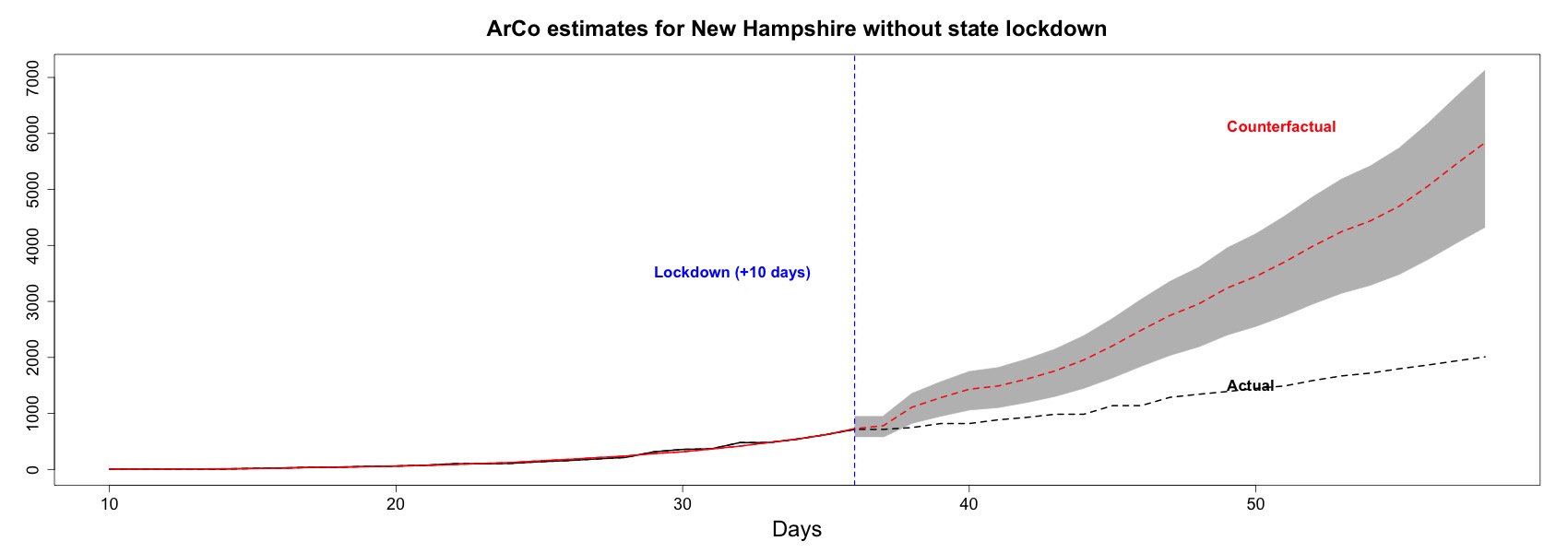}
    \label{fig:cs_all_NY}
\end{figure}

\begin{figure}[!htbp]
    \centering
    \caption{ArCo Estimates for New York  (Cumulative Cases)}
    \includegraphics[width=\textwidth]{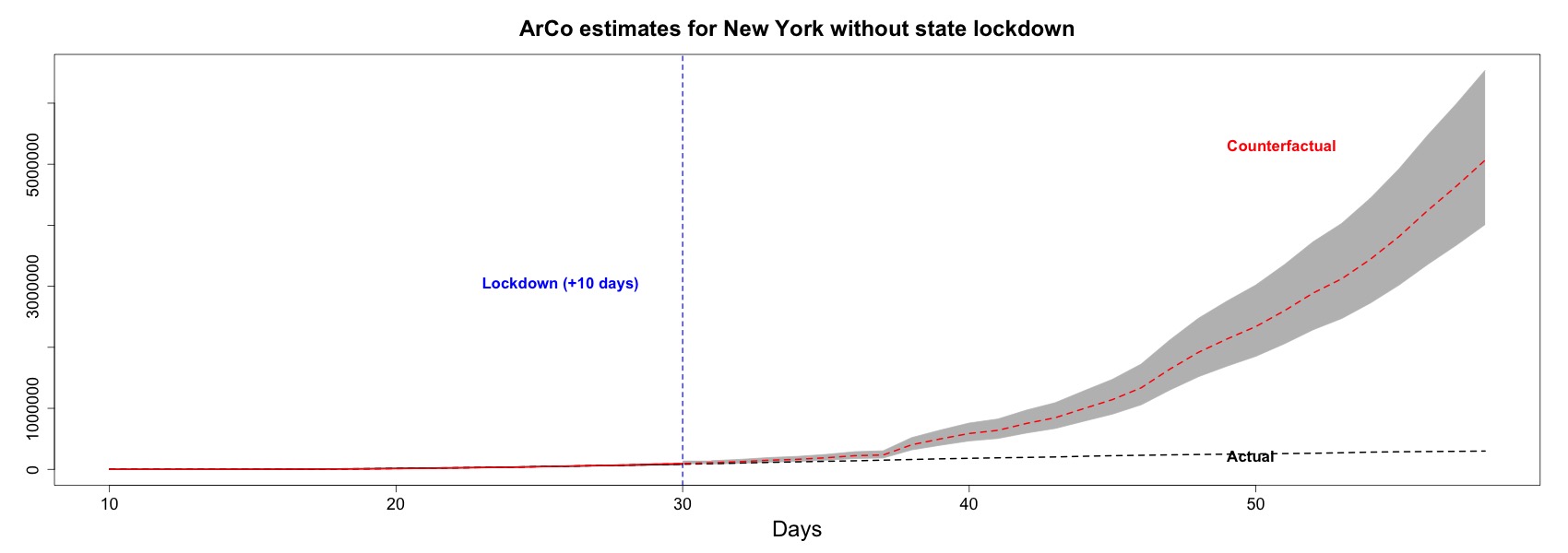}
    \label{fig:cs_all_NY}
\end{figure}

\subsection{Cumulative deaths: ArCo estimates for treated states} \label{app:deaths}

\begin{figure}[!htbp]
    \centering
    \caption{ArCo Estimates for Alabama  (Cumulative Deaths)}
    \includegraphics[width=\textwidth]{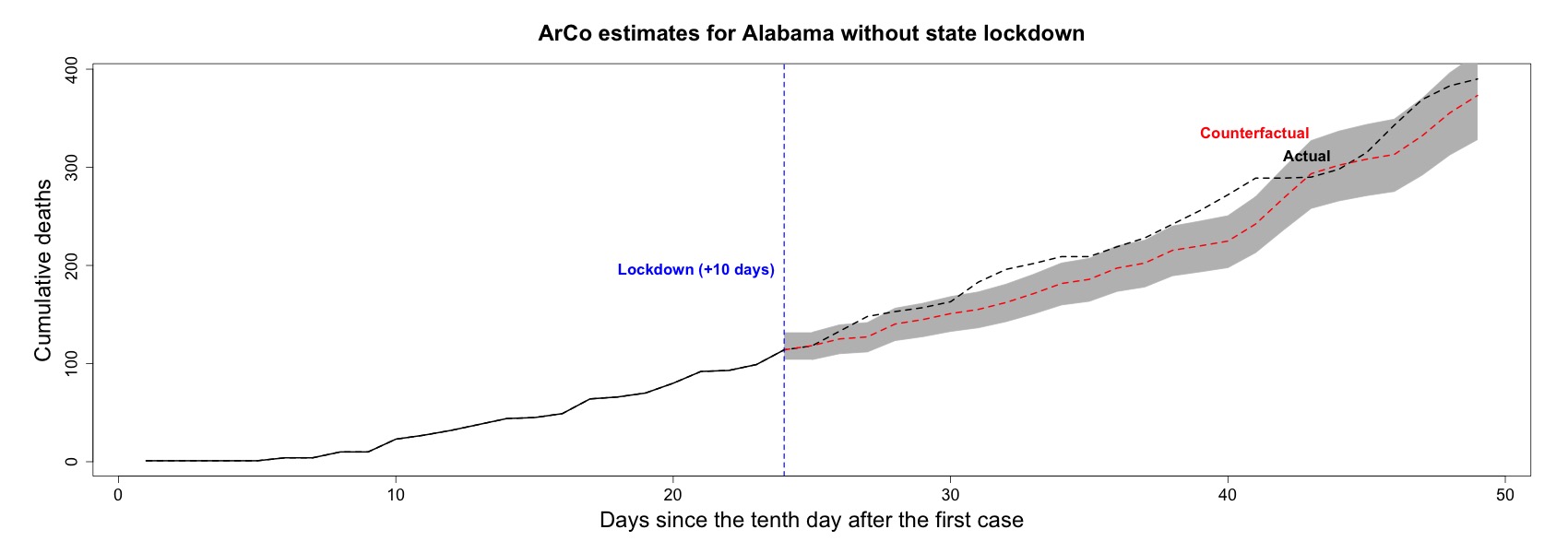}
    \label{fig:deaths_AL}
\end{figure}

\begin{figure}[!htbp]
    \centering
    \caption{ArCo Estimates for Colorado (Cumulative Deaths)}
    \includegraphics[width=\textwidth]{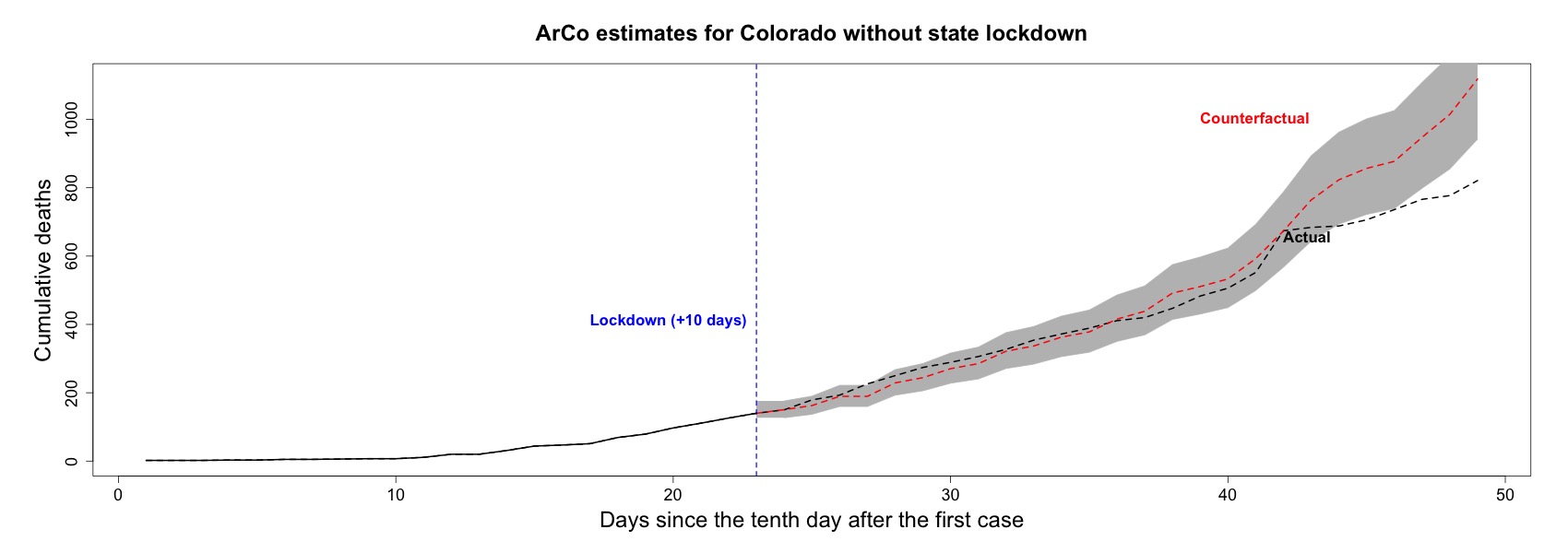}
    \label{fig:deaths_CO}
\end{figure}

\begin{figure}[!htbp]
    \centering
    \caption{ArCo Estimates for Florida (Cumulative Deaths)}
    \includegraphics[width=\textwidth]{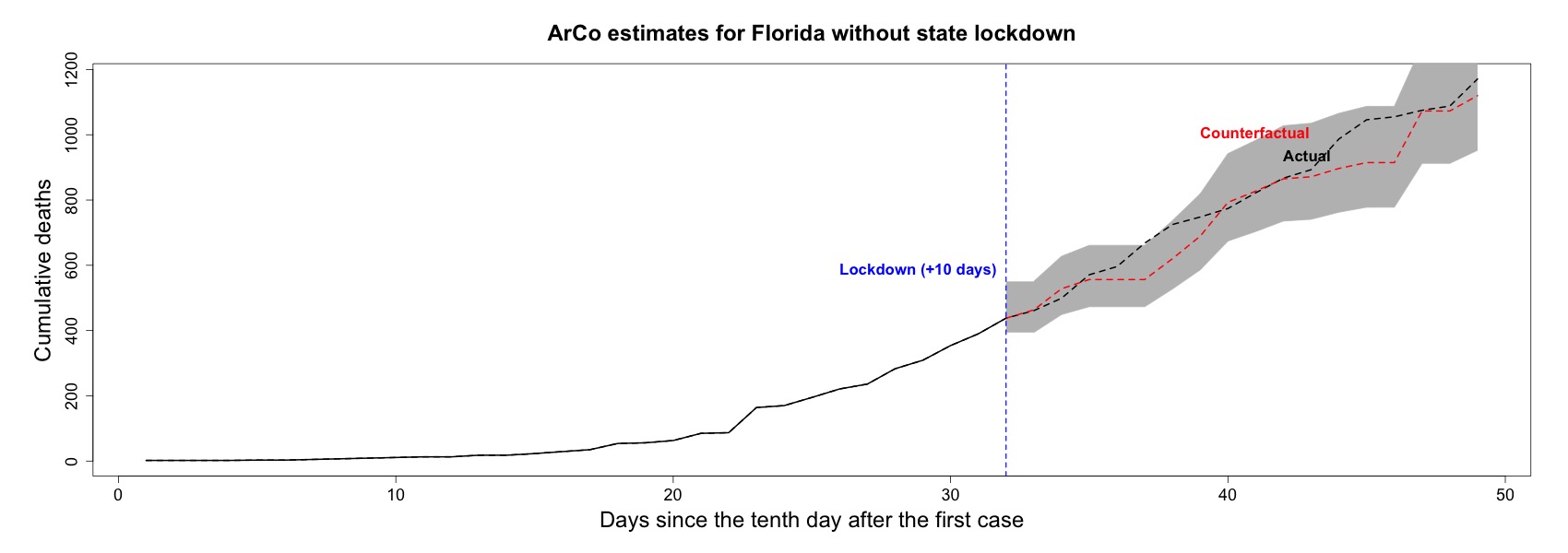}
    \label{fig:deaths_FL}
\end{figure}

\begin{figure}[!htbp]
    \centering
    \caption{ArCo Estimates for Georgia (Cumulative Deaths)}
    \includegraphics[width=\textwidth]{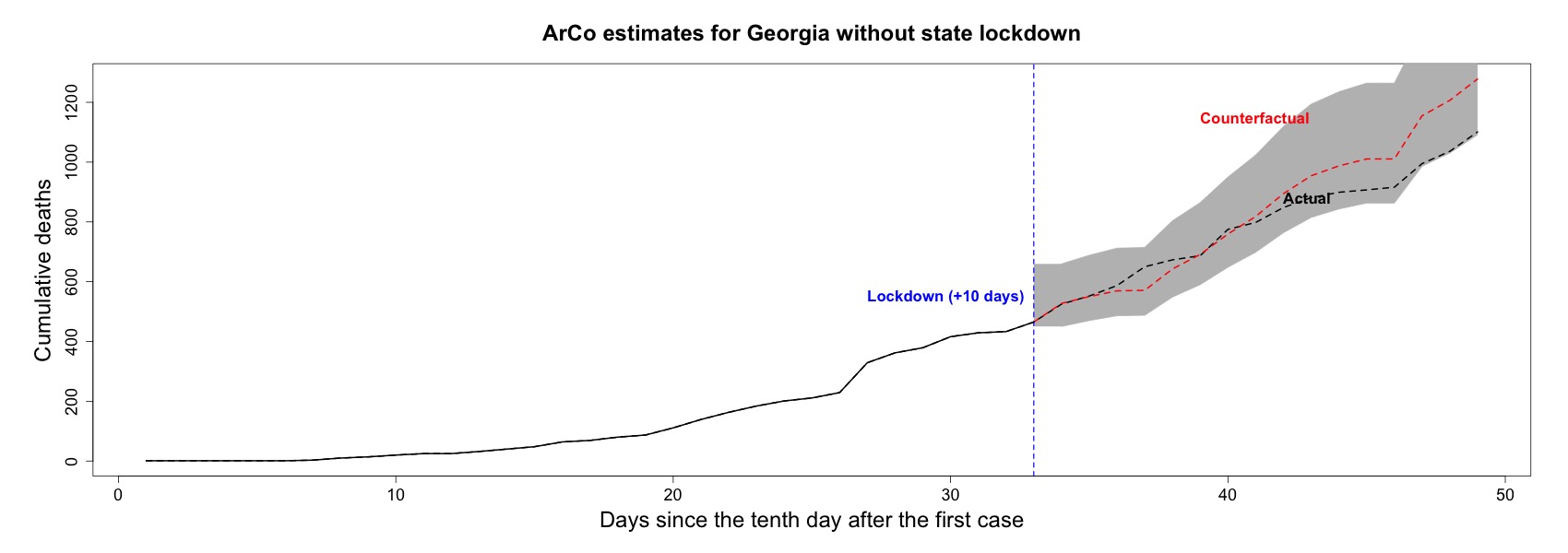}
    \label{fig:deaths_GA}
\end{figure}

\begin{figure}[!htbp]
    \centering
    \caption{ArCo Estimates for Kansas (Cumulative Deaths)}
    \includegraphics[width=\textwidth]{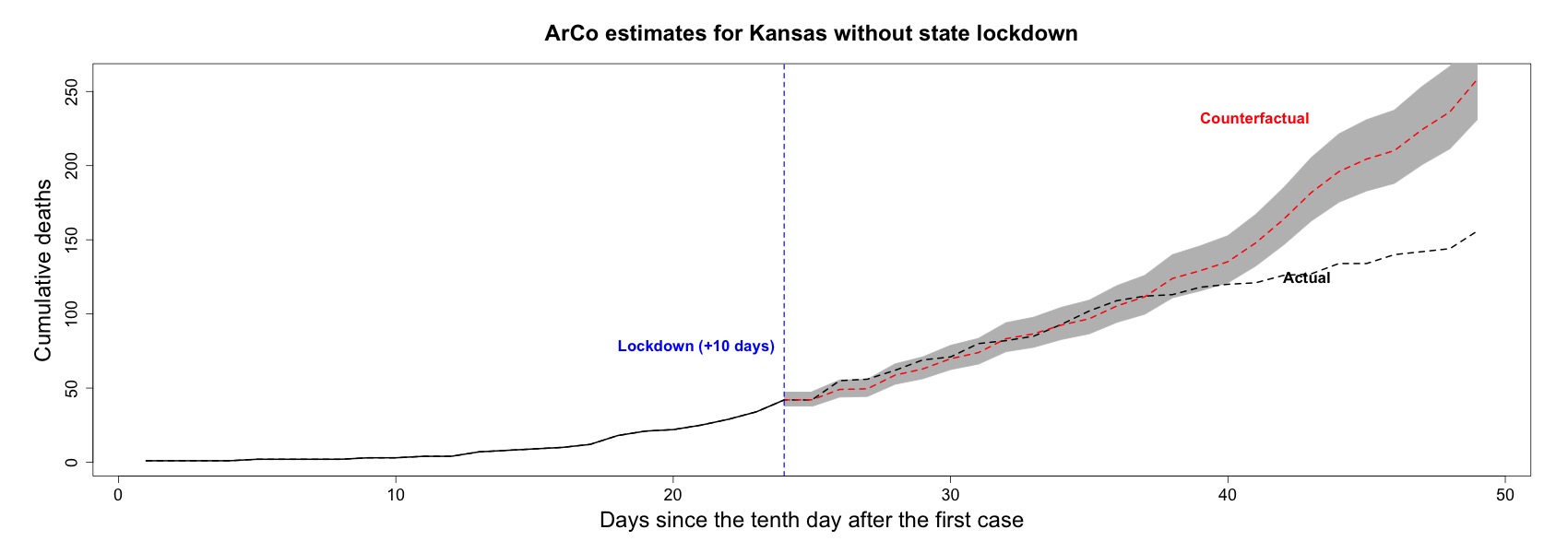}
    \label{fig:deaths_KS}
\end{figure}

\begin{figure}[!htbp]
    \centering
    \caption{ArCo Estimates for Kentuky (Cumulative Deaths)}
    \includegraphics[width=\textwidth]{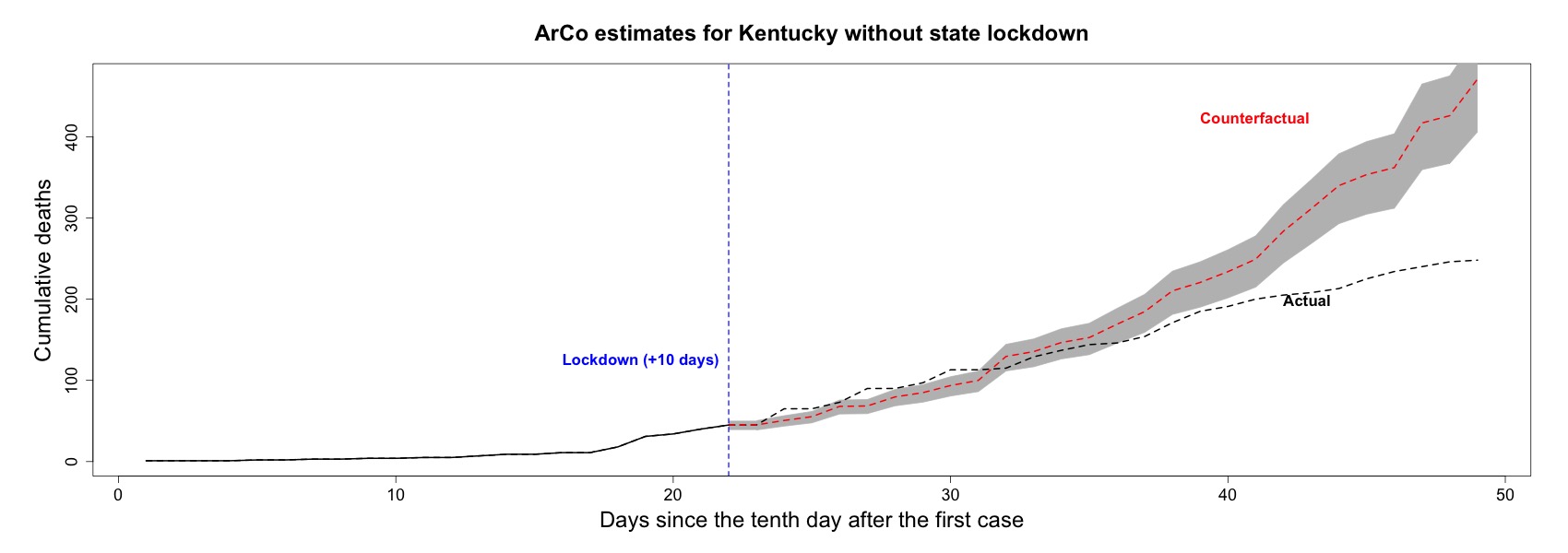}
    \label{fig:deaths_KY}
\end{figure}

\begin{figure}[!htbp]
    \centering
    \caption{ArCo Estimates for Maryland (Cumulative Deaths)}
    \includegraphics[width=\textwidth]{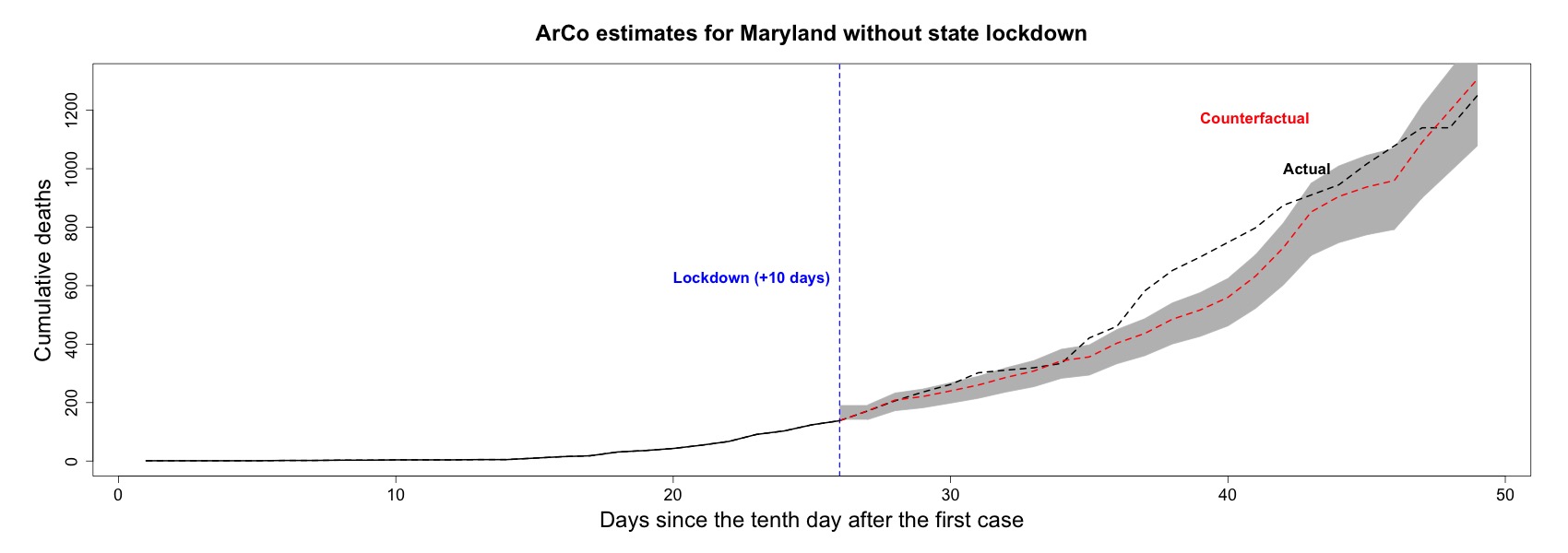}
    \label{fig:deaths_MD}
\end{figure}

\begin{figure}[!htbp]
    \centering
    \caption{ArCo Estimates for Maine (Cumulative Deaths)}
    \includegraphics[width=\textwidth]{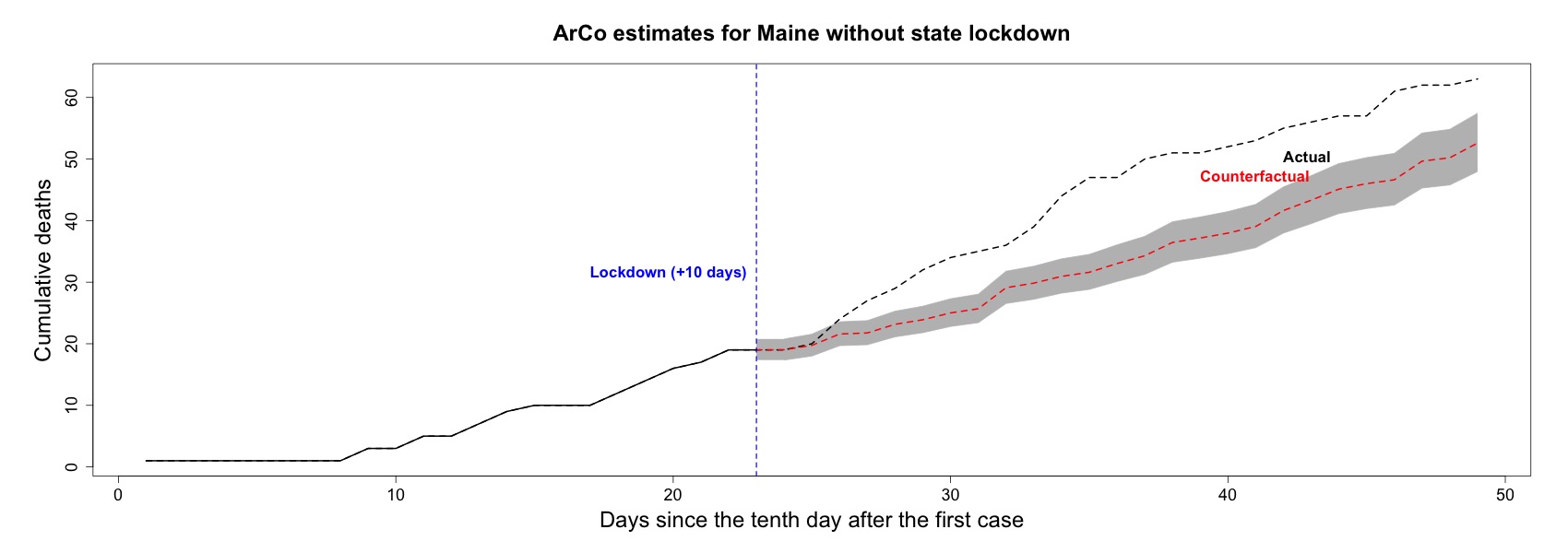}
    \label{fig:deaths_ME}
\end{figure}

\begin{figure}[!htbp]
    \centering
    \caption{ArCo Estimates for Missouri (Cumulative Deaths)}
    \includegraphics[width=\textwidth]{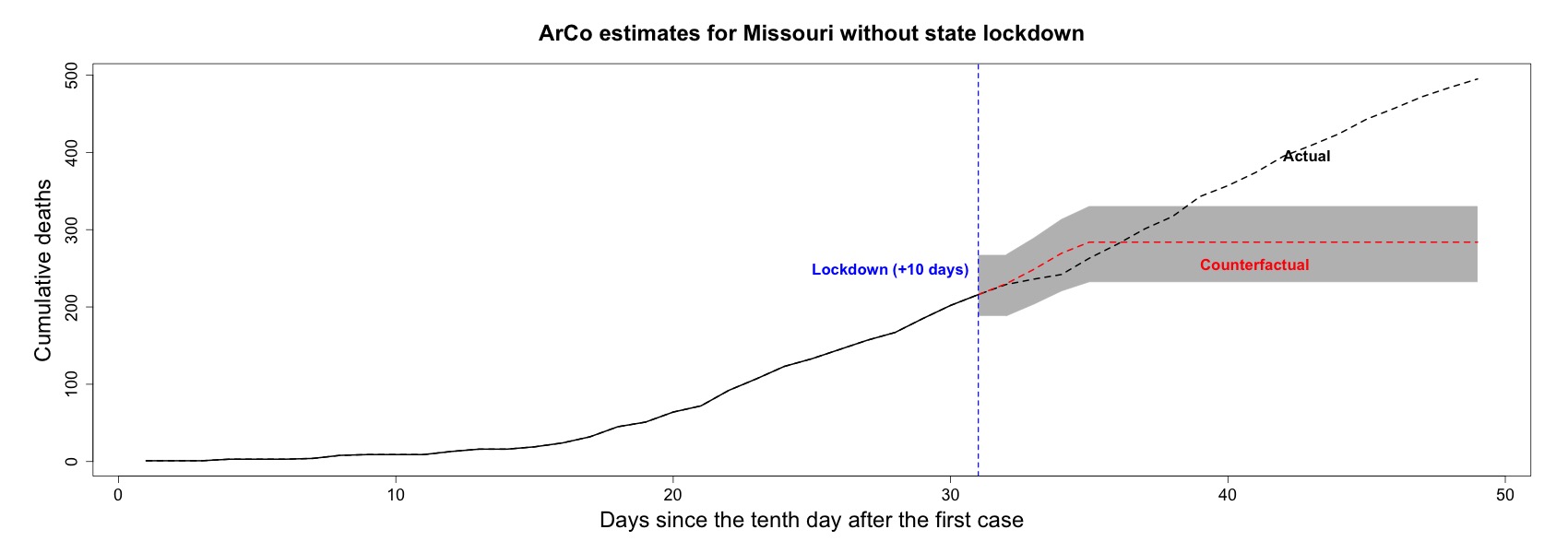}
    \label{fig:deaths_MO}
\end{figure}

\begin{figure}[!htbp]
    \centering
    \caption{ArCo Estimates for Mississipi (Cumulative Deaths)}
    \includegraphics[width=\textwidth]{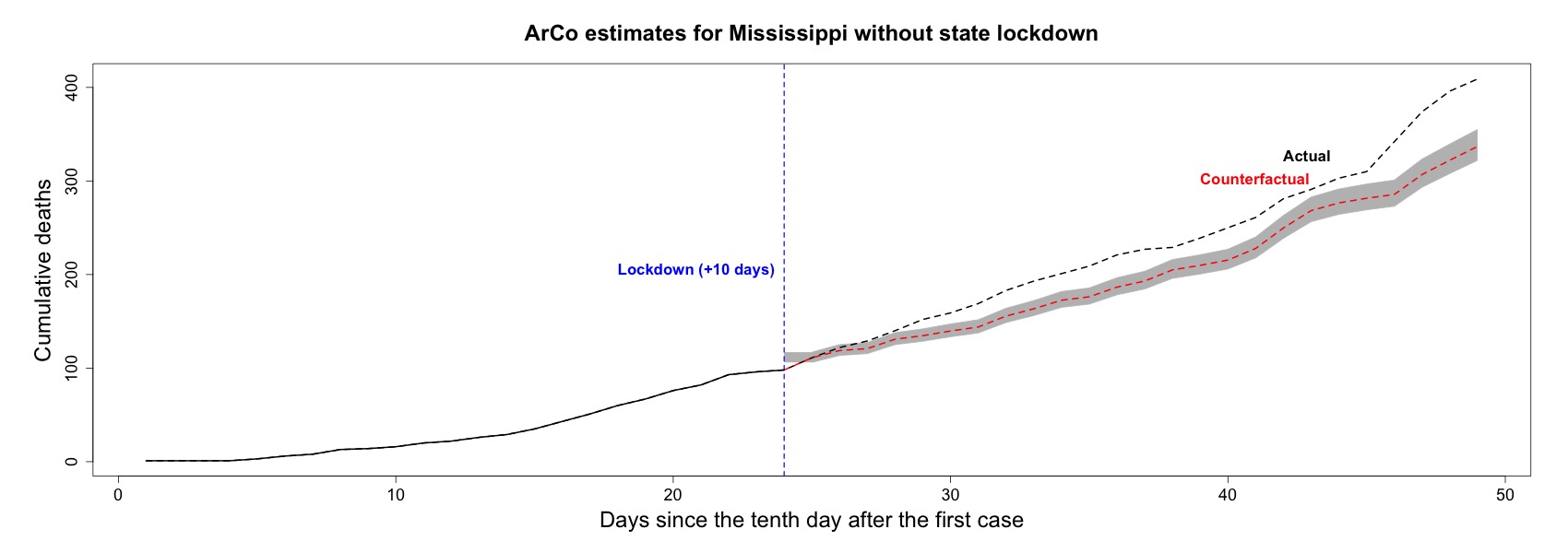}
    \label{fig:deaths_MS}
\end{figure}

\begin{figure}[!htbp]
    \centering
    \caption{ArCo Estimates for North Carolina (Cumulative Deaths)}
    \includegraphics[width=\textwidth]{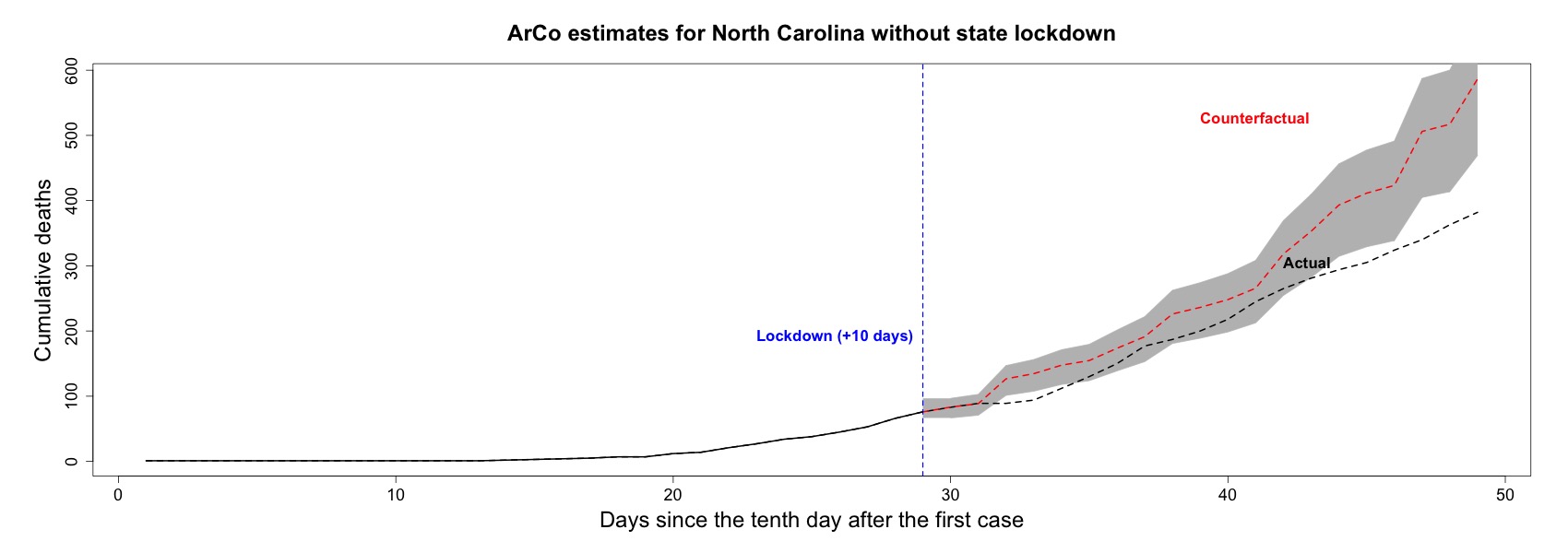}
    \label{fig:deaths_NC}
\end{figure}

\begin{figure}[!htbp]
    \centering
    \caption{ArCo Estimates for New Hampshire (Cumulative Deaths)}
    \includegraphics[width=\textwidth]{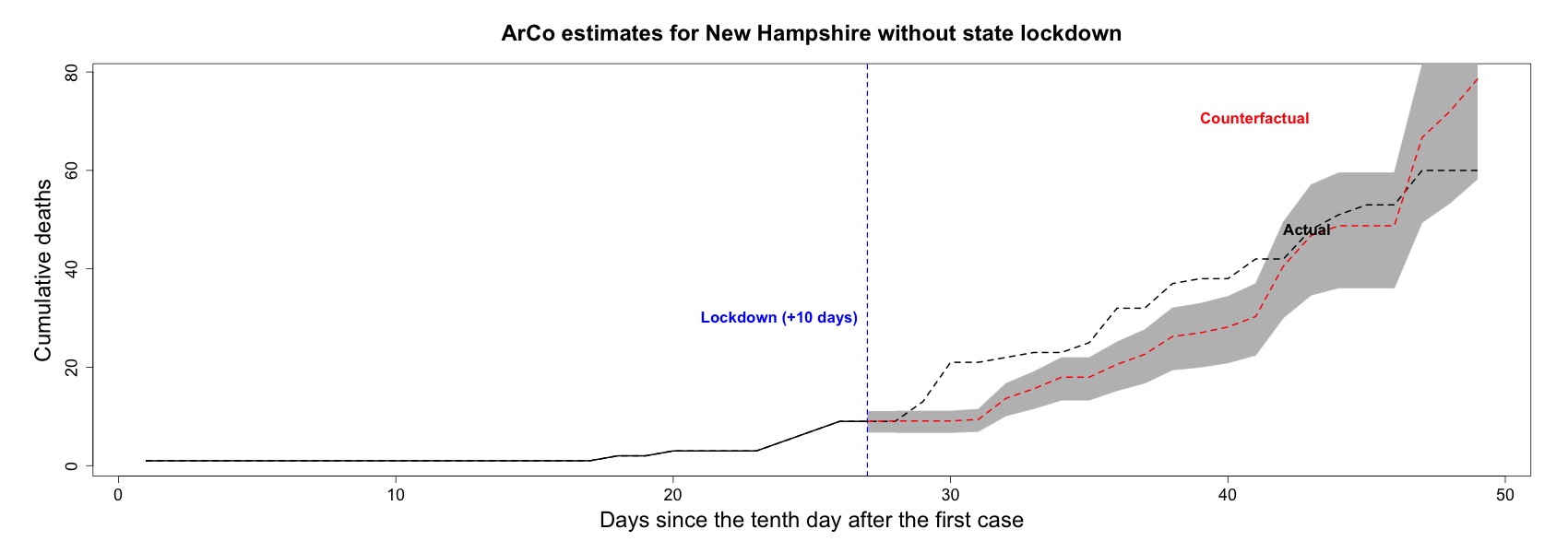}
    \label{fig:deaths_NH}
\end{figure}

\begin{figure}[!htbp]
    \centering
    \caption{ArCo Estimates for Nevada (Cumulative Deaths)}
    \includegraphics[width=\textwidth]{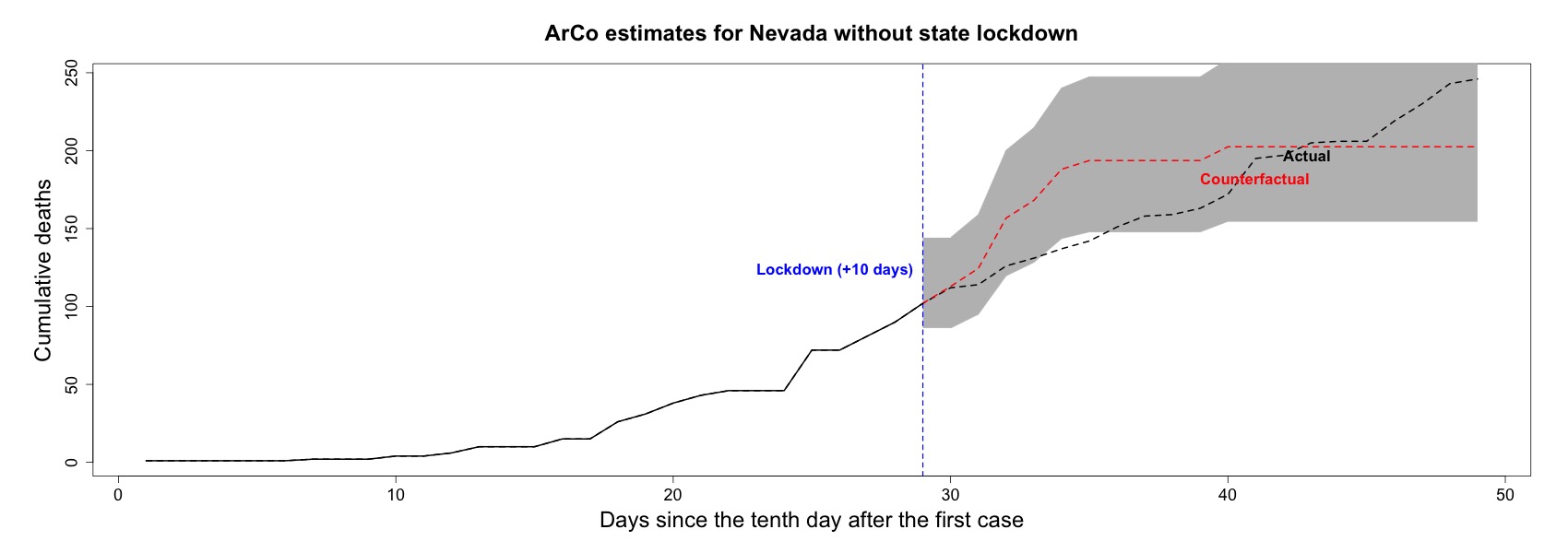}
    \label{fig:deaths_NV}
\end{figure}

\begin{figure}[!htbp]
    \centering
    \caption{ArCo Estimates for New York (Cumulative Deaths)}
    \includegraphics[width=\textwidth]{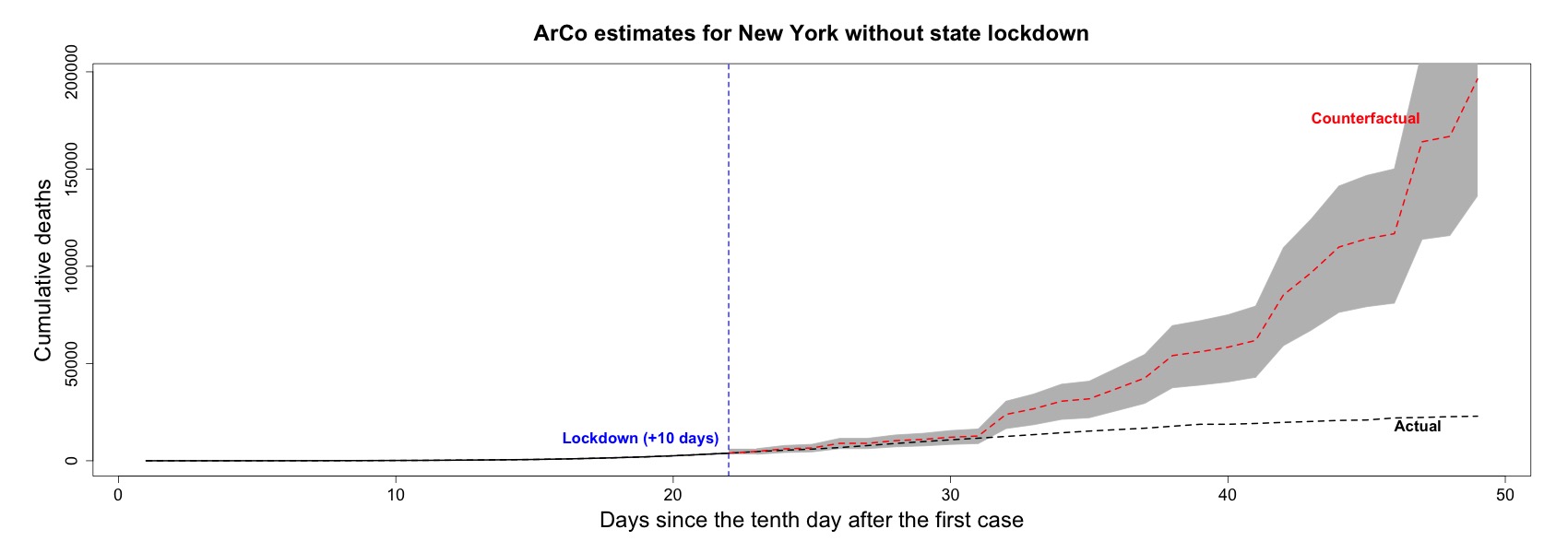}
    \label{fig:deaths_NY}
\end{figure}

\begin{figure}[!htbp]
    \centering
    \caption{ArCo Estimates for Oregon (Cumulative Deaths)}
    \includegraphics[width=\textwidth]{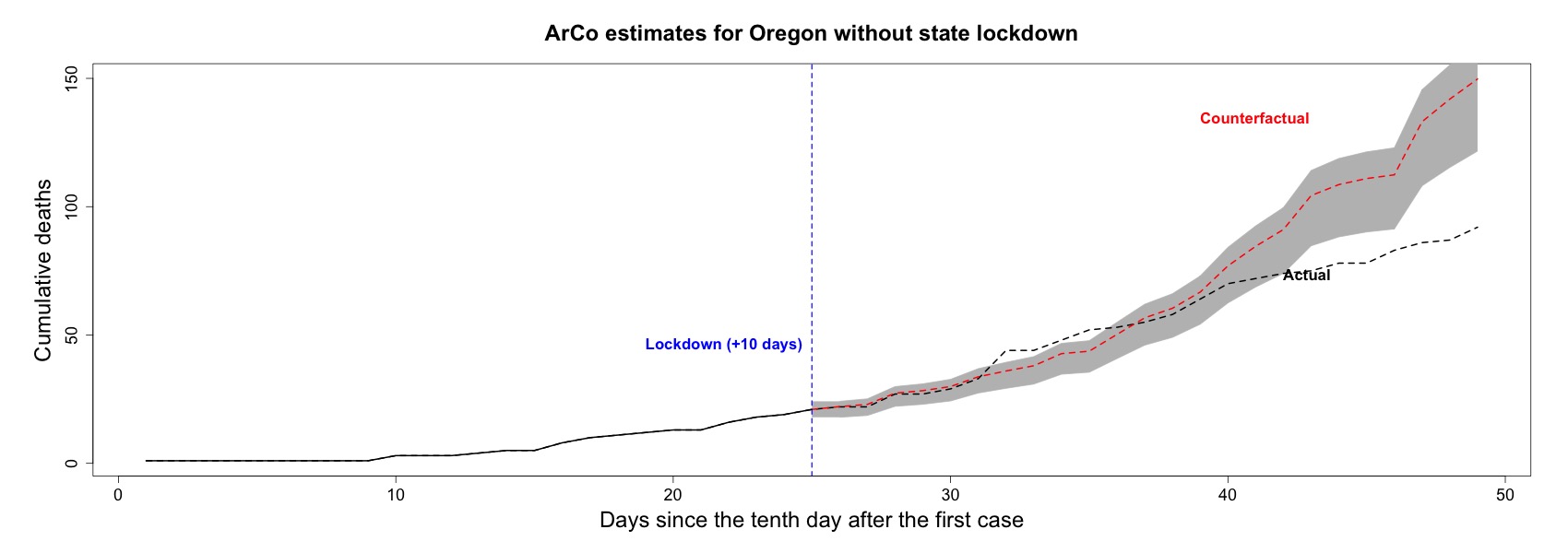}
    \label{fig:deaths_OR}
\end{figure}

\begin{figure}[!htbp]
    \centering
    \caption{ArCo Estimates for Pennsylvania (Cumulative Deaths)}
    \includegraphics[width=\textwidth]{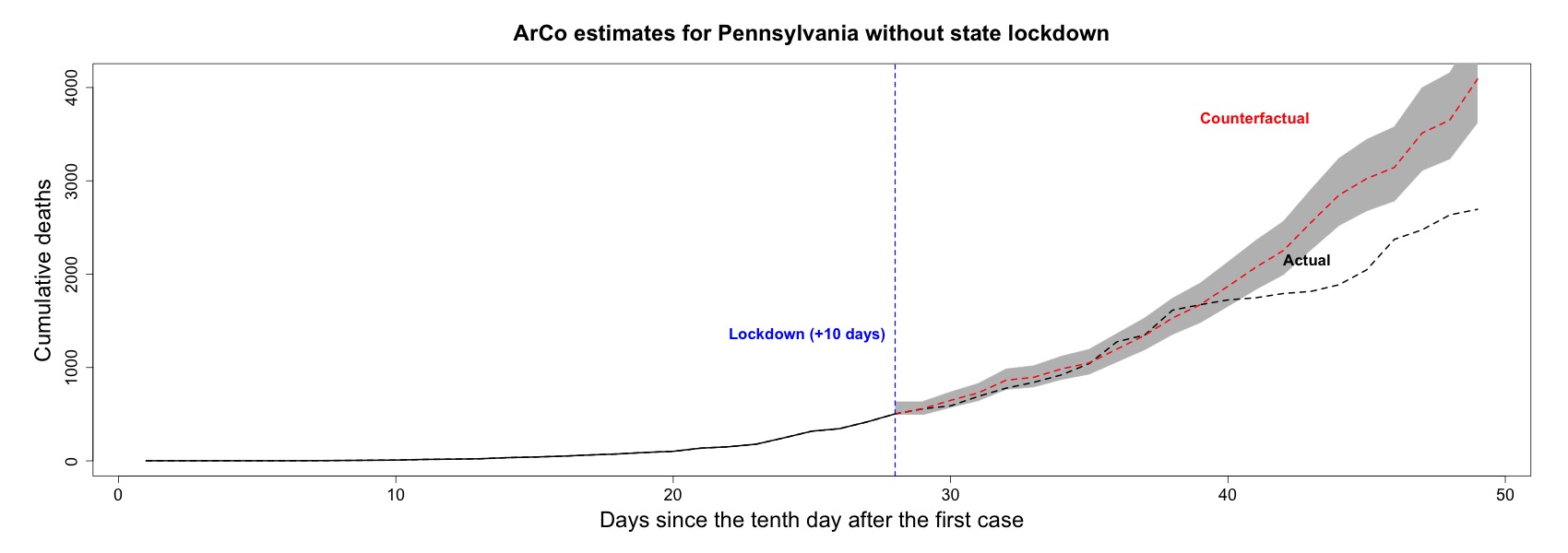}
    \label{fig:deaths_PA}
\end{figure}

\begin{figure}[!htbp]
    \centering
    \caption{ArCo Estimates for Rhode Island (Cumulative Deaths)}
    \includegraphics[width=\textwidth]{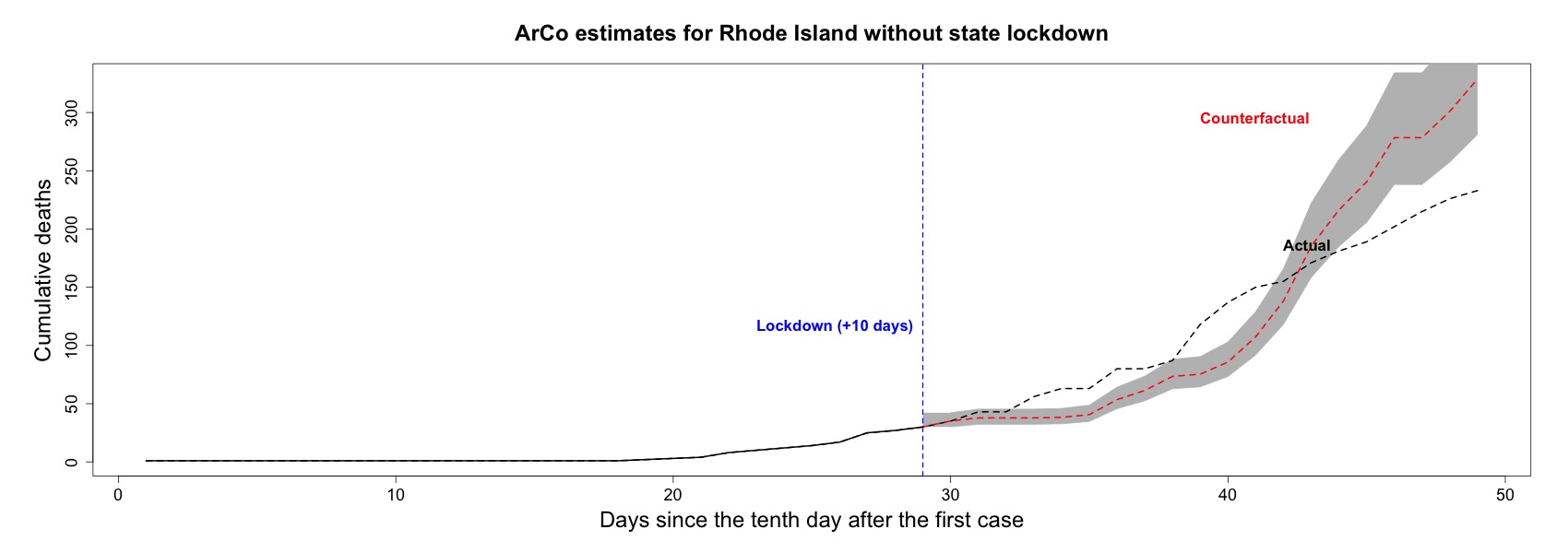}
    \label{fig:deaths_RI}
\end{figure}

\begin{figure}[!htbp]
    \centering
    \caption{ArCo Estimates for South Carolina (Cumulative Deaths)}
    \includegraphics[width=\textwidth]{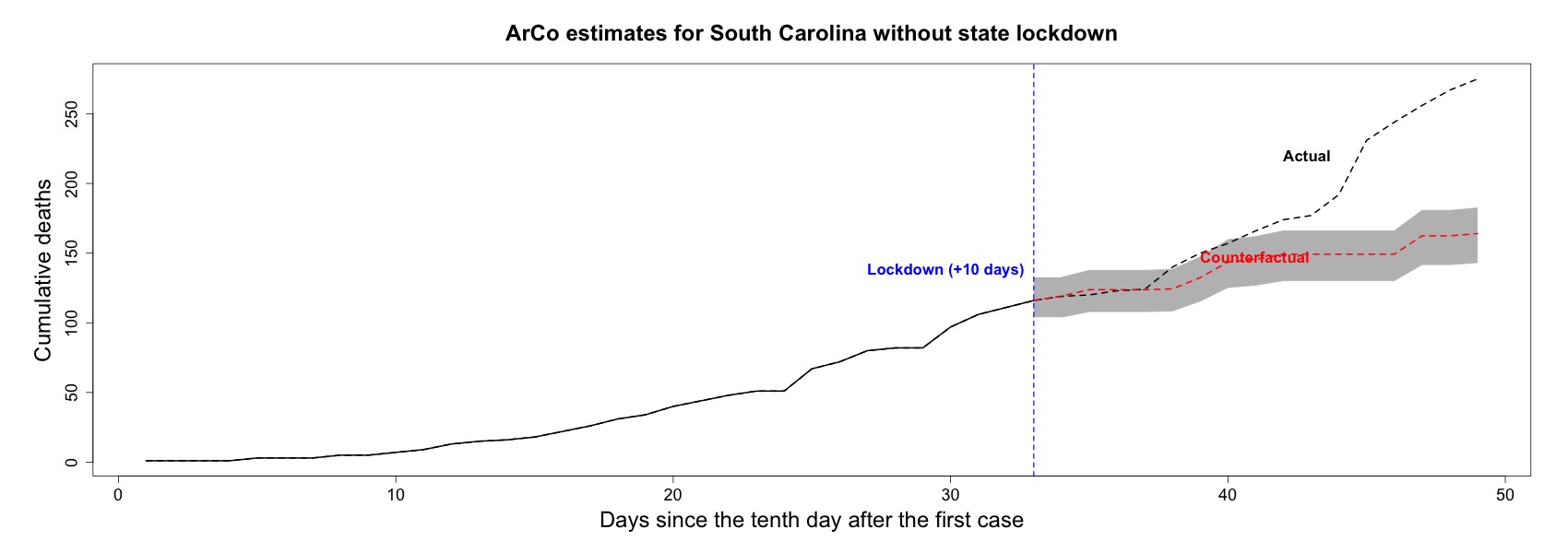}
    \label{fig:deaths_SC}
\end{figure}

\begin{figure}[!htbp]
    \centering
    \caption{ArCo Estimates for Tennessee (Cumulative Deaths)}
    \includegraphics[width=\textwidth]{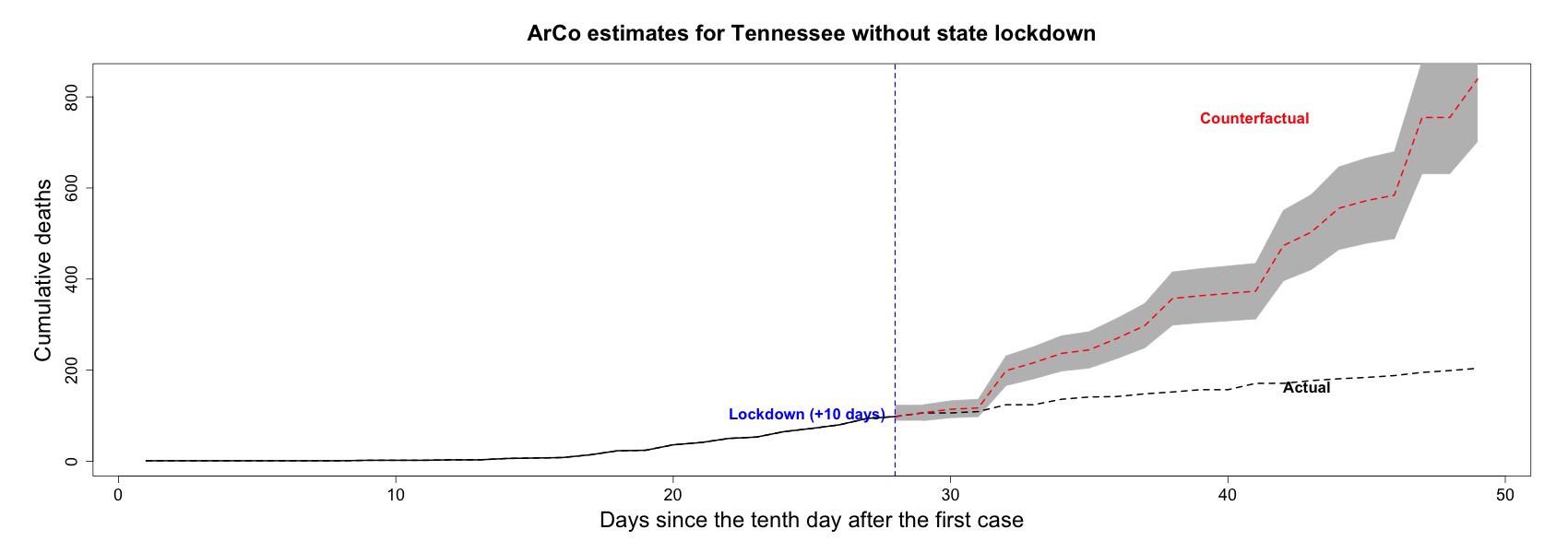}
    \label{fig:deaths_TN}
\end{figure}

\begin{figure}[!htbp]
    \centering
    \caption{ArCo Estimates for Texas (Cumulative Deaths)}
    \includegraphics[width=\textwidth]{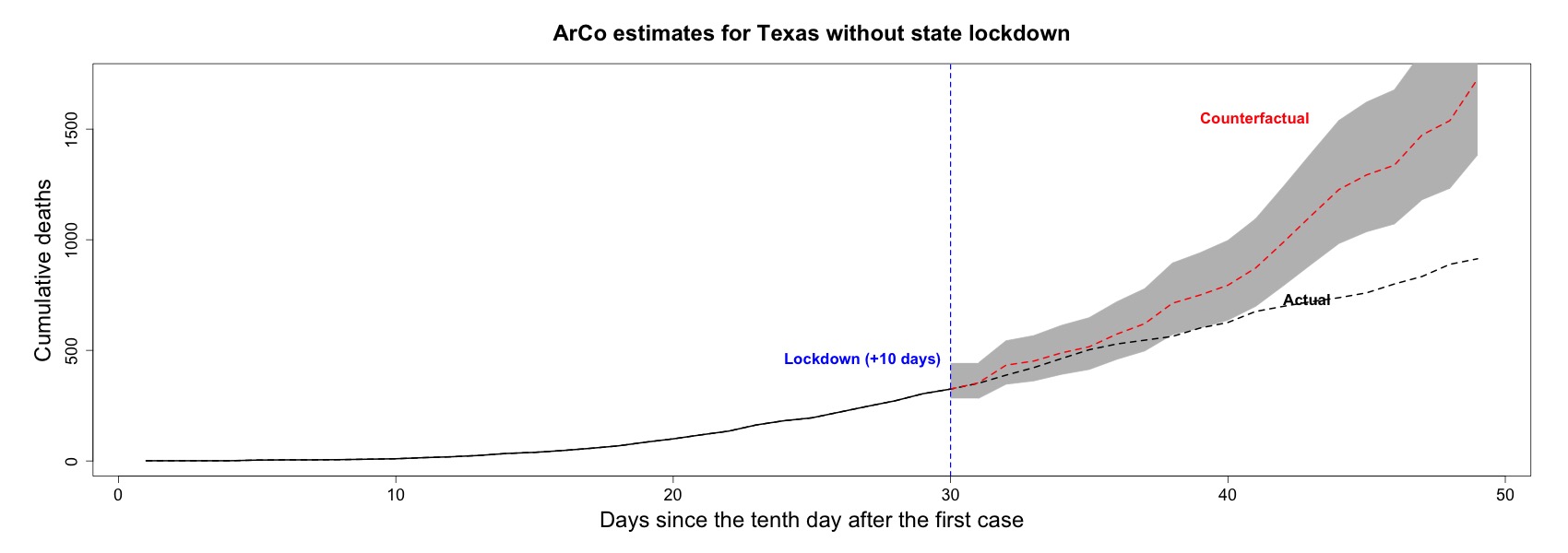}
    \label{fig:deaths_TX}
\end{figure}

In this section, in Figures \ref{fig:deaths_AL}--\ref{fig:deaths_TX}, we report the counterfactual estimates for cumulative deaths based on the methodology described in Section \ref{sec:counter_deaths}. As we discuss in the main text, although for some states, the counterfactuals exhibit similar shapes to those regarding cumulative cases, for many other states, they are not statistically significant at least for the first days after the policy implementation.

\newpage

\subsection{Google mobility data}
\label{ssec:google_mobility}

We know that lockdowns affect the Covid-19 dynamics by imposing social distancing and mobility restrictions. To help understand the results described in this paper, we analyze the mobility data available at Google Mobility Reports (https://www.google.com/covid19/mobility/).

Google mobility data show how visits and length of stay at different places change compared to a baseline, before the outbreak of the pandemic. In particular, the baseline is the median value, for the corresponding day of the week, during the five weeks between January 3rd and February 6th 2020.

In order to understand how the population in each group (treated and control states) is behaving during the Covid-19 crisis, we compute the median of mobility changes across our sample period, i.e. the 48 days following the tenth day after the first confirmed case in each state. Also, the data concern mobility changes for six categories, being five of them related to outdoor activities. Namely, grocery \& pharmacy, transit stations, parks, retail \& recreation, and workplaces. The remaining one concerns indoor activities, namely, residential.

\begin{figure}[H]
    \centering
    \includegraphics[trim={0 0 0 1cm}, clip, scale = 0.48]{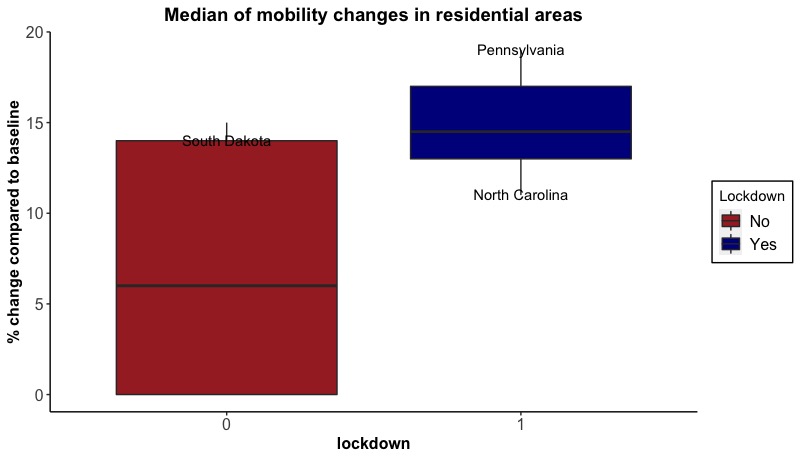}
    \caption{Median of mobility changes in residential areas}
    \label{fig:median_residential}
\end{figure}

\begin{figure}[H]
    \centering
    \includegraphics[trim={0 0 0 1cm}, clip, scale = 0.48]{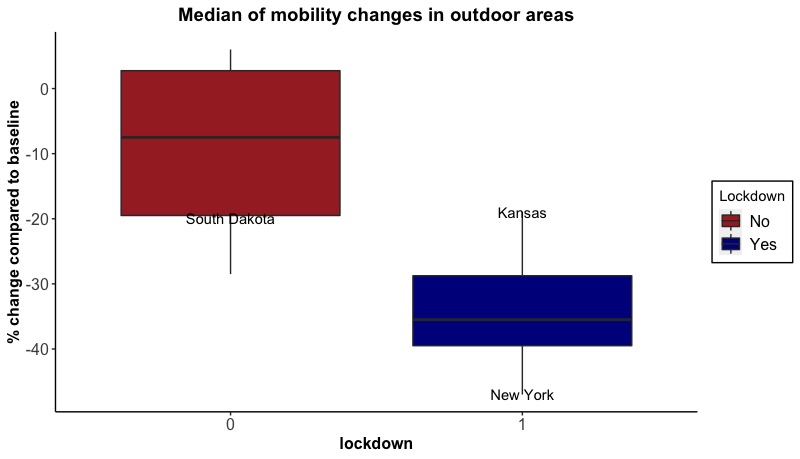}
     \caption{Median of mobility changes in outdoor areas}
    \label{fig:median_outdoors}
\end{figure}

Hence, to capture an idea of outdoor mobility changes, we aggregate the aforementioned five categories into a single one defined as the median of the original five categories. In contrast, mobility changes in residential areas capture indoor mobility changes. The two boxplots in Figures \ref{fig:median_residential} and \ref{fig:median_outdoors} present the median of mobility changes in all analyzed states both in residential and in outdoor areas, respectively. We report results for treated and control states separately.

Regarding mobility changes in residential areas, on average, residents from every state analyzed spent more time in these areas after the pandemic outbreak. However, those from treated states spent even more time indoor. Nevertheless, there are outliers. For instance, residents from South Dakota spent a lot more time in residential areas than before the pandemic, which helps understand the results found for this state in the placebo test.

We found similar results for mobility changes in outdoor areas. Clearly, residents from treated states remained in outside areas less often than residents from controls (always compared to the period before the pandemic). In New York, for example, there was a 50\% decrease of outdoor mobility. Once more,  South Dakota is an outlier for the control group, reinforcing the thesis that its population voluntarily decided to stay more at home. Indeed, residents from South Dakota  spent almost 20\% less time in outside areas, while those from the median state for the control group spent nearly 8\% less.

\end{document}